\documentclass[11pt]{article}
\pdfoutput=1
\usepackage{draft}
\usepackage{comment}
\usepackage[normalem]{ulem}
\usepackage{hyperref}
\hypersetup{
 colorlinks=true,
 urlcolor=blue,
 anchorcolor=blue,
 citecolor=blue,
 filecolor=blue,
 linkcolor=blue,
 menucolor=blue,
 pagecolor=blue,
 linktocpage=true,
 pdfa=true
}
\usepackage{graphicx,color,subfig}
\usepackage{cite}
\usepackage{empheq}

\DeclareFontEncoding{U}{}{}
\DeclareFontFamily{U}{bbold}{}
\DeclareFontShape{U}{bbold}{m}{n}
 {  <5> <6> <7> <8> <9> gen * bbold
   <10> <10.95> bbold10
  <12> <14.4> bbold12
 <17.28> <20.74> <24.88> bbold17
  }{}
\DeclareSymbolFont{bbold}{U}{bbold}{m}{n}
\DeclareSymbolFontAlphabet{\mathbbold}{bbold}

\usepackage[usenames,dvipsnames,table]{xcolor}
\graphicspath{{./figures/}}

\def\hol{holomorphic }
\def\rar{\rightarrow}
\def\ahol{anti-holomorphic }
\def\ab{\bar\alpha}
\def\b{\beta}
\def\cF{\mathcal{F}}

\def\d{\delta}
\def\c{\cite}
\def\a{\alpha}

\def\zb{\bar z}

\def\cN{\mathcal{N}}

\def\c{\cite}
\def\1{{\rm 1-loop}}

\def\o{\over}
\def\g{\gamma}
\def\D{\Delta}

\def\eqr{\eqref}
\def\O{{\cal O}}
\def\ra{\rangle}
\def\la{\langle}

\def\i{\infty}
\def\foot{\footnote}
\newcommand{\es}[2] {\begin{equation} \label{#1} \begin{split} #2 \end{split} \end{equation}}
\newcommand{\e}[2] {\begin{equation} \label{#1} #2 \end{equation}}

\newcommand{\NN}{\mathbbold{Z}_{\geq 0}}
\newcommand{\ZZ}{\mathbbold{Z}}
\newcommand{\RR}{\mathbbold{R}}
\renewcommand{\Re}{\operatorname{Re}}
\renewcommand{\Im}{\operatorname{Im}}
\DeclareMathOperator*{\Res}{Res}
\DeclareMathOperator*{\dRes}{dRes}
\newcommand{\kernel}{\mathbbold{S}}
\newcommand{\id}{\mathbbold{1}}
\DeclareMathOperator{\dDisc}{dDisc}
\newcommand{\op}{\mathcal{O}}
\DeclareMathOperator{\Tr}{Tr}

\def\ie{\begin{equation}\begin{aligned}}
\def\fe{\end{aligned}\end{equation}}

\begin{document}

\unitlength = .8mm

\begin{titlepage}
\rightline{ CALT-TH 2018-051}
\begin{center}

\hfill \\
\hfill \\
%\vskip 1cm

\title{Quantum Regge Trajectories\\and the Virasoro Analytic Bootstrap}

\author{Scott Collier,$^{\Upsilon_b}$ Yan Gobeil,$^{\gamma_b}$ Henry Maxfield,$^{\gamma_b,\Gamma_b}$ Eric Perlmutter$^{S_b}$}

\address{
$^{\Upsilon_b}$Jefferson Physical Laboratory, Harvard University, 
Cambridge, MA 02138, USA
\\
$^{\gamma_b}$Department of Physics, McGill University,
Montreal, QC H3A 2T8, Canada\\
$^{\Gamma_b}$Department of Physics, University of California,
Santa Barbara, CA 93106, USA\\
$^{S_b}$Walter Burke Institute for Theoretical Physics, Caltech,
Pasadena, CA 91125, USA
}

\email{scollier@g.harvard.edu, yan.gobeil@mail.mcgill.ca, hmaxfield@physics.ucsb.edu, perl@caltech.edu}

\end{center}

\abstract{Every conformal field theory (CFT)  above two dimensions contains an infinite set of Regge trajectories of local operators which, at large spin, asymptote to ``double-twist'' composites with vanishing anomalous dimension. In two dimensions, due to the existence of local conformal symmetry, this and other central results of the conformal bootstrap do not apply. We incorporate exact stress tensor dynamics into the CFT$_2$ analytic bootstrap, and extract several implications for AdS$_3$ quantum gravity. Our main tool is the Virasoro fusion kernel, which we newly analyze and interpret in the bootstrap context. The contribution to double-twist data from the Virasoro vacuum module defines a ``Virasoro Mean Field Theory'' (VMFT); its spectrum includes a finite number of discrete Regge trajectories, whose dimensions obey a simple formula exact in the central charge $c$ and external operator dimensions. We then show that VMFT provides a baseline for large spin universality in two dimensions: in every unitary compact CFT$_2$ with $c > 1$ and a twist gap above the vacuum, the double-twist data approaches that of VMFT at large spin $\ell$. Corrections to the large spin spectrum from individual non-vacuum primaries are exponentially small in $\sqrt{\ell}$ for fixed $c$. We analyze our results in various large $c$ limits. Further applications include a derivation of the late-time behavior of Virasoro blocks at generic $c$; a refined understanding and new derivation of heavy-light blocks; and the determination of the cross-channel limit of generic Virasoro blocks. We deduce non-perturbative results about the bound state spectrum and dynamics of quantum gravity in AdS$_3$.}

\vfill

\end{titlepage}

\eject

\begingroup
\hypersetup{linkcolor=black}

\renewcommand{\baselinestretch}{0.93}\normalsize
\tableofcontents
\renewcommand{\baselinestretch}{1.0}\normalsize

\endgroup

\section{Introduction and summary}

Recent years have seen enormous progress in understanding generic conformal field theories (CFTs), largely through conformal bootstrap methods. Some of the most powerful results exist at large $N$ or in an expansion in large spin, where analytic methods reveal features of the operator product expansion (OPE) that are universal to all CFTs. In two dimensions, the enhancement of spacetime symmetry to the infinite-dimensional Virasoro algebra would seem to aid efforts to analytically explore the space of CFTs and, via holography, the properties of AdS$_3$ quantum gravity. While this has proven true in certain kinematic and parametric limits, Virasoro symmetry has been frustratingly difficult to harness for theories deep in the irrational regime at finite central charge, without any small parameters.

In this paper, we will combine some of the maxims of recent analytic bootstrap studies in higher dimensions with the power of Virasoro symmetry to uncover universal properties of irrational CFTs at finite central charge, and their implications for AdS$_3$ quantum gravity. These results represent the complete, exact summation of the stress tensor contributions to certain OPE data. To do this, we will leverage the power of an underexploited tool, the Virasoro fusion kernel.

\subsection{Motivation by inversion}
A key conceptual and technical tool in the modern conformal bootstrap is the Lorentzian inversion formula for four-point functions \cite{Caron-Huot2017} (see also \cite{Simmons-Duffin2018a,Kravchuk2018}). This is a transform which acts on a four-point correlation function to extract the spectral data of intermediate states, providing an inverse to the conformal block expansion. The Lorentzian inversion formula (as opposed to a Euclidean inversion formula\cite{Karateev2018}) makes manifest the analyticity in spin of CFT OPE data: in particular, it shows that CFT operators live in analytic families -- ``Regge trajectories'' -- which asymptote at large spin to the double-twist (or higher multi-twist) operators whose existence was first discovered by the lightcone bootstrap \cite{Komargodski2013,Fitzpatrick2013,Fitzpatrick2014} (see \cite{Kaviraj:2015cxa, Kaviraj:2015xsa, Alday:2015ewa, Alday:2016njk, Simmons-Duffin2017} for subsequent developments prior to \cite{Caron-Huot2017}). In this way, the Lorentzian inversion formula goes beyond the large spin expansion, and implies a rigid structure of the OPE in any CFT.

Inversion may be carried out block-by-block, inverting T-channel conformal blocks to find the dual spectrum and OPE coefficients in the S-channel. These data are encoded in the poles and residues of the $6j$ symbol for the conformal group $SO(d+1,1)$, also known as the crossing kernel, as it decomposes blocks in one channel in the cross-channel \cite{Liu2018}. Inverting the contribution of the unit operator gives the OPE data of double-twist operators of Mean Field Theory (MFT), schematically of the form $:\!\op_i \Box^n \partial^\ell \op_j\!:$ with traces subtracted. In many cases of interest -- most notably, at large $N$ or in the lightcone limit -- the exchange of the unit operator parametrically dominates the correlation function in a particular channel. Including other T-channel operators gives corrections and additional contributions to this. Analysis in the lightcone limit implies that the spectrum of any $d>2$ (unitary, compact) CFT approaches that of MFT at large spin, with anomalous dimensions suppressed by inverse powers of spin $\ell^{-2h_t}$, where $h_t$ is the twist ($2h_t=\Delta_t-\ell_t$) of some T-channel operator. These are the double-twist Regge trajectories present in every CFT in $d>2$, dual to towers of two-particle states in AdS, which are well separated, and hence non-interacting, at large spin. 

What about two dimensions? From the above perspective, it is well-appreciated that two-dimensional CFTs are exceptional. Unitarity does not impose a gap in twist above the vacuum, and in particular, every operator in the Virasoro vacuum module -- namely, the stress tensor and its composites -- has zero twist. Since these operators all contribute to the large spin expansion at leading order, the analysis of higher dimensions is not valid. So the question remains, in a two-dimensional CFT at finite central charge, what do the double-twist Regge trajectories look like? Phrasing the same question in dual quantum gravitational language: in AdS$_3$ quantum gravity coupled to matter, what is the spectrum of two-particle states? Because the gravitational potential does not fall off at large distance in three dimensions, the interactions do not become weak, even at large spin. By solving this problem, one might also hope to infer lessons from the two-dimensional case for summing stress tensor effects on Regge trajectories in higher dimensions. In $d>2$, stress tensor dynamics are not universal, as the $TT$ OPE can contain arbitrary primaries consistent with the symmetries. Even when there is Einstein gravity in the IR thanks to a large higher spin gap in the CFT \cite{Heemskerk2009}, summing stress tensor dynamics to access Planckian processes in gravity is out of reach.

 It is not possible to solve this problem by global inversion of a Virasoro block because no simple expression is known for the block. Moreover, the Lorentzian inversion formula does not incorporate Virasoro symmetry, as its output is the OPE data for the primaries under the finite-dimensional global conformal algebra (quasiprimaries), rather than under the Virasoro algebra.\footnote{Here and throughout we take ``primary'' to mean Virasoro primary, and ``quasiprimary'' to mean $\mathfrak{sl}(2)$ primary.} While some perturbative results exist in which the stress tensor is treated as a quasiprimary \cite{Cardona:2018dov, Liu2018, Kraus:2018zrn, Gopakumar:2018xqi, Cardona:2018qrt, Sleight:2018epi, Sleight:2018ryu}, what we are after is the full incorporation of the two-dimensional stress tensor dynamics, which are completely determined by the Virasoro algebra, into the above picture.

Notwithstanding the important qualitative differences between two and higher dimensions, we will achieve this by adopting the same ``inversion'' strategy of branching T-channel conformal blocks into S-channel data. Remarkably, formulas for a Virasoro $6j$ symbol, the object that we refer to as the \emph{fusion kernel} (also called the crossing kernel), has been known for some time \cite{Ponsot1999,Ponsot2001,Teschner2001}. We will also apply the fusion kernel to various other problems in AdS$_3$/CFT$_2$, demonstrating its power and versatility.

\subsection{The Virasoro fusion kernel}
For this section, and the majority of the paper, we will focus on four-point functions of two pairs of operators $\op_1,\op_2$,
\begin{equation}
	\langle \op_1(0)\op_2(z,\bar{z})\op_2(1)\op_1(\infty) \rangle=\sum_s C_{12s}^2 \mathcal{F}_S(\alpha_s)\bar{\mathcal{F}}_S(\bar{\alpha}_s) = \sum_t C_{11t}C_{22t} \mathcal{F}_T(\alpha_t)\bar{\mathcal{F}}_T(\bar{\alpha}_t),
\end{equation}
where we have written the conformal block expansion in the S-channel, involving the $\op_1\op_2$ OPE, and the T-channel, involving the OPE of identical pairs of operators. In this expression, $\mathcal{F}(\alpha)$ denotes a holomorphic Virasoro block in the indicated channel, with intermediate holomorphic conformal weights labelled by $\alpha$, in a parameterisation that we will introduce presently. Dependence on the dimensions of external operators is suppressed, along with kinematic dependence on locations of operators. In a diagonal basis $\langle\op_i\op_j\rangle\propto \delta_{ij}$ (possible given our assumption that the theory is compact), the identity operator will appear in the T-channel.

As we will soon see, it is natural to express the central charge $c$ and conformal weight $h$ in terms of parameters $b$ and $\alpha$, respectively, as
\begin{equation} \label{params}
	c=1+6Q^2,\quad Q=b+b^{-1},\quad h(\alpha)=\alpha(Q-\alpha),
\end{equation}
along with antiholomorphic counterpart $\bar{\alpha}$ related in the same way to $\bar{h}=h+\ell$, where $\ell$ is the spin. We call $\alpha$ the ``momentum,'' in analogy with the terminology of the Coulomb gas or linear dilaton theory, or Liouville theory, for which $\alpha$ is related to a target-space momentum (perhaps most familiar from vertex operators of the free boson at $c=1$). For irrational unitary theories with $c>1$ and $h,\bar{h}\geq 0$ for all operators, highest-weight representations of the Virasoro algebra fall into two qualitatively different ranges, depending on whether $h$ lies above or below the threshold $\left(\frac{Q}{2}\right)^2=\frac{c-1}{24}$. For $h<\frac{c-1}{24}$, we can choose $0<\alpha<\frac{Q}{2}$, but for $h>\frac{c-1}{24}$, we have complex $\alpha \in \frac{Q}{2}+i\RR$; we will denote these the `discrete' and `continuum' ranges, respectively, for reasons that will become clear imminently. 

The fusion kernel, which we denote by $\kernel_{\alpha_s\alpha_t}$, decomposes a T-channel Virasoro block in terms of S-channel blocks:
\begin{equation}\label{fusionKernel}
	\mathcal{F}_T (\alpha_t) = \int_C \frac{d\alpha_s}{2i} \kernel_{\alpha_s\alpha_t} \mathcal{F}_S(\alpha_s)
\end{equation}
We will use an explicit closed-form expression for the kernel $\kernel_{\alpha_s\alpha_t}$ due to Ponsot and Teschner \cite{Ponsot1999,Ponsot2001}, presented in the next section. Apart from early work \cite{Teschner2001} that used the kernel to show that Liouville theory solves the bootstrap equation through the DOZZ formula \cite{Dorn1994a,Zamolodchikov1996}, its utility for analytic conformal bootstrap has only recently been appreciated \cite{Jackson:2014nla,Chang2016,Chang2016b,Esterlis2016,He2017,Mertens2017}.

\subsubsection*{Support of the fusion kernel}

Without giving the explicit formula for the kernel itself, let us briefly summarize an important property for our purposes, namely its support, i.e.\ the set of S-channel representations that appear in the decomposition of a T-channel block. 

If the external operators are sufficiently heavy that $\Re(\alpha_1+\alpha_2)> \frac{Q}{2}$, then the contour $C$ in \eqref{fusionKernel} can be chosen to run along the vertical line $\alpha_s=\frac{Q}{2}+i\RR$, meaning that the T-channel block is supported on S-channel blocks in the continuum range of $\alpha_s$:
\begin{equation}\label{intblocks}
	\mathcal{F}_T (\alpha_t) = \int_0^\infty dP \; \kernel_{\alpha_s\alpha_t} \mathcal{F}_S(\alpha_s=\tfrac{Q}{2}+iP)
\end{equation}
However, $\kernel_{\alpha_s\alpha_t}$ is a meromorphic function of $\alpha_s$, and has poles at
\begin{equation}
	\alpha_m := \alpha_1+\alpha_2+ m b
\end{equation}
that, for sufficiently light external operators, may cross the contour $\alpha_s=\frac{Q}{2}+i\RR$. In the case of unitary external operator weights, when $\alpha_1+\alpha_2< \frac{Q}{2}$ ($\alpha_i$ necessarily real, in the `discrete' range), the integral acquires additional contributions given by the residues of these poles:
\begin{equation}\label{sumintblocks}
	\mathcal{F}_T (\alpha_t) = \sum_m\; -2\pi \mkern-6mu \Res_{\alpha_s= \,\alpha_m} \left\{\kernel_{\alpha_s\alpha_t} \mathcal{F}_S(\alpha_s)\right\} + \int_0^\infty dP \; \kernel_{\alpha_s\alpha_t} \mathcal{F}_S(\alpha_s=\tfrac{Q}{2}+iP)
\end{equation}
The sum runs over nonnegative integers $m$ such that the location of the corresponding pole satisfies $\a_m<\frac{Q}{2}$.\foot{If $c\leq25$, for which $b$ is a pure phase, only the $m=0$ term can be present; this ensures the reality of various results to follow. For $c>25$ we have chosen the convention $0<b<1$.} The distinct ways in which light and heavy operators appear in this decomposition is why we have dubbed these ranges of dimensions `discrete' and `continuous' respectively. 

For $\alpha_t=0$, corresponding to the exchange of the identity multiplet, $\kernel_{\alpha_s\id}$ has simple poles, so the blocks $\mathcal{F}_S(\a_m)$ appear with coefficient given by the residue of the kernel. In other cases ($\alpha_t\neq 0$), they are instead double poles, so the residue of the kernel times the block includes a term proportional to the derivative of the block with respect to $\alpha_s$, evaluated on the pole.

\subsection{Summary of physical results}

The main technical work of analysing the fusion kernel using the integral formula is in section \ref{kernelAnalysis}. This includes its analytic structure and residues of poles contributing to \eqref{sumintblocks}; a closed-form, non-integral expression for T-channel vacuum exchange, given in \eqr{vackernel}; and various pertinent limits of the kernel. This also leads straightforwardly to a derivation of the cross-channel behavior of Virasoro blocks. Subsequent sections will use the resulting formulas for various physical applications to CFT$_2$ and AdS$_3$ quantum gravity, the main results of which we now summarise.

\subsubsection*{Quantum Regge trajectories and stress tensor corrections to MFT}

In every compact CFT, the vacuum Verma module contributes to the T-channel expansion of the four-point function under consideration. Inverting the vacuum Virasoro block by taking $\a_t=0$, the kernel $\kernel_{\alpha_s \id}$ (times its antiholomorphic counterpart) is the corresponding ``OPE spectral density'' of Virasoro primaries in the S-channel, henceforth referred to simply as the spectral density. This spectral density is the finite central charge deformation of MFT double-twist data which takes into account contributions from all Virasoro descendants of the identity. This motivates the term ``Virasoro Mean Field Theory'' (VMFT) to refer to the OPE data resulting from inversion of the Virasoro vacuum block (including both left- and right-moving halves). Note that, unlike the case for MFT, this does not correspond to a sensible correlation function; in particular the spectrum of VMFT is continuous and contains non-integral spins, so VMFT refers only to a formal, though universal, set of OPE data. The results summarised here are discussed in more detail in section \ref{VMFT}

\begin{figure}[t]
\includegraphics[scale=0.54]{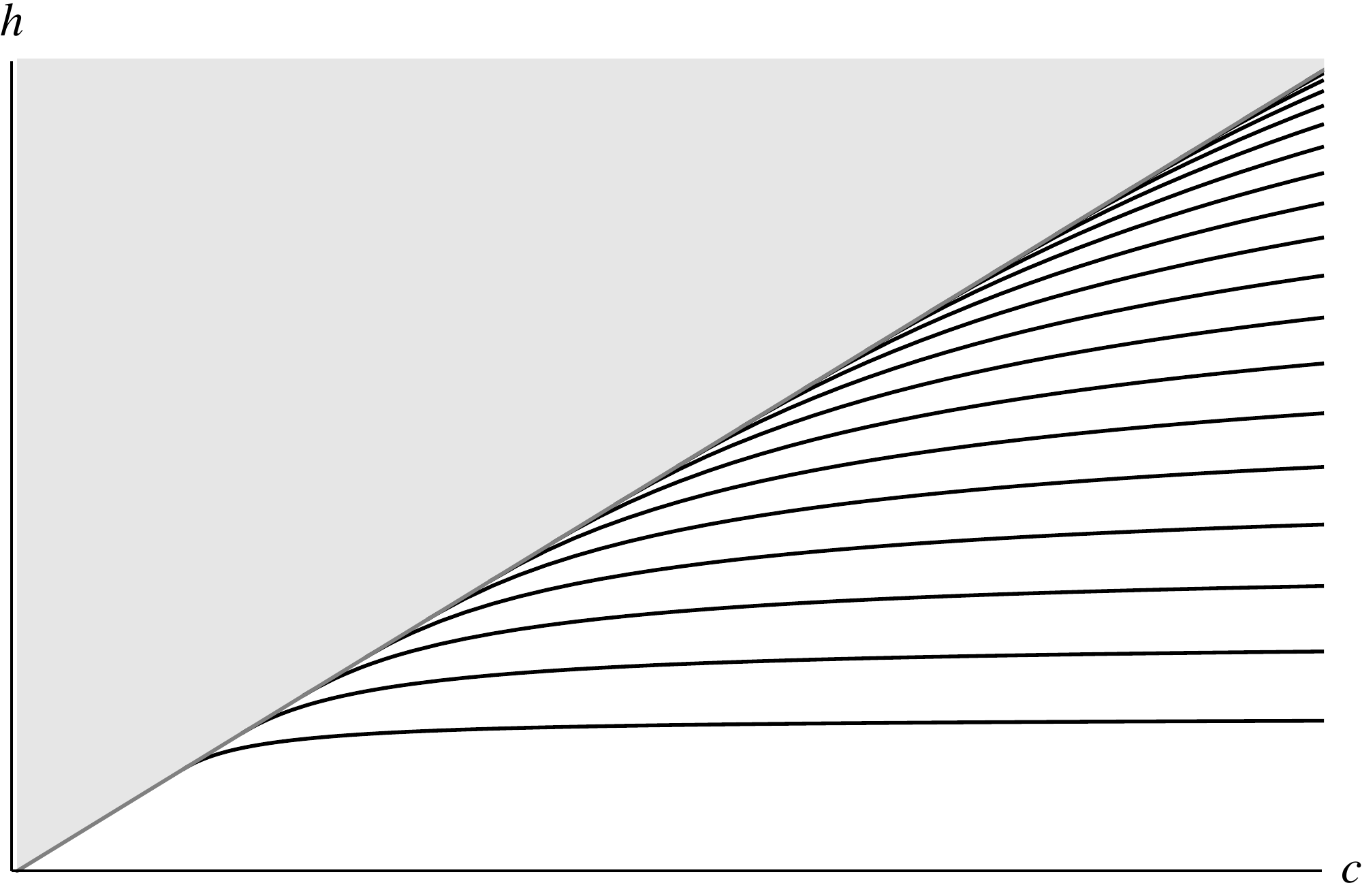}
\centering
\caption{The twist spectrum of Virasoro double-twist operators arising from inversion of the \hol vacuum block, i.e.\ the spectrum of VMFT. The twists of the external operators are fixed while $c$ varies. For a given central charge $c$, there is a discrete number of operators, shown as the solid lines, below the continuum at $h>\frac{c-1}{24}$, shaded in grey. The exact formula for the twists is given in \eqr{anomdim}. Each line, when combined with an anti-\hol component, forms a quantum Regge trajectory that is exactly linear in spin. At large $c$, one recovers the integer-spaced operators of MFT.\label{fig:regge}}
\end{figure}

Because the vacuum block factorises, we can describe the spectrum in terms of chiral operators, with the understanding that the full two-dimensional spectrum is obtained from products of holomorphic and antiholomorphic operators. This chiral spectrum is the support of the fusion kernel described above. This departs qualitatively from the infinite tower of evenly spaced Regge trajectories in MFT. There is a \emph{finite} discrete part of the spectrum, coming from the poles in \eqref{sumintblocks}, and a continuous spectrum above $h=\frac{c-1}{24}$. The discrete contributions appear at
\begin{equation}\label{eq:VirasoroDoubleTwist}
\alpha_m = \alpha_1 + \alpha_2 + m b \, , \quad \text{for } m=0,1,\ldots, \left\lfloor b^{-1}\left(\tfrac{Q}{2}-\alpha_1-\alpha_2\right)\right\rfloor.
\end{equation}
For the discrete trajectories, inclusion of the Virasoro descendants therefore has the remarkably simple effect that the MFT additivity of dimensions $h$ is replaced by additivity of momenta $\alpha$. 
The dependence of this spectrum on central charge is pictured in figure \ref{fig:regge}.

Writing the result \eqref{eq:VirasoroDoubleTwist} in terms of the twist\footnote{In what follows we will use the term `twist' (referring to $\tau = \Delta - \ell = 2\min(h,\bar h)$) almost interchangeably with the holomorphic conformal weights $h$. Likewise, we refer to $\delta h$ as the `anomalous twist'.}, we find
\begin{equation}\label{anomdim}
h_m = h_1 + h_2+ m + \delta h_m \, ,\quad \text{where}\quad \delta h_m = -2(\alpha_1+mb)(\alpha_2+mb)+m(1+m)b^2<0.
\end{equation}
The departure $\delta h_m$ from the corresponding MFT dimension is the exact anomalous twist due to summation of all multi-traces built from the stress tensor. Note that if we take $c\to\infty$ ($b\to 0$) with $h_{1,2}$ fixed, $\delta h_m$ goes to zero, and the maximal value of $m$ goes to infinity, recovering MFT. In addition, $\delta h_m$ is always negative, so the twist is reduced when compared to those of MFT, seen by the monotonicity in figure \ref{fig:regge}.

Including both chiral halves, the discrete set of twists \eqref{anomdim} form what we call ``quantum Regge trajectories,'' so named to reflect the finite-$c$ summation and their duality to two-particle states in AdS$_3$ quantum gravity with $G_N$ finite in AdS units, as we discuss momentarily. These trajectories are exactly linear in spin. We emphasize the distinction with the analogous problem in $d>2$, where two-particle dynamics at Planckian energies remain inaccessible.

The data of VMFT is modified by inclusion of other, non vacuum operators in the T-channel. The spectrum is shifted by ``anomalous twists'', coming from the double poles in the fusion kernel. These anomalous twists from individual operators, as well as anomalous OPE coefficients, can be formally computed from the coefficients of these poles, with the result given in \eqref{deltaa0}.  This is much the same as inversion of global conformal blocks for non-unit operators, which give corrections to MFT.

\subsubsection*{Large spin universality}

The inversion of the T-channel vacuum block, giving the spectrum of VMFT discussed above, is rather formal and not immediately obvious how it relates to physical data of actual CFTs, but in some limits it is in fact universal. Just as MFT governs the large spin OPE of $d>2$ CFTs, there is a VMFT universality governing the spectrum of $d=2$ CFTs:
\begin{quote}
\begin{em}
	In a unitary compact CFT$_2$ with $c>1$ and a positive lower bound on twists of non-vacuum primaries, the OPE spectral density approaches that of VMFT at large spin.
\end{em}
\end{quote}
This is made more precise in section \ref{sec:largeSpin}. In particular, it means that there are Regge trajectories of double-twist operators with twist approaching the discrete values in \eqref{anomdim} at large spin, and that Regge trajectories with twist approaching a different value of $h<\frac{c-1}{24}$ have parametrically smaller OPE coefficients. The continuum for $h>\frac{c-1}{24}$ requires an infinite number of Regge trajectories with $h$ accumulating to $\frac{c-1}{24}$ at large spin, such that any given interval of twists above this value contains an infinite number of operators with spectral density approaching that of VMFT.

 We also compute the rate at which the asymptotic twist is approached by including an additional T-channel operator with momenta $(\alpha_t,\bar\a_t)$, given by the formula \eqref{nonholope}. At large spin $\ell = \bar{h}-h$, this scales as\footnote{Throughout this paper, we use the notation $x\sim y$ to denote that ${x\over y}\to 1$ in the limit of interest, while we use $x\approx y$ to specify the leading scaling, denoting in particular that $|\log(x)-\log(y)|$ does not grow in the relevant limit.}
 \begin{equation}\label{deltahlargeell}
	\delta h_m^{(\a_t,\bar\a_t)} \approx \exp\left({-2\pi \bar\alpha_t \sqrt{\ell}}\right).
\end{equation}
This decays much faster than the power-law suppression $\ell^{-2\bar{h}_t}$ one obtains in $d>2$, or by ignoring the stress tensor in $d=2$ (for example, in computing correlation functions of a QFT in a fixed AdS$_3$ background). We also compute $\delta h_m^{(\a_t,\bar\a_t)}$ to leading order in a semiclassical regime of ``Planckian spins'', taking $\ell\to\infty $ and $c\to\infty$ with $\ell\o c$ and external operator dimensions fixed. The result, given in \eqr{eq:PlanckianSpinGeneral}, has the simple dependence
\begin{equation}
	\delta h_m^{(\alpha_t,\bar\alpha_t)}\propto \left({c\over 6\pi}\cosh(\pi \bar p_s)\right)^{-2\bar h_t}, \quad \text{with }\quad \bar p_s \sim {1\over 2}\sqrt{{24\ell\over c}-1}~.
\end{equation}
This interpolates between the exponential behavior \eqr{deltahlargeell} at $\ell\gg c$ and the power $\ell^{-2\bar h_t}$ at $\ell\ll c$, where the latter is the original result of the lightcone bootstrap \cite{Komargodski2013,Fitzpatrick2013}. We give this a gravitational interpretation in section \ref{sec:anomTwistsGravity}.

For T-channel exchanges between identical operators $\O_1=\O_2$ obeying $\a_t<2\a_1$, upon including the coefficient in \eqr{deltahlargeell}, the anomalous twist of the leading Regge trajectory $\delta h_0^{(\a_t,\bar\a_t)}$ is negative. This can be thought of as a Virasoro version of Nachtmann's theorem \cite{Nachtmann1973,Komargodski2013}: the leading large spin correction to the twist of the first Regge trajectory is negative, so this trajectory is convex. 

Because the arguments above ultimately rely on the form of the {\it holomorphic} fusion kernel, they  imply similar results about OPE asymptotics in limits of large conformal weight, rather than large spin. As an explicit application of this, in section \ref{sec:largedim} we give the asymptotic average density of ``light-light-heavy'' OPE coefficients in any unitary compact 2D CFT with $c>1$ and a dimension gap above the vacuum. 

\subsubsection*{Cross-channel Virasoro blocks}
While we derive these results purely from the crossing kernel, without direct reference to the four-point function itself, we can relate this to methods of the original derivations of the lightcone bootstrap \cite{Fitzpatrick2013,Komargodski2013}, which solved crossing in the lightcone limit $z\to 1$ of the correlation function. See appendix \ref{app:OldLC} for a review of the `old-fashioned' lightcone bootstrap. This requires analysis of the `cross-channel' limit of Virasoro blocks, which we provide in section \ref{sec:crosschannel} for both S- and T-channel blocks.

We note that the lightcone bootstrap is one example of a large class of arguments determining asymptotic OPE data from dominance of the vacuum in a kinematic limit, and crossing symmetry or modular invariance \cite{Kraus2017a,Cardy2017,Das2017c,Das2017}. The same strategy of using an appropriate fusion or modular kernel could be applied to streamline these arguments. For this purpose, we note that there also exists a modular kernel for primary one-point functions on the torus \cite{Ponsot1999,Ponsot2001,Teschner2003,Nemkov2015,Nemkov2017}, and along with the fusion kernel this is sufficient to encode any other example \cite{Moore:1988uz}. A particularly simple example is Cardy's formula for the asymptotic density of primary states \cite{Cardy1986a,Keller2015}.

\subsubsection*{Global limit and $1/c$ corrections}

In section \ref{sec:global} we consider the global limit, in which we fix all conformal weights while taking $c\to \infty$ ($b\to 0$). This decouples the Virasoro descendants, and the Virasoro algebra contracts to its global $\mathfrak{sl}(2)$ subalgebra. In this limit, the number of discrete Regge trajectories is of order $c$, and the $m$th VMFT trajectory becomes the $m$th MFT trajectory, with twists accumulating to $ h_m = h_1 + h_2 + m$ with $m\in\NN$. This provides a novel method for the computation of MFT OPE data, including at subleading twist $m\neq 0$. At subleading orders in $1/c$, one can systematically extract the large $c$ expansion of double-twist OPE data due to non-unit operators, by performing the small $b$ expansion near the $m$th pole. As a check, we recover the OPE data of MFT by taking the global limit of the residues of the vacuum kernel, as well as some known results for double-twist anomalous dimensions due to non-unit operators. These matches follow from a correspondence between the double-twist poles of the Virasoro fusion kernel and a `holomorphic half' of the global $6j$ symbol computed in \cite{Liu2018}, the precise statement of which can be found in \eqref{kernellim}--\eqref{kernellim2}. 

The data obtained in the $1/c$ expansion is useful for the study of correlation functions of light operators in theories which admit weakly coupled AdS$_3$ duals, especially if the CFT has a sparse light spectrum, whereupon the number of exchanges is parametrically bounded. Expansion of the VMFT OPE data to higher orders in $1/c$ may be performed as desired, for example to extract the anomalous dimension due to multi-graviton states. 

\subsubsection*{AdS$_3$ interpretation}

The previous results all have an interpretation in AdS$_3$ quantum gravity, which we discuss in section \ref{sec:gravity.1}. The discrete quantum Regge trajectories are dual to two-particle bound states, while the large spin continuum at $h={c-1\o 24}$ corresponds to spinning black holes. Heuristically, this dichotomy reflects the threshold for black hole formation at $h={c-1\o 24}$, including the quantum shift $c \to c-1$ not visible in the classical regime \c{Benjamin2016}. The finiteness of the tower of discrete trajectories may be viewed as a kind of quantum gravitational exclusion principle, reflecting the onset of  black hole formation. The negativity $\delta h_m<0$ of the VMFT anomalous twist given in \eqr{anomdim} translates into a negative binding energy in AdS$_3$, thus reflecting the attractive nature of gravity at the quantum level. The corrections $\delta h_m^{(\a_t,\ab_t)}$ to the VMFT twists $h_m$ are dual to contributions to the two-particle binding energy due to bulk matter. In higher dimensions, the decay $\ell^{-2h}$ of anomalous dimensions reflects the exponential falloff of the $h$-mediated interaction between two particles, with orbit separated by a distance of order $\log\ell$. The $d=2$ result \eqref{deltahlargeell} actually has precisely the same interpretation, with the apparent discrepancy coming from a gravitational screening effect explained in section \ref{sec:anomTwistsGravity}. The particles orbiting in AdS come with a dressing of boundary gravitons, and at very large spin this dressing carries most of the energy and angular momentum; this can be removed by a change of conformal frame, corresponding to removing descendants to form a Virasoro primary state.

\begin{figure}[t]
\includegraphics[scale=0.36]{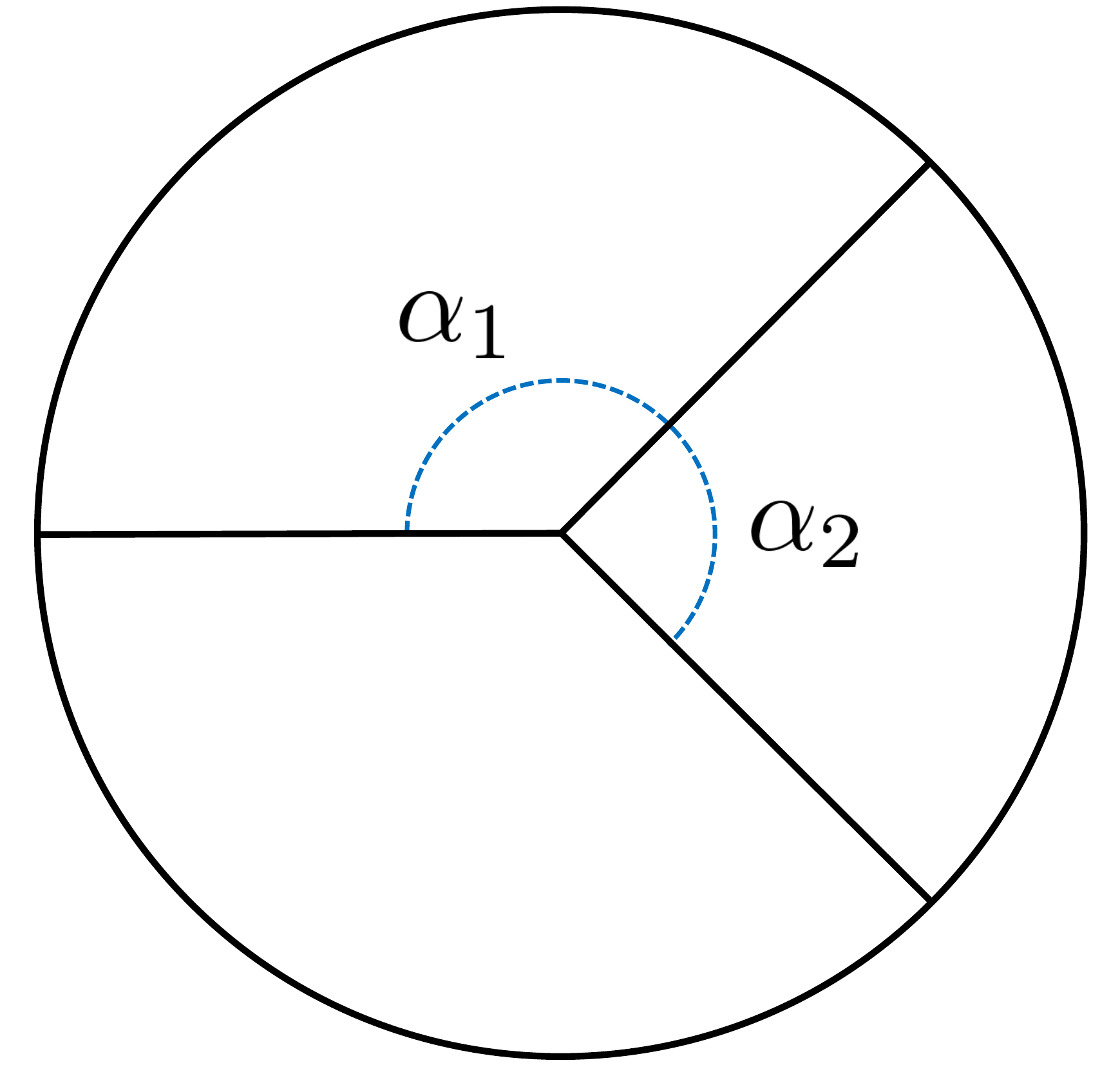}
\centering
\caption{In the semiclassical regime, the additivity rule \eqr{eq:VirasoroDoubleTwist} for discrete momenta of large spin double-twist operators translates into additivity of conical defect angles in AdS$_3$, where the deficit angle is $\Delta\phi = \frac{4\pi}{Q} \alpha$. The leading-twist operator, whose dual conical defect is depicted as the sum of two constituent defects, has momentum $\a_1+\a_2$. We have suppressed the $\tfrac{4\pi}{Q}$ for clarity.\label{defectfig}}
\end{figure}

The addition of momentum $\alpha$ in VMFT has a simple geometric realisation in the semiclassical regime in which $\O_1$ and $\O_2$ are dual to bulk particles that backreact to create conical defect geometries. The deficit angle created by a particle, proportional to its classical mass, is $\Delta\phi_i = \frac{4\pi}{Q}\alpha_i$. The spectrum of discrete twists has the elegant bulk interpretation that the bulk masses, and hence deficit angles, simply add according to \eqr{eq:VirasoroDoubleTwist}. This is depicted in figure \ref{defectfig}.

\subsubsection*{Heavy-light semiclassical limit}

In section \ref{HHLL}, we study the fusion kernel in the large $c$ heavy-light limit of \cite{Fitzpatrick2014,Fitzpatrick2015a}, in which the dimension of one pair of external operators scales with $c$, while the other pair have dimensions fixed. This leads to two new derivations of heavy-light Virasoro blocks, one for the vacuum block (recalled in \eqref{eq:xInt}) for heavy operators above the black hole threshold at $\alpha=\frac{Q}{2}$, and another for non-vacuum heavy-light blocks when the heavy operator is below the black hole threshold. The result is obtained by actually performing the sums \eqref{sumintblocks} over S-channel blocks, using knowledge of $\kernel_{\a_s\a_t}$ and a simplification of the S-channel blocks in this limit. The derivation also gives a new understanding of the emergence of the ``forbidden singularities'' of the heavy-light blocks, and relates their resolution to the analytic stucture of the S-channel blocks.

\subsubsection*{Late time}
One result that follows easily from our analysis is an analytic derivation of the behaviour of the heavy-light Virasoro vacuum block at late Lorentzian times. This block was found numerically to decay exponentially at early times, followed by a $t^{-3/2}$ decay at late times \cite{Fitzpatrick2017a,Chen2017}. The idea here is as follows: upon using the fusion kernel to write a T-channel block as an integral over S-channel blocks, the time evolution gives a simple phase in the S-channel, and a saddle point computation then yields the $t^{-3/2}$ behaviour. This follows from the existence of a double zero of the fusion kernel $\kernel_{\a_s\id}$ at $\alpha_s={Q\o 2}$, which is the relevant saddle point for this computation. The more complete expression for the behaviour of the block is in \eqref{fulllatetime}. In fact, this power-law falloff is universal for Virasoro blocks with ${\rm Re}(\alpha_1+\alpha_2)>{Q\over 2}$; neither large $Q$ nor the semiclassical limit is required. In the heavy-light case, we also derive the crossover time between the exponential and power law.

\centerline{\rule{8cm}{0.4pt}}

Looking forward, we anticipate an array of uses for the results herein, and for the fusion kernel more generally. The validity of our results at finite $c$ provides strong motivation to understand these features of AdS$_3$ quantum gravity directly in the bulk. To begin, it would be nice to perform bulk calculations that match the $1/c$ expansion of the anomalous twist \eqr{anomdim} for light external operators (e.g. using Wilson lines \c{Fitzpatrick2017, Besken:2017fsj, Besken:2018zro, Hikida:2017ehf, Kraus:2018zrn} or proto-fields \c{Anand:2017dav}). Likewise, the effective theory in \c{Cotler:2018zff} may also be capable of reproducing our results. One would also like to reproduce the all-orders result from the bulk: namely, to find a gravitational calculation at large $c$ and finite $h_i/c$, possibly with $m$ of order $c$, that reproduces the complete $\delta h_m$. As we pursue an improved understanding of irrational two-dimensional CFT, perhaps the overarching question suggested by our results is the following: with Virasoro symmetry under more control, can we build a better bootstrap in two dimensions? These developments sharpen the need for an explicit realization of an irrational compact CFT$_2$ to serve as a laboratory for the application of these ideas --- a ``3D Ising model for 2D.''\footnote{Not to be confused with the 2D Ising model. We point out that the IR fixed point of the coupled Potts model \cite{Dotsenko1999} is a potential candidate for such a theory, but its low-lying spectrum has not been conclusively pinned down.}
	
\subsubsection*{Note added:}
While this work was in preparation, the paper \cite{Kusuki2018c} appeared, which also derives the cross-channel limit of the Virasoro blocks using the analytic structure of the crossing kernel in $\alpha_s$, and thereby infers the leading accumulation points in the spectrum of twists at large spin for two-dimensional CFTs at finite central charge.

\section{Analyzing the fusion kernel}\label{kernelAnalysis}

We begin by considering the most general four-point function of primary operators $\op_{1,2,3,4}$ in a two-dimensional CFT, 
\begin{equation}
	G(z,\bar{z}) = \langle \op_1(0,0)\op_2(z,\bar{z})\op_3(1,1)\op_4(\infty,\infty) \rangle,
\end{equation}
with conformal cross ratios ($z,\zb$). By appropriately taking the OPE between pairs of operators, or, equivalently, inserting complete sets of states in radial quantisation, we can write this as a sum over products of three-point coefficients $C_{ijk}$ with intermediate primary operators, times the Virasoro conformal blocks. There are several choices of which operators to pair in this process, which must give the same result; equating the expansion in the S- and T-channels results in the crossing equation
\begin{equation}\label{crossingEq}
	G(z,\bar{z})=\sum_s C_{12s}C_{s34} \mathcal{F}^{21}_{34}\left(\alpha_s\middle|z\right)\bar{\mathcal{F}}^{21}_{34}\left(\bar{\alpha}_s\middle|\bar{z}\right) = \sum_t C_{14t}C_{t32} \mathcal{F}^{23}_{14}\left(\alpha_t\middle|1-z\right)\bar{\mathcal{F}}^{23}_{14}\left(\bar{\alpha}_t\middle| 1-\bar{z}\right),
\end{equation}
which imposes strong constraints on the OPE data of the CFT.
The block $\mathcal{F}^{21}_{34}(\alpha|z)$ is the contribution to $G(z,\zb)$ of holomorphic Virasoro descendants of a primary of weight $h(\alpha)$ in the OPE taken between operators $(12),(34)$, normalised such that %
\begin{equation}
	\mathcal{F}^{21}_{34}(\alpha|z) \sim z^{-h_1-h_2+h(\alpha)} \quad \text{as}\quad z\to 0.
\end{equation}
In this paper, we will not study this crossing equation directly, but rewrite it as a direct relation between T- and S-channel OPE data.

We will henceforth employ  the parameterization in terms of the ``background charge" $Q$ or ``Liouville coupling" $b$ (defined by $c=1+6Q^2$, $Q=b+b^{-1}$), and ``momentum" $\alpha$ (defined by $h=\alpha(Q-\alpha)$) introduced in \eqref{params}.\footnote{There are many reasons why this parameterisation is most natural. Minimal model values of $c$ correspond to negative rational values of $b^2$, and for any $c$, degenerate representations of the Virasoro algebra occur at $\alpha_{r,s} = -\tfrac{1}{2}(r b+s b^{-1})$, with $r,s\in \NN $.} Note that $h(\a) = h(Q-\a)$ and $Q(b) = Q(b^{-1})$.  We will fix the choice of $b$ by taking $0<b<1$ if $c>25$, and by taking $b$ to lie on the unit circle in the first quadrant if $1<c<25$. For $c>1$, unitarity of Virasoro highest-weight representations requires that $h\geq 0$. Our parameterisation naturally splits up this range of dimensions into two distinct pieces:% 
\begin{align}
	0<h<\frac{c-1}{24} &\longleftrightarrow 0<\alpha<\tfrac{Q}{2} \qquad \text{(discrete)}\label{disc} \\
	h\ge\frac{c-1}{24} &\longleftrightarrow \alpha \in \tfrac{Q}{2}+i \RR \qquad \text{(continuous)}\label{cont}
\end{align}
We call these the ``discrete'' and ``continuous'' ranges because, as we will see, the analytic structure of the crossing kernel implies that T-channel blocks have support on S-channel blocks for a discrete set of dimensions in \eqref{disc}, but over the whole continuum \eqref{cont}. This also echoes terminology used in AdS$_3$/CFT$_2$. In Liouville theory, these correspond to dimensions of non-normalisable and normalisable vertex operators respectively.

The defining identity for the fusion kernel is
\begin{equation}\label{eq:TBranchIntoS}
	\mathcal{F}^{23}_{14}(\alpha_t|1-z) = \int_C \frac{d\alpha_s}{2i} \kernel_{\alpha_s\alpha_t}\!\!\begin{bmatrix}\alpha_2 & \alpha_1 \\ \alpha_3 & \alpha_4 \end{bmatrix} \mathcal{F}^{21}_{34}(\alpha_s|z),
\end{equation}
for which the contour of integration $C$ will be discussed shortly. It is not obvious that such an object should even exist, but nonetheless a closed form expression for $\kernel$ has been written down by Ponsot and Teschner \cite{Ponsot1999,Ponsot2001,Teschner2001}, which we present (without derivation) in a moment. (See also \c{Teschner2014,Teschner2015}, and \cite{Ponsot2004} for a compact summary.) \eqr{eq:TBranchIntoS} is formally defined when all operators have momenta in the continuum.  As is the case in studies of the global crossing kernel, we will analytically continue away from the continuum to infer OPE data about four-point functions of arbitrary highest weight representations.

Perhaps a better interpretation is to view the fusion kernel not in terms of blocks, but as a map between S-channel and T-channel OPE data. To do this, define the OPE spectral densities\footnote{The $\delta$-functions supported at imaginary $\alpha$ may be unfamiliar, but make sense in a space of distributions dual to holomorphic test functions (the only requirement being that the blocks are contained in this space).} in S- and T-channels,
\begin{equation}
	c^{21}_{34}(\alpha_s,\bar{\alpha}_s) := \sum_p C_{12p}C_{p34} \delta(\alpha_s-\alpha_p)\delta(\bar{\alpha}_s-\bar{\alpha}_p), \quad c^{23}_{14}(\alpha_t,\bar{\alpha}_t) := \sum_p C_{14p}C_{p32} \delta(\alpha_t-\alpha_p)\delta(\bar{\alpha}_t-\bar{\alpha}_p),
\end{equation}
which allows us to write the correlation function $G$ as the spectral density integrated against the conformal blocks, in either channel. Replacing the T-channel block using the fusion kernel, then stripping away the S-channel blocks, leads to the following reexpression of the crossing equation:
\begin{equation}\label{eq:OPEDensity}
	c^{21}_{34}(\alpha_s,\bar{\alpha}_s) = \int_C {d\alpha_t\over 2i}\int_C {d\bar{\alpha}_t \over 2i}\;  \kernel_{\alpha_s \alpha_t} \; \bar{\kernel}_{\bar{\alpha}_s\bar{\alpha}_t} \; c^{23}_{14}(\alpha_t,\bar{\alpha}_t).
\end{equation}
This can be thought of as an expression for the S-channel spectrum as a linear operator acting on the T-channel spectrum $c_s=\kernel\cdot c_t$ (where we suppress the holomorphic-antiholomorphic factorisation)\footnote{The notation $\kernel_{\a_s\a_t}$ is chosen as it gives the matrix elements of this linear operator. The blocks (at fixed $z$) can be thought of as elements of the dual space, which explains why the indices in (\ref{eq:TBranchIntoS}) are transposed relative to
 what might have been expected. For minimal model values of $c$ and $h$, $\kernel$ becomes an ordinary finite-dimensional matrix.}. In particular, including just a single block in the T-channel, as we will be doing for most of the paper, $\kernel_{\alpha_s \alpha_t} \; \bar{\kernel}_{\bar{\alpha}_s\bar{\alpha}_t}$ simply gives the corresponding S-channel spectral density.

\subsection{Integral form of the kernel}
The closed-form expression for the fusion kernel requires the introduction of some special functions, particularly $\Gamma_b(x)$, which we define in appendix \ref{app:Special}, accompanied by a discussion of some of its properties. The salient information is that $\Gamma_b(x)$ is a meromorphic function with no zeros, and poles at $x=-m b-n b^{-1}$ for $m,n\in\NN$. In this sense, $\Gamma_b$ can be thought of as analogous to the usual $\Gamma$-function, but adapted to the lattice of points made by nonnegative integer linear combinations of $b,b^{-1}$ rather than just the integers. 
Using this function, along with
\begin{equation}
	S_b(x):= \frac{\Gamma_b(x)}{\Gamma_b(Q-x)},
\end{equation}
the kernel can be written as
\ie\label{eq:CrossingKernel}
 \kernel_{\alpha_s\alpha_t}\!\!\begin{bmatrix}\alpha_2 & \alpha_1 \\ \alpha_3 & \alpha_4 \end{bmatrix} = P(\alpha_i;\alpha_s,\alpha_t)P(\alpha_i;Q-\alpha_s,Q-\alpha_t)\int_{C'}\frac{ds}{i} \prod_{k=1}^4 \frac{S_b(s+U_k)}{S_b(s+V_k)},
\fe
where
\es{}{&P(\alpha_i;\alpha_s,\alpha_t)\\
=&\frac{\Gamma_b(\alpha_s+\alpha_3-\alpha_4)\Gamma_b(\alpha_s+Q-\alpha_4-\alpha_3)\Gamma_b(\alpha_s+\alpha_2-\alpha_1)\Gamma_b(\alpha_s+\alpha_1+\alpha_2-Q)}{\Gamma_b(\alpha_t+\alpha_1-\alpha_4)\Gamma_b(\alpha_t+Q-\alpha_1-\alpha_4)\Gamma_b(\alpha_t+\alpha_2-\alpha_3)\Gamma_b(\alpha_t+\alpha_2+\alpha_3-Q)}\frac{\Gamma_b(2\alpha_t)}{\Gamma_b(2\alpha_s-Q)}}
and we define $U_k,V_k$ as follows:
\ie
	\begin{split}
		U_1&=\alpha_1-\alpha_4\\
		U_2&=Q -\alpha_1-\alpha_4 \\
		U_3&= \alpha_2+\alpha_3-Q \\
		U_4&=\alpha_2-\alpha_3
	\end{split}
	\qquad
	\begin{split}
		V_1 &= Q-\alpha_s+\alpha_2-\alpha_4\\
		V_2 &=  \alpha_s+\alpha_2-\alpha_4 \\
		V_3 &= \alpha_t \\
		V_4 &= Q-\alpha_t
	\end{split}
\fe
The contour of integration $C'$ runs from $-i\infty$ to $i\infty$, passing to the right of the towers of poles at $s = -U_i-mb-nb^{-1}$ and to the left of the poles at $s = Q-V_j+mb+nb^{-1}$, for $m,n\in\NN$.

\begin{figure}[h!]
\centering
\includegraphics[width=.45\textwidth]{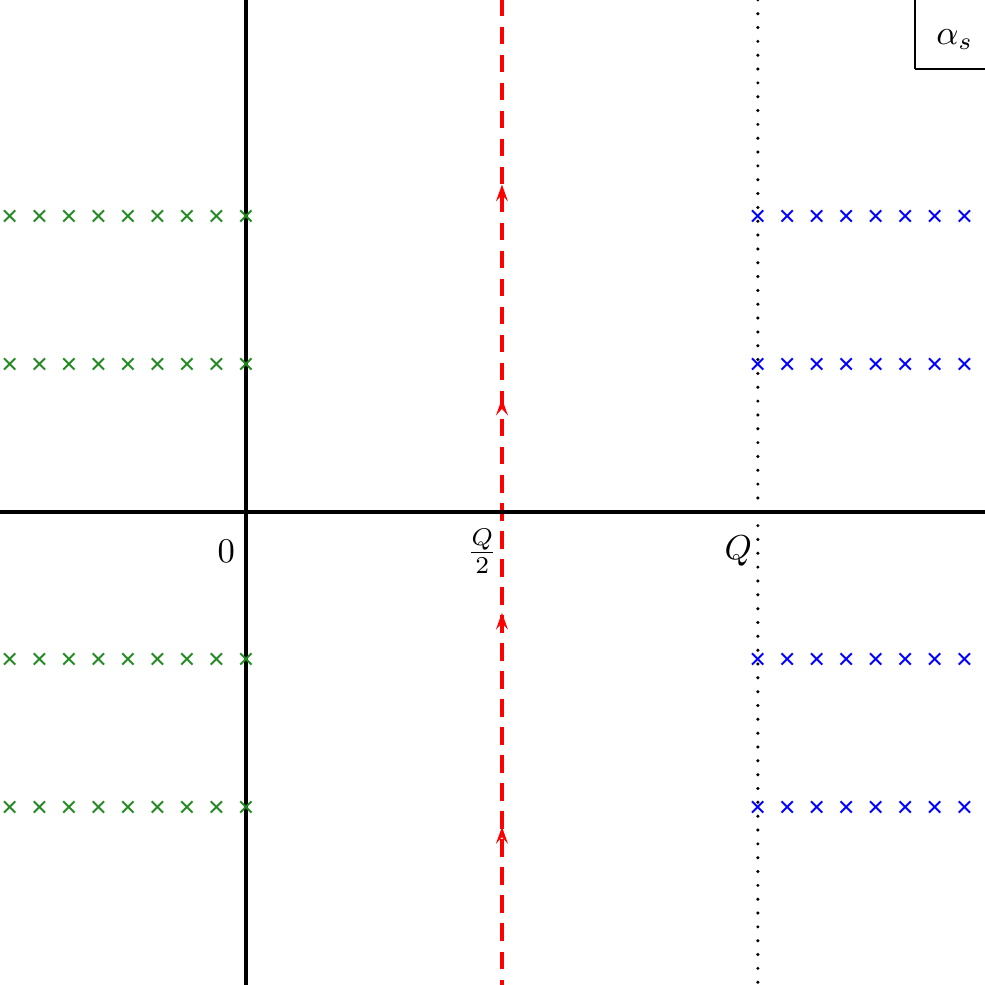}
\caption{For sufficiently heavy external pairwise identical operators (${\rm Re}(\alpha_1 + \alpha_2)>{Q\over 2}$), the fusion kernel has four semi-infinite lines of poles extending in either direction. In the case of vacuum exchange in the T-channel, these poles are simple, otherwise they are double poles. Here we show an example of this in the case that all external operators have weights in the continuum, with $0<b<1$. The dashed red curve denotes the contour of integration in the decomposition of the T-channel Virasoro block into S-channel blocks, while the blue and green crosses denote the poles of the fusion kernel.\label{fig:VPSKernelPlane}}
\end{figure}

\begin{figure}[h!]
\centering
\includegraphics[width=.8\textwidth]{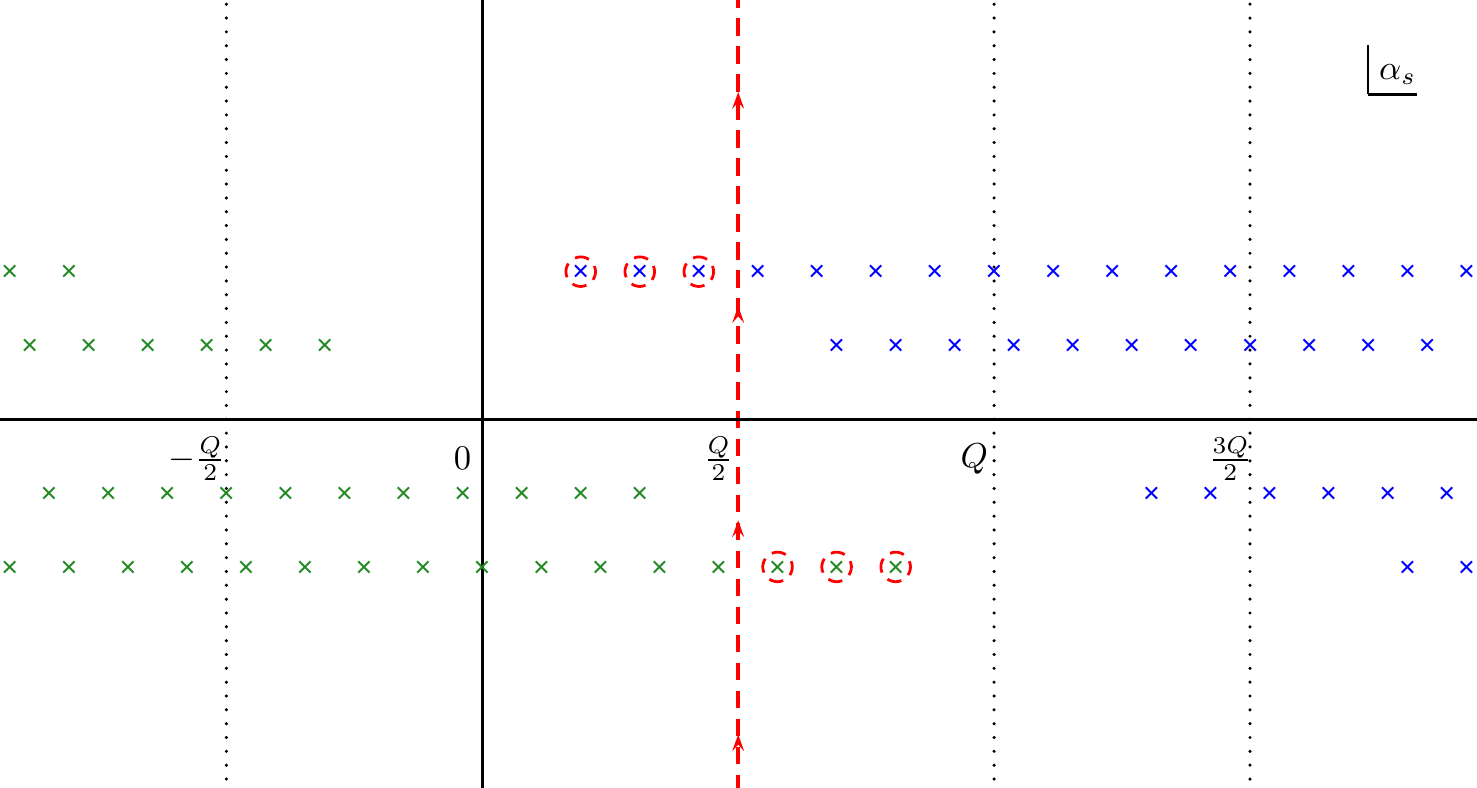}
\caption{Here we plot the poles of the fusion kernel as a function of $\alpha_s$ in the case of pairwise identical external operators and $\Re(\alpha_1 + \alpha_2)<\frac{Q}{2}$, with $0<b<1$. In this case, the poles at $\alpha_s = \alpha_1 + \alpha_2 + mb< \frac{Q}{2}$ (and their reflections) cross the contour of integration and give discrete residue contributions to the T-channel Virasoro block. Note that despite the fact that the external operators have weights lying in the discrete range rather than the continuum we have given $\alpha_i$ small imaginary parts for the purpose of presentation.\label{fig:VDTKernelPlane}}
\end{figure}

The analytic structure of the fusion kernel as a function of $\alpha_s$ will play an important role in our analysis. This is depicted in figures \ref{fig:VPSKernelPlane} and \ref{fig:VDTKernelPlane}. For generic external dimensions and operators in the T-channel, the kernel has simple poles in $\alpha_s$ organized into eight semi-infinite lines extending to the right, and another eight semi-infinite lines extending to the left:
\ie
&\text{poles at }\alpha_s = \alpha_0 + mb+nb^{-1},~Q-\alpha_0-mb-nb^{-1},~\text{for }m,n\in\ZZ_{\ge 0}\\
&\text{with }\alpha_0 = \alpha_1+\alpha_2,~\alpha_3+\alpha_4\text{ (+ six permutations under reflections $\alpha_i\to Q-\alpha_i$)}.
\fe
Schematically, half of these poles come directly from the special functions in the prefactor, with the other half coming from singularities of the integral. The latter occur when poles of the integrand coincide and pinch the contour of integration between them, namely when $V_j-U_i-Q=mb+nb^{-1}$. 
In the important case of pairwise identical external operators, the eight semi-infinite lines of poles in each direction degenerate to four such lines of \emph{double} poles extending in either direction. A notable exception occurs when the internal dimension becomes degenerate ($\alpha_t,Q-\alpha_t = -{1\over 2}(mb+nb^{-1}))$ for $m,n\in\NN$), which requires external dimensions consistent with the fusion rules \cite{Belavin:1984vu}. For us, this will be important when $\alpha_t = 0$ with pairwise identical external operators, relevant for vacuum exchange, in which case the kernel has only simple poles.

For external operators with weights in the continuous range ($\alpha\in\frac{Q}{2}+i\RR$), the towers of poles extending to the right and left all begin on the line $\Re(\alpha_s) = Q$ and the imaginary axis respectively. In this case the contour $C$ can be taken to run along the line $\alpha_s = {Q\over 2}+i\RR$ between them, so that only continuum S-channel blocks appear in the decomposition of the T-channel block, illustrated in figure \ref{fig:VPSKernelPlane} . This demonstrates \eqref{intblocks}.

If some of the external operators have weights in the discrete range $h_i <\frac{c-1}{24}$, certain lines of poles move inward towards the integration contour, and for $\alpha_1 + \alpha_2 <\frac{Q}{2}$ or $\alpha_3 + \alpha_4 <\frac{Q}{2}$, some of these poles cross the contour of integration. To maintain analyticity in the parameters, we must deform the contour to include portions surrounding the relevant poles, contributing a residue. This leads to a finite, discrete sum of S-channel blocks appearing in the decomposition of the T-channel block in addition to the continuum starting at $h_s = {c-1\over 24}$, as in \eqref{sumintblocks}. For unitary values of the weights, the only momenta that can contribute to this finite sum are
\begin{equation}
	\alpha_s = \alpha_1+\alpha_2+mb~,\quad \alpha_3+\alpha_4+m'b, \quad\text{with } m,m'\in\NN
\end{equation}
when these values are less than $\frac{Q}{2}$.\footnote{A similar phenomenon occurs in the analytic continuation of four-point functions in Liouville theory,  whose conformal block decomposition takes the form of the DOZZ structure constants \cite{Dorn1994a,Zamolodchikov1996} integrated against the Virasoro conformal blocks, when poles cross the contour of integration over intermediate dimensions
\cite{Lin2017,Collier2018,Balasubramanian2017}. However, in that case the poles correspond to scalar operator dimensions, and not chiral dimensions separately.}  The poles at reflected values of $\alpha_s$ give identical contributions, and the other lines of poles can never cross the contour for unitary external operators. The contour in this scenario is illustrated in figure \ref{fig:VDTKernelPlane}.

Henceforth, we specialise to the case of pairwise identical operators $\alpha_4 = \alpha_1, \alpha_3 = \alpha_2$, and for notational brevity omit the labels for the external operators, using the condensed notation
\begin{equation}
	\mathcal{F}_S(\alpha_s):=\mathcal{F}^{21}_{21}(\alpha_s|z),\quad \mathcal{F}_T(\alpha_t):=\mathcal{F}^{22}_{11}(\alpha_t|1-z), \quad \kernel_{\alpha_s\alpha_t}:= \kernel_{\alpha_s\alpha_t}\!\!\begin{bmatrix}\alpha_2&\alpha_1\\ \alpha_2 & \alpha_1\end{bmatrix}.
\end{equation}
We also record some results for the cross-channel kernel $\widetilde{\kernel}_{\alpha_t\alpha_s}:=\kernel_{\alpha_t\alpha_s}\!\!\begin{bmatrix}\alpha_2&\alpha_2\\ \alpha_1 & \alpha_1\end{bmatrix}$, which is the inverse operator to $\kernel_{\alpha_s\alpha_t}$, in appendix \ref{appd}.

\subsection{Computing properties of the kernel}

In the remainder of the section, we outline how various properties and limits of the fusion kernel are computed, and give the main results, with additional formulas given in appendix \ref{app:moreKernel}. Readers interested only in the physical application of these results may skip to section \ref{sec3}.

\subsubsection{Vacuum kernel}

With external operators identical in pairs, the fusion rules allow $\alpha_t=0$, which we will use extensively for exchange of the identity operator, and the corresponding fusion kernel, denoted $\kernel_{\alpha_s \id}$, greatly simplifies. The integral in \eqref{eq:CrossingKernel} becomes singular as $\alpha_t\to 0$, since the contour must pass between a double pole at $s=0$ and a single pole at $s=\alpha_t$; the singular piece can be evaluated simply by the residue of the latter. This gives a simple pole in $\alpha_t$ (an additional zero at $s=-\alpha_t$ reduces the strength of the singularity), which is cancelled by a zero in the prefactor. This leaves the following simplified expression for the vacuum fusion kernel:
\begin{equation}\label{vackernel}
	\kernel_{\alpha_s \id} = \frac{\Gamma_b(2Q)}{\Gamma_b(Q)}\frac{\Gamma_b(\alpha_1+\alpha_2-\alpha_s)\times (7 \text{ terms with } \alpha\leftrightarrow Q-\alpha )}{\Gamma_b(Q)^2\Gamma_b(Q-2\alpha_s)\Gamma_b(2\alpha_s-Q)\Gamma_b(2\alpha_1)\Gamma_b(2Q-2\alpha_1)\Gamma_b(2\alpha_2)\Gamma_b(2Q-2\alpha_2)}.
\end{equation}
The seven terms not written comprise all possible combinations of reflections of the three momenta $\alpha_{1,2,s}$. This simple expression makes the polar structure manifest.\foot{This idea was also used in \cite{Esterlis2016} to compute a piece of the vacuum kernel.}

\subsubsection{Singularities}

The poles in the fusion kernel encode the coefficients of the discrete sum of blocks appearing in \eqref{sumintblocks}; in this section we compute these coefficients.

For the identity kernel, we can use the expression \eqref{vackernel} to evaluate the residue of the simple poles at $\alpha_s=\alpha_1+\alpha_2+m b$, using identities written in appendix \ref{app:Special} to simplify the result. The result for general $m$ is written in \eqref{eq:HigherResidues}, and for $m=0$ gives
\ie\label{vacleading}
\Res_{\alpha_s= \alpha_1+\alpha_2} \kernel_{\alpha_s \id} =&-{1\over 2\pi}{\Gamma_b(2Q)\Gamma_b(Q-2\alpha_1)\Gamma_b(Q-2\alpha_2)\Gamma_b(2Q-2\alpha_1-2\alpha_2)\over \Gamma_b(Q)\Gamma_b(2Q-2\alpha_1)\Gamma_b(2Q-2\alpha_2)\Gamma_b(Q-2\alpha_1-2\alpha_2)}\\
=&-\frac{1}{2\pi} \frac{(Q-2\alpha_1)(Q-2\alpha_2)}{Q(Q-2\alpha_1-2\alpha_2)} \left(\frac{\Gamma(1+b^2-2b\alpha_1)\Gamma(1+b^2-2b\alpha_2)}{\Gamma(1+b^2)\Gamma(1+b^2-2b(\alpha_1+\alpha_2))} \times (b\leftrightarrow b^{-1})\right).
\fe

In the case of pairwise identical external operators with a non-vacuum primary propagating in the T-channel, the fusion kernel has \emph{double} poles in $\alpha_s$. To compute the behaviour at the poles $\a_s = \a_1+\a_2+mb+nb^{-1}$, it is easiest to make use of the kernel's reflection symmetry as follows. As written in (\ref{eq:CrossingKernel}), the contour integral contributes only a simple pole at $\alpha_s = \alpha_1+\alpha_2 + mb +nb^{-1}$; combining with the simple pole from the prefactor, we would need to be able to compute the finite part of the contour integral in order to determine the residue. However, the contour integral also contributes a double pole at $\alpha_s = Q+\alpha_1-\alpha_2+mb+nb^{-1}$. Since the kernel is invariant under reflections $\alpha_i\to Q-\alpha_i$, we can simply send $\alpha_2\to Q-\alpha_2$ so that the prefactor is regular at $\alpha_s = \alpha_1+\alpha_2+mb+nb^{-1}$ and isolate the singularities of the contour integral, much like the computation of the vacuum kernel. In this way, we find
\begin{align}\label{nonvackernel}
	\kernel_{\alpha_s\alpha_t} =& \frac{\Gamma_b(\alpha_1+\alpha_2-\alpha_s)^2\Gamma_b(Q-\alpha_1+\alpha_2-\alpha_s)^2\Gamma_b(Q+\alpha_1-\alpha_2-\alpha_s)^2\Gamma_b(2Q-\alpha_1-\alpha_2-\alpha_s)^2}{\Gamma_b(2Q-2\alpha_s)\Gamma_b(Q-2\alpha_s)} \nonumber\\
&\times\frac{\Gamma_b(2\alpha_t)\Gamma_b(\alpha_t+\alpha_s-\alpha_1-\alpha_2)}{\Gamma_b(\alpha_t)^2\Gamma_b(\alpha_t+\alpha_1+\alpha_2-\alpha_s)\Gamma_b(\alpha_t+Q-2\alpha_1)\Gamma_b(\alpha_t+Q-2\alpha_2)} \times(\alpha_t\to Q-\alpha_t) \nonumber\\
&+(\text{regular at }\alpha_s=\alpha_1+\alpha_2).
\end{align}
This allows the coefficients of the double and simple poles to be read off upon expanding the divergent factor in the numerator $\Gamma_b(\alpha_1+\alpha_2-\alpha_s)^2$. A formula that captures the singularities of the non-vacuum kernel at the subleading poles is given in equation (\ref{nonvackernel2}).

The coefficient of the non-vacuum kernel at the leading ($m=0$) double pole is given by
\ie\label{dresnonvac}
\dRes_{\alpha_s=\alpha_1+\alpha_2}\kernel_{\alpha_s\alpha_t}&=\frac{\Gamma_b(Q)^2\Gamma_b(Q-2\alpha_1)^2\Gamma_b(Q-2\alpha_2)^2\Gamma_b(2Q-2\alpha_1-2\alpha_2)}{(2\pi)^2\Gamma_b(Q-2\alpha_1-2\alpha_2)}\\
&\quad\times\left(\frac{\Gamma_b(2\alpha_t)}{\Gamma_b(\alpha_t)^2\Gamma_b(Q-2\alpha_1+\alpha_t)\Gamma_b(Q-2\alpha_2+\alpha_t)}\times(\alpha_t\to Q-\alpha_t)\right).
\fe
Here we have introduced the facetious notation $\dRes$ to denote the coefficient of a double pole. The result for general $m$ is given in \eqref{eq:HigherResidues2}. The residue at the leading pole may be written in terms of a $b$-deformed digamma function, the logarithmic derivative of $\Gamma_b$
\begin{equation}
	\psi_b(z) := \frac{\Gamma_b'(z)}{\Gamma_b(z)} \sim -\frac{1}{z}-\gamma_b +\cdots \text{ as }z\to 0,
\end{equation}
where the $z\to 0$ limit defines a $b$-deformed version of the Euler-Mascheroni constant, $\gamma_b$.
We find
\begin{equation}\label{eq:NonVacuumResidue}
\frac{\displaystyle \Res_{\alpha_s=\alpha_1+\alpha_2}\kernel_{\alpha_s\alpha_t}}{\displaystyle \dRes_{\alpha_s=\alpha_1+\alpha_2}\kernel_{\alpha_s\alpha_t}} = 2(\psi_b(\alpha_t)+\psi_b(Q-\alpha_t)+\gamma_b+\psi_b(Q-2\alpha_1-2\alpha_2)-\psi_b(Q-2\alpha_1)-\psi_b(Q-2\alpha_2))
\end{equation}
The result for the residue of the kernel at the subleading poles is given in (\ref{eq:NonVacHigherResidues}).

\subsubsection{Large dimension limit}

We here present the asymptotics of the fusion kernel in the limit of large S-channel internal weight, with details of the calculations appearing in appendix \ref{app:largeweight}. These results are important for applications to large spin in particular, as explained in section \ref{sec:largeSpin}.

Our main tool in this analysis is the following asymptotic formula for the special function $\Gamma_b(x)$ at large argument $|x|\to\infty$, derived in appendix \ref{app:LargeWeightDerivation}:
\ie\label{eq:GammaAsymptotics}
\log\Gamma_b(x)= -{1\over 2}x^2\log x+{3\over 4}x^2+{Q\over 2}x\log x-{Q\over 2}x-{Q^2+1\over 12}\log x +\log\Gamma_0(b)+\mathcal{O}\left(x^{-1}\right),
\fe
Using this along with the expression \eqref{vackernel} for the vacuum block, the asymptotic form of the vacuum kernel at large internal weight $h_s = {Q^2\over 4}+P^2$ is
\begin{equation}\label{largePvac}
	\kernel_{\alpha_s \id} \sim 2^{-4P^2}e^{\pi \sqrt{\frac{c-1}{6}}P} P^{4 (h_1+ h_2)-\frac{c+1}{4}} 2^{\frac{c+5}{36}} \Gamma_0(b)^6 \frac{\Gamma_b(2Q)}{\Gamma_b(Q)^3\Gamma_b\left(2\alpha_1)\Gamma_b(2Q-2\alpha_1\right)\Gamma_b\left(2\alpha_2\right)\Gamma_b(2Q-2\alpha_2)}.
\end{equation}
The leading exponential piece exactly cancels a similar factor in the large dimension asymptotics of the blocks ($\mathcal{F}\approx (16 q)^h$, for $q=e^{\pi i\tau}$, $\tau = i{ {}_2F_1({1\over 2},{1\over 2};1;1-z)\over {}_2F_1({1\over 2},{1\over 2};1;z)}$), as necessary for correct convergence properties. The formula has a direct interpretation as the asymptotics of `light-light-heavy' OPE coefficients, discussed in section \ref{sec:largedim}.

The computation of the asymptotic form of the non-vacuum kernel is similar, but requires a careful evaluation of the integral appearing in (\ref{eq:CrossingKernel}) in this limit, discussed in appendix \ref{app:largeweight}. For $h_t<{c-1\over 24}$ (i.e. $\a_t<{Q\o 2}$), we have
\ie\label{eq:AsymptoticRatioOfKernels}
	\frac{\kernel_{\alpha_s \alpha_t}}{\kernel_{\alpha_s \id}} \sim& e^{-2\pi \alpha_t P}
	\frac{\Gamma_b\left(2\alpha_1)\Gamma_b(2Q-2\alpha_1\right)}{\Gamma_b(2\alpha_1-\alpha_t)\Gamma_b(2Q-2\alpha_1-\alpha_t)}
	\frac{\Gamma_b\left(2\alpha_2\right)\Gamma_b(2Q-2\alpha_2)}{\Gamma_b(2\alpha_2-\alpha_t) \Gamma_b(2Q-2\alpha_2-\alpha_t)}
	\\
	 &\times\frac{\Gamma_b(2Q-2\alpha_t) \Gamma_b(Q-2\alpha_t) \Gamma_b(Q)^3
	}
	{\Gamma_b(2Q)
	 \Gamma_b(Q-\alpha_t)^4
	}.
\fe
For heavier operators $h_t\ge {c-1\over 24}$ an additional contribution, given by $\alpha_t \rightarrow Q-\alpha_t$, must be added.

\subsubsection{Large central charge limits}\label{sec:scGammab}

In sections \ref{sec:global} and \ref{sec:gravity}, we will study properties of the fusion kernel in global and semiclassical limits, with large central charge. This requires expansions of the special function $\Gamma_b$ in limits as $b\to 0$. In appendix \ref{app:SemiClass}, we derive an all-orders asymptotic series \eqref{eq:GammabSemiclassical} for $\log\Gamma_b$, with argument scaling as $b^{-1}$. This improves on the semiclassical form of $\Upsilon_b(x):=\frac{1}{\Gamma_b(x)\Gamma_b(Q-x)}$ derived in \cite{Harlow2011}. The leading result reads
\begin{equation}\label{eq:scGammab}
	\log\Gamma_b(b^{-1}x+\tfrac{b}{2})  = \frac{1}{2b^2} \left(\frac{1}{2}-x\right)^2\log b+\frac{2x-1}{4b^2}\log(2\pi)-\frac{1}{b^2}\int_{\frac{1}{2}}^x dt\log \Gamma(t) +O(b^2).
\end{equation}
To determine the behaviour of $\log\Gamma_b$ in the global limit, in which the argument scales like $b$, one can use this formula with $x = 1+\mathcal{O}(b^2)$ in conjunction with the recursion relation (\ref{eq:GammaShift}).

\subsubsection{Virasoro double-twist exchanges}

The kernel also simplifies in the case that the T-channel momentum matches the ``Virasoro double twists'' of the relevant external operators, namely $\a_t=2\a_1+m b$ or $\a_t=2\a_2+m b$. Here, we take the leading double twist $\alpha_t=2\alpha_2$ (with corresponding results for $m> 0$ given in appendix \ref{app:SubleadingVDTs}). The simplification is much the same as for the vacuum kernel given in \eqref{vackernel}: the prefactor $P(\alpha_i;Q-\alpha_s,Q-\alpha_t)$ vanishes at $\alpha_t = 2\alpha_2$, but the integral contributes a singularity, so we need only evaluate a pole of the integrand, giving
\ie\label{nonvacdouble}
\kernel_{\alpha_s,2\alpha_2} =& \frac{\Gamma_b(4\alpha_2)}{\Gamma_b(2\alpha_2)^4}\frac{\Gamma_b(\alpha_2+\alpha_1-\alpha_s)^2\Gamma_b(\alpha_2+Q-\alpha_{1}-\alpha_s)^2\Gamma_b(\alpha_1+\alpha_2+\alpha_s-Q)^2\Gamma_b(\alpha_2-\alpha_{1}+\alpha_s)^2}{\Gamma_b(Q)\Gamma_b(2\alpha_2+Q-2\alpha_{1})\Gamma_b(2\alpha_2+2\alpha_1-Q)\Gamma_b(2\alpha_s-Q)\Gamma_b(Q-2\alpha_s)}.
\fe
Note that the kernel still has double poles at the locations of the S-channel Virasoro double-twists, and thus the T-channel Virasoro double-twists contribute to anomalous momenta and double-twist OPE data in the S-channel. This is in contrast with a property of $d>2$ Lorentzian inversion, for which T-channel double-twists give vanishing contribution to the (analytic part of) S-channel OPE data. However, the kernel decays much more rapidly at large dimension, with the leading term in \eqref{eq:AsymptoticRatioOfKernels} replaced by $e^{-2\pi(Q-2\alpha_2)\sqrt{h_s}}$.

\subsection{Cross-channel limit of Virasoro blocks}\label{sec:crosschannel}
Our results immediately allow us to read off the behaviour of the T-channel block in the cross-channel limit $z\to 0$, because the fusion kernel expresses it in terms of S-channel blocks, which have simple power law behaviour $z^{-h_1-h_2+h_s}$. The leading order behaviour is determined by the smallest weight $h_s$ on which the T-channel block has support in the S-channel, which depends on whether the external dimensions are light enough for the discrete dimensions to be present.  Here, we compute these limits for operators identical in pairs, for both S- and T-channel blocks. For the T-channel blocks, there is a qualitative difference between vacuum exchange and other operators if the external operators are sufficiently light that $\alpha_s=\alpha_1+\alpha_2$ dominates, since non-vacuum exchange gives a double pole, and hence an additional logarithm.

  For $\Re(\alpha_1+\alpha_2)<\frac{Q}{2}$, the leading behaviour is controlled by $\alpha_s = \alpha_1+\alpha_2$
\begin{align}
	\text{Vacuum}:~~ \mathcal{F}^{22}_{11}(0|1-z)~\stackrel{z\to 0}{\sim}~&-\Big( \Res_{\alpha_s= \alpha_1+\alpha_2}\,2\pi\,\kernel_{\a_s\id}\Big)\,z^{-2\a_1\a_2}\\
\text{Non-vacuum}:~~\mathcal{F}^{22}_{11}(\a_t|1-z)~\stackrel{z\to 0}{\sim}~&- \Big( \dRes_{\alpha_s= \alpha_1+\alpha_2}\,2\pi\,\kernel_{\a_s\a_t}\Big)(Q-2(\alpha_1+\alpha_2))z^{-2\a_1\a_2}\log z,
\end{align}
where Res and dRes are given in \eqref{vacleading} and \eqref{dresnonvac}. 

For ${\rm Re}(\alpha_1+\alpha_2)>{Q\over 2}$, the leading behaviour is controlled by the bottom of the continuum, $\alpha_s=\frac{Q}{2}$. By performing a saddle point analysis of the continuum integral over $\alpha_s$ as $z\to 0$, we find
\ie\label{crosscont}
\mathcal{F}^{22}_{11}(\alpha_t|1-z)~~\stackrel{z\to 0}{\sim}~~ -{\sqrt{\pi}\over 8}\left.\partial_{\alpha_s}^2\kernel_{\alpha_s\alpha_t}\right|_{\alpha_s={Q\over 2}}z^{{Q^2\over 4}-h_1-h_2}\left(\log{1\over z}\right)^{-{3\over 2}}.
\fe
In the case of vacuum exchange, this coefficient can be computed explicitly\footnote{The borderline case $\alpha_1+\alpha_2=\frac{Q}{2}$ must be treated separately. We point out that the S-channel Virasoro block with $\alpha_1+\alpha_2 = \frac{Q}{2},\alpha_s = \frac{Q}{2}$ is equal to the T-channel block with $\alpha_t=2\alpha_2=Q-2\alpha_1$, given simply by a (chiral half of a) Coulomb gas correlation function $z^{-2\alpha_1\alpha_2}(1-z)^{-2\alpha_2^2}$, and hence crossing-invariant on its own \cite{Esterlis2016,Collier2018}. Consistent with this, in the limit approaching $\alpha_1+\alpha_2=\frac{Q}{2}$ the kernel \eqref{nonvacdouble} becomes a $\delta$-function at $\alpha_s=\frac{Q}{2}$.} 
\ie\label{crosscont2}
\left.\partial_{\alpha_s}^2\kernel_{\alpha_s\id}\right|_{\alpha_s={Q\over 2}} =& -32\pi^2{\Gamma_b(2Q)\Gamma_b^2({3Q\over 2}-\alpha_1-\alpha_2)\Gamma_b({Q\over 2}+\alpha_1-\alpha_2)^2\Gamma_b({Q\over 2}-\alpha_1+\alpha_2)^2\Gamma_b(\alpha_1+\alpha_2-{Q\over 2})^2\over\Gamma_b(Q)^5\Gamma_b(2Q-2\alpha_1)\Gamma_b(2\alpha_1)\Gamma_b(2Q-2\alpha_2)\Gamma_b(2\alpha_2)}.
\fe
As a check, in \ref{app:Zamolodchikovc25} we explicitly verify \eqref{crosscont} with $c=25$ ($b=1$) and $h_1=h_2 = 15/16$ ($\alpha_i = {3\over 4}$), with arbitrary internal dimension, using the exactly known expression for the relevant conformal blocks \cite{Zamolodchikov1986}. 

We can use similar methods to determine the cross-channel limit of the S-channel Virasoro blocks by making use of the decomposition 
\ie
\mathcal{F}^{21}_{21}(\alpha_s|z) =& \int_C \frac{d\alpha_t}{2i}\widetilde\kernel_{\alpha_t\alpha_s}\mathcal{F}^{22}_{11}(\alpha_t|1-z),
\fe
where we have introduced $\widetilde\kernel_{\alpha_t\alpha_s} = \kernel_{\alpha_t\alpha_s}\begin{bmatrix}\alpha_2 & \alpha_2 \\ \alpha_1 & \alpha_1\end{bmatrix}$. The analytic structure of $\widetilde\kernel_{\alpha_t\alpha_s}$ as a function of $\alpha_t$ is slightly different than that of $\kernel_{\alpha_s\alpha_t}$ as a function of $\alpha_s$; in particular it only has simple poles at $\alpha_t = 2\alpha_1+mb+nb^{-1}$, $2\alpha_2+mb+nb^{-1}$ and others obtained by reflections $\alpha_t\to Q-\alpha_t$.  There are also quadruple poles at $\alpha_t = Q+mb+nb^{-1}$ and $\alpha_t = -mb -nb^{-1}$. The cross-channel limit of these blocks is, in the case that at least one of $\alpha_{1,2}<\frac{Q}{4}$, controlled by the leading pole at $\alpha_t = 2\min(\alpha_1,\alpha_2)$
\ie\label{eq:SChannelAsymptotics}
\mathcal{F}^{21}_{21}(\alpha_s|z) \stackrel{z\to 1}{\sim}& -\left(\Res_{\alpha_t = 2\alpha_1}2\pi \widetilde\kernel_{\alpha_t\alpha_s}\right)(1-z)^{h(2\alpha_1)-2h_2};~\alpha_1<{Q\over 4},\alpha_2\\
\mathcal{F}^{21}_{21}(\alpha_s|z) \stackrel{z\to 1}{\sim}& -\left(\Res_{\alpha_t = 2\alpha_2}2\pi \widetilde\kernel_{\alpha_t\alpha_s}\right)(1-z)^{h(2\alpha_2)-2h_2};~\alpha_2<{Q\over 4},\alpha_1,
\fe
with the relevant residues recorded in \ref{appd}. If neither $\alpha_1$ nor $\alpha_2$ is less than $Q\over 4$, the leading behaviour in the cross-channel limit is again controlled by the bottom of the continuum
\ie\label{eq:SChannelAsymptotics2}
\mathcal{F}^{21}_{21}(\alpha_s|z) \stackrel{z\to 1}{\sim}& -{\sqrt{\pi}\over 8}\left.\partial^2_{\alpha_t}\widetilde\kernel_{\alpha_t\alpha_s}\right|_{\alpha_t={Q\over 2}}(1-z)^{{Q^2\over 4}-2h_2}\left(\log{1\over 1-z}\right)^{-{3\over 2}};~\alpha_1,\alpha_2>{Q\over 4}.
\fe

The analytic structure of the kernel as a function of $\alpha_s$ manifestly explains the recent numerical observations \cite{Kusuki2018b,Kusuki2018a} that Virasoro blocks have drastically different cross-channel asymptotics depending on whether the external operators are sufficiently heavy (for example, for identical operators, the observed threshold at large central charge was $h\sim {c\over 32}$, corresponding to $\alpha \sim \frac{Q}{4}$).

\section{Extracting CFT data}\label{sec3}
In this section we discuss some implications of the crossing kernel results from the point of view of ``Virasoro Mean Field Theory'' outlined in the introduction and corrections to it, and then apply this to give universal results for the spectrum at large spin.

\subsection{Quantum Regge trajectories and a ``Virasoro Mean Field Theory''}\label{VMFT}
As discussed in the introduction, our results for the inversion of the Virasoro {\it vacuum} block lead us to coin the nickname Virasoro Mean Field Theory (VMFT) for the non-perturbative incorporation of the stress tensor into ordinary MFT double-trace data. In MFT, correlation functions are computed by Wick contractions of pairs of identical operators (so connected correlation functions are defined to vanish), which for $G=\langle\op_1\op_2\op_2\op_1\rangle$ gives
\begin{equation}
	G_\text{MFT}(z,\bar{z}) = \frac{1}{((1-z)(1-\bar{z}))^{\Delta_2}}.
\end{equation}
Decomposing this in the S-channel, we find a spectrum of quasiprimaries comprising double-trace operators\footnote{In terms of individual operators (`scaling blocks'), rather than quasiprimaries, the MFT double trace data was derived in 1665 \cite{newton}.} $[\op_1\op_2]_{n,\ell}$ for each spin $\ell$ and each $n\in\NN$. In two dimensions, because the $z$ and $\bar{z}$ dependence of $G_\text{MFT}$ factorises, we can think of the double-traces as products of chiral operators $[\op_1\op_2]_m$ for $m\in\NN$, with equally spaced twists $h_{m} = h_1 + h_2 + m$, and their antiholomorphic counterparts labelled by $\bar{m}$:
\begin{equation}
	[\op_1\op_2]_{n,\ell} 
	= [\op_1\op_2]_m
	[\bar{\op}_1 \bar{\op}_2]_{\bar{m}}
	\quad n=\min(m,\bar m), \quad\ell = |m-\bar{m}|
\end{equation}

\begin{figure}
\includegraphics[scale=0.5]{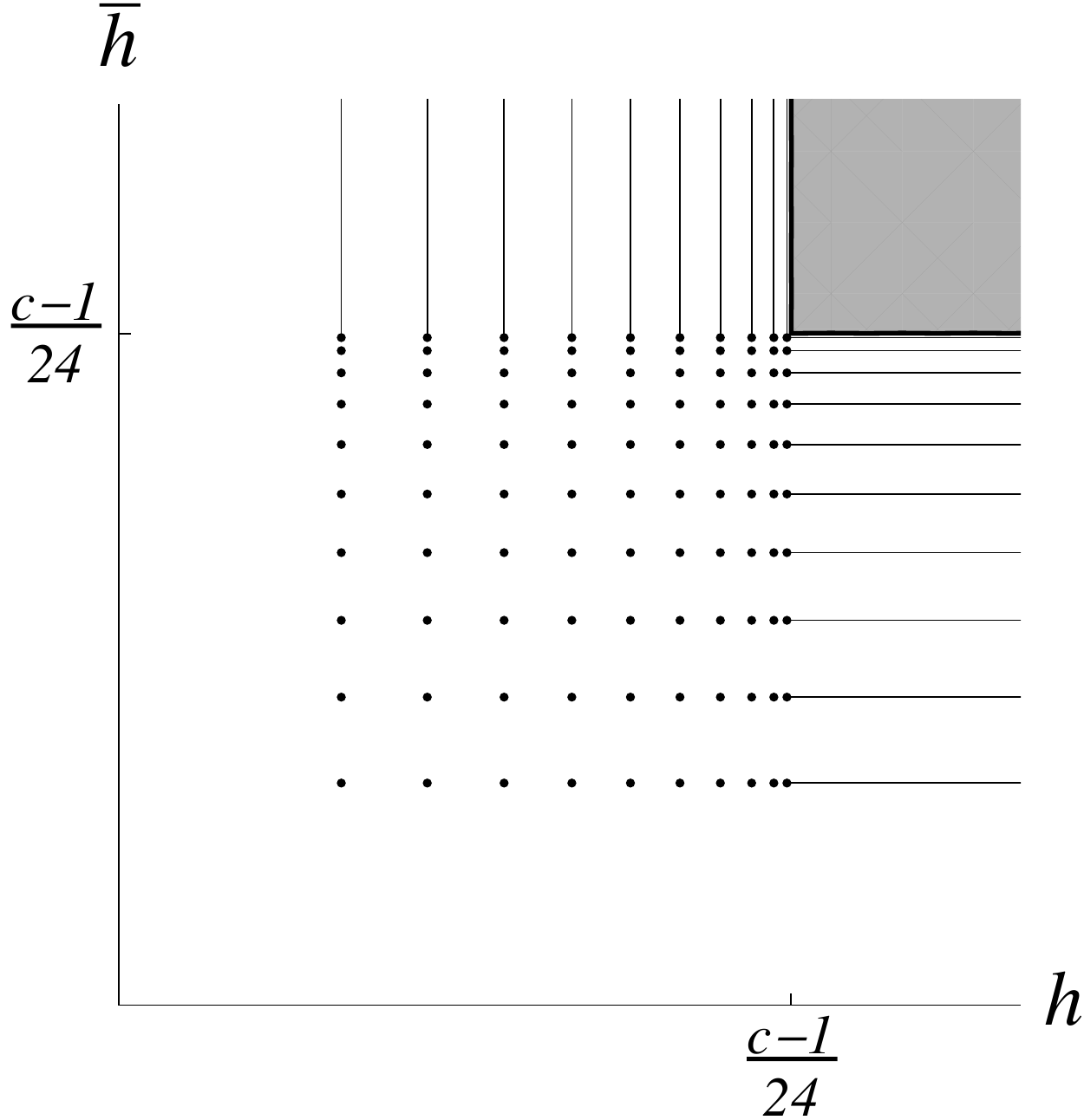}
\centering
\caption{The VMFT spectrum, combining \hol and \ahol sectors.}
\label{fig:kernelSupport}
\end{figure}

For VMFT, we replace the exchange of the unit operator in the T-channel by the full Virasoro vacuum block:
\begin{equation}
	G_\text{VMFT}(z,\bar{z}) = \mathcal{F}^{22}_{11}(0|1-z) \bar{\mathcal{F}}^{22}_{11}(0|1-\bar {z})
\end{equation}
While this does not make sense as a correlation function of a sensible theory\footnote{For a proposal on how this might be upgraded to a full correlation function, see \cite{Maloney2017}.} -- in particular, it is not single-valued on the Euclidean plane -- we can still discuss it at the level of the S-channel data required to reproduce $G_\text{VMFT}$ (on the first sheet). The factorisation property remains, so for most of this subsection we will just discuss the \hol half of the OPE data, with the understanding that the full VMFT data is constructed from products of left- and right-movers, as in figure \ref{fig:kernelSupport}.
 The spectrum of VMFT is the support of the vacuum kernel $\kernel_{\alpha_s\id}$. The resulting Virasoro primary spectrum contains a continuum starting at $\alpha=Q/2$ and, for operators with $\a_1+\a_2<Q/2$, the following discrete set: 
\e{}{\lbrace \O_1\O_2\rbrace_m~,~\text{with}~ \a = \alpha_m:=\alpha_1+\alpha_2+m b~,}
with $m\in \NN$ small enough that $\alpha_m<\frac{Q}{2}$. The latter are natural analogues of the usual MFT double-twists. Summarizing so far, there are three main properties of VMFT that differ from MFT:
\begin{itemize}
\item There is a continuum of double-twists for $h>\frac{c-1}{24}$. 
\item  The discrete part of the spectrum is truncated, and is entirely absent when ${\rm Re}(\a_1+\a_2) >Q/2$. The spins are generically non-integral.
\item Whereas MFT is additive in the twists $h_i$, VMFT is additive in the momenta $\a_i$.
\end{itemize}

The discrete set of \hol VMFT double-twist operators $\lbrace \O_1\O_2\rbrace_m$ acquire anomalous twists relative to the MFT operators $[\O_1\O_2]_m$. Writing $h_m(\a_1+\a_2) = h_1+h_2+m +\delta h_m$, as in \eqr{anomdim}, the additivity of momenta implies that $\lbrace \O_1\O_2\rbrace_m$ has anomalous twist
\begin{equation}\label{anomtwist}
	\delta h_m = -2 (\a_1+mb) (\a_2+mb)+m(m+1)b^2 \,.
\end{equation}
We note that this is always negative. The three-point couplings between the (holomorphic) VMFT double-twist operators and the external operators, $C_{\O_1\O_2\lbrace \O_1\O_2\rbrace_m}$, are given by the residues of the vacuum kernel at the double-twist locations:
\begin{equation}\label{holvmftope}
	C_{\mathcal{O}_1\mathcal{O}_2\{\mathcal{O}_1\mathcal{O}_2\}_m} = -2\pi\Res_{\alpha_s=\,\alpha_m}\kernel_{\alpha_s\id}
\end{equation}
 The explicit expression for the residue of the vacuum kernel is given in \eqref{eq:HigherResidues}. 

\subsubsection*{Perturbing VMFT}

Starting from MFT, one can perturb the theory by adding non-vacuum operators to the T-channel, leading to anomalous dimensions $\g_{n,\ell}$ of the MFT double-twist operators. From the $6j$ symbol, this arises because it acquires double poles at the double-twist locations (and their shadows) \cite{Liu2018}. The situation for discrete trajectories in VMFT is entirely analogous. As seen in \eqref{nonvackernel}, $\kernel_{\a_s\a_t}$ acquires double poles at $\alpha_s=\alpha_m$ for non-vacuum block inversion ($\a_t\neq 0$). This leads to a derivative of S-channel blocks with respect to momentum $\alpha_s$, gives an additional logarithm in the cross-channel behaviour of the T-channel block, and thus generates anomalous \emph{momenta} for $\lbrace \O_1\O_2\rbrace_m$, i.e.
\e{}{\alpha_s = \alpha_m+\delta\alpha_m.}
If we formally consider the inversion of a single holomorphic non-vacuum block, we can define a holomorphic anomalous momentum as\foot{Since all corrections to VMFT come from inverting non-vacuum Virasoro blocks which have both holomorphic and antiholomorphic parts, the full result for the anomalous momentum also depends on the \ahol kernel, $\bar \kernel_{\ab_s\ab_t}$. In the next subsection, we will be more explicit about putting left- and right-movers together to get complete results.}
 \begin{equation}\label{deltaa0}
	\delta\a_m = C_{11t}C_{22t} \Res_{\alpha_s=\,\alpha_m} \left(\frac{\kernel_{\alpha_s\alpha_t}}{\kernel_{\alpha_s\id}}\right),
\end{equation}
since this is the coefficient of the double pole of $\kernel_{\alpha_s\alpha_t}$ divided by the VMFT OPE coefficient, as follows from \eqref{holvmftope}. (See section 3.3 of \cite{Liu2018} for the analogous method of computation of anomalous dimensions from the global $6j$ symbol.) The double pole does not, of course, change the momentum by a finite amount, but the derivative of the block with respect to $\alpha_s$ is interpreted as the first term in a Taylor series, whose coefficient we identify as the anomalous momentum (higher terms must come from sums over infinitely many T-channel operators; see \cite{Alday:2016njk, Simmons-Duffin2017} for example). Note that, from \eqref{vacleading} and \eqref{dresnonvac}, the leading twist coefficients (for $m=0$, $\alpha_s=\alpha_1+\alpha_2<\frac{Q}{2}$) are sign-definite:
\begin{equation}
	\Res_{\alpha_s=\alpha_0} \left(\frac{\kernel_{\alpha_s\alpha_t}}{ \kernel_{\alpha_s\id}} \right)<0
\end{equation} 
It follows that for identical operators, $\delta\alpha_0<0$. Even if $\O_1\neq\O_2$, the anomalous momentum is negative for any T-channel operator which couples with the same sign to $\O_1$ and $\O_2$.

These results nicely generalize properties of global inversion and OPE data of MFT. In section \ref{sec:global}, we show that the MFT results are recovered from the large $c$ limit of VMFT for fixed operator dimensions $h$.

\subsubsection*{Comparison to $d>2$}
In $d>2$ CFTs, the MFT Regge trajectories are the same as those in $d=2$: an infinite tower of flat trajectories with twists $\tau=\D_1+\D_2+2n$. Incorporating the stress tensor $T_{\mu\nu}$, of twist $d-2$, produces anomalous dimensions $\g_{n,\ell}$ which behave as $\ell^{-{d+2}}$ at large spin $\ell\gg n$. In the opposite regime of large twist, $n\gg\ell$, they scale as \c{p1, p2, p3}
\e{}{\g_{n,\ell}\Big|_{T} \approx {n^{d-1}\o c_T}~~\text{as}~~n\gg\ell\gg1}
where $c_T$ is the normalization of the stress-tensor two-point function. In gravitational variables, $c_T \approx 1/G_N \approx M_{\rm pl}^{d-1}$. Therefore, the perturbative expansion of $\g_{n,\ell}$ breaks down at or below the Planck scale, $n\lesssim M_{\rm pl}$. Arguments from perturbative unitarity \cite{Katz} and causality \cite{Camanho:2014apa} place even stronger bounds on this breakdown. However, we do not know how to obtain quantitative understanding of what actually happens to the classical Regge trajectories at Planckian energies. This is due both to technical difficulty, and to non-universality of the $TT$ OPE. The latter implies that the all-orders stress tensor contribution to the Regge trajectories depends on the details of the CFT: in particular, unlike in $d=2$, the three-point coefficients $\la TT\O\ra$, where $\O$ is $T$ itself or any multi-$T$ composite, are sensitive to the rest of the CFT data \c{HM,Camanho:2014apa}. These comments underscore the value of the two-dimensional setting -- non-trivial, yet computable -- in giving insight into Planckian processes in AdS quantum gravity.

\subsection{Large spin} \label{sec:largeSpin}

The OPE data of VMFT has so far been a formal construction, but in this section we will see that it governs a universal sector in physical theories at large spin, applying to all unitary CFTs with $\mathfrak{sl}(2)$ invariant ground state, and no conserved currents besides the vacuum Virasoro module.\footnote{More precisely, we require that non-vacuum Virasoro primaries have twist bounded away from zero, since it is a logical possibility to have an infinite tower of operators with $\bar{h}$ accumulating to zero at large $h$. While we can't rule this out, this seems unlikely to happen in theories of interest. For example, the CFT dual to $\mathrm{AdS}_3\times S^3 \times T^4$ at the symmetric orbifold point has infinitely many higher-spin currents, but when perturbing away from the orbifold point, the anomalous dimensions acquired by the currents seem to grow logarithmically with spin \cite{Gaberdiel2015}. This suggests that away from the orbifold point, the theory has a finite twist gap above the superconformal vacuum descendants.} This means that for any two primary operators, there are associated towers of double-twist operators which asymptote to the VMFT twists \eqr{anomdim} at large spin. Corrections to this come from including T-channel primary operators of positive twist, giving a systematic large spin perturbation theory.

Like analogous results in higher dimensions, this can be derived from solving crossing in the lightcone limit, as briefly reviewed in appendix \ref{app:OldLC}, but we will argue more directly from \eqref{eq:OPEDensity}. The contribution to the S-channel spectral density from any given T-channel operator is simply the fusion kernel $\kernel_{\alpha_s\alpha_t}\bar{\kernel}_{\bar{\alpha}_s\bar{\alpha}_t}$. Now, we take the limit of large spin in the S-channel, at fixed twist:
\begin{equation}
	\alpha_s ~~\text{fixed},\quad\bar{\alpha}_s=\frac{Q}{2}+i\bar{P}\quad \text{with} \quad \bar{P}\sim\sqrt{\ell_s}\to\infty
\end{equation}
The relative importance of T-channel operators in this limit is encoded by \eqref{eq:AsymptoticRatioOfKernels}:
\begin{equation}\label{eq:RatioOfKernels}
	{\bar{\kernel}_{\bar{\alpha}_s\alpha_t}\over \bar{\kernel}_{\bar{\alpha}_s\id}}\approx e^{-2\pi\bar{\alpha}_t \bar{P}}
\end{equation}
This shows that the contribution of operators with $\Re(\bar{\alpha}_t)>0$ is suppressed relative to the vacuum. This suggests that, in a theory with $\bar{\alpha}_t$ bounded away from zero, the S-channel density is dominated by the inversion of the T-channel vacuum -- that is, VMFT -- at large spin. (Recall that negative real $\bar\alpha_t$ is forbidden by unitarity.) We make a more careful argument in a moment, after discussing the consequences. We separately discuss large-spin discrete trajectories and the large-spin continuum, before making further comments.\foot{These two large spin sectors correspond to VFMT operators in the discrete $\times$ continuum and continuum $\times$ continuum representations, respectively; the discrete $\times$ discrete operators are bounded in spin due to the upper bound on $m$, cf. \eqr{eq:VirasoroDoubleTwist}, so they are not part of the large spin universality.}

\subsubsection{Discrete trajectories}
Consider fixing the twist $\alpha_s$ in the discrete range as we take large spin, with $\Re(\alpha_1+\alpha_2)<\frac{Q}{2}$. Let us first review what we found earlier: for each $m\in\NN$ with $\alpha_m=\alpha_1+\alpha_2+m b<\frac{Q}{2}$, there must be a Regge trajectory of operators with $\alpha_s\to \alpha_m$ as $\ell_s\to\infty$: these are the Virasoro double-twist families $\{\mathcal{O}_1\mathcal{O}_2\}_{m,\ell_s}$ of VMFT. The asymptotics of the corresponding OPE coefficients are determined by the vacuum fusion kernel:
\begin{equation}\label{eq:VMFTOPE}
	\rho_m^{{(0)}}(\bar P) = -2\pi\bar\kernel_{\bar\alpha_s\id} \Res_{\alpha_s=\,\alpha_m} \!\kernel_{\alpha_s\id}, \quad \text{with } \bar{\alpha}_s = {Q\over 2}+i\bar{P} ~~\text{and}~~ \ell_s\sim \bar{P}^2 \to \infty.
\end{equation}
Here, $\rho_m$ is a spectral density of the $m$th Regge trajectory, in terms of $\bar{P}$, so it contributes to the correlation function as
\begin{equation}\label{eq:doubletwistcontribution}
	G \supset \int_0^\infty d\bar{P} \:\rho_m^{(0)}(\bar P) \,\mathcal{F}_S(\alpha_m)\bar{\mathcal{F}}_S(\tfrac{Q}{2}+i\bar{P}).
\end{equation}
The superscript denotes that this is the VMFT density. The explicit expression for the residue of the vacuum kernel is given in \eqref{vacleading} for $m=0$, and \eqref{eq:HigherResidues} for higher Regge trajectories, and the large internal weight limit of the kernel is in \eqref{largePvac}.

Additional T-channel operators give corrections to this, adding a spin-dependent anomalous momentum $\delta\alpha_m^{(\a_t,\bar\a_t)}$ to $\alpha_m$, as well as a correction $\delta\rho_m^{(\a_t,\bar\a_t)}$ to $\rho_m^{(0)}$. Expanding \eqref{eq:doubletwistcontribution} to first order generates a derivative of the block proportional to $\delta\alpha_m^{(\a_t,\bar\a_t)}$, which can be matched to the coefficient of the double pole in the fusion kernel $\kernel_{\alpha_s\alpha_t}$ at $\alpha_s=\alpha_m$. The anomalous OPE density $\delta\rho_m^{(\a_t,\bar\a_t)}$ can be read off from the residue at the same point.\footnote{Note that $\delta\bar{\rho}$ does not translate immediately into a change of spectral density in terms of spin, which is more directly related to anomalous OPE coefficients of individual operators. This picks up a Jacobian factor $2\bar{P}-\delta h'(\bar{P})$ from the anomalous twist, reflecting the fact that the Regge trajectories are no longer exactly linear.}  Altogether, \eqr{eq:OPEDensity} translates into the following corrections to OPE data:
\begin{align}\label{nonholope}
	\delta\alpha_m^{(\a_t,\bar\a_t)}(\bar P) &= \left.C_{11t}C_{22t} \frac{\bar{\kernel}_{\bar\alpha_s\bar\alpha_t}}{\bar{\kernel}_{\bar\alpha_s\id}}  \Res_{\alpha_s=\,\alpha_m} \left(\frac{\kernel_{\alpha_s\alpha_t}}{ \kernel_{\alpha_s\id}}\right)\right|_{\bar\alpha_s = {Q\over 2}+i\bar P}\\
\delta\rho_{m}^{(\a_t,\bar\a_t)}(\bar P) &= -\left.2\pi C_{11t}C_{22t} \bar\kernel_{\bar\alpha_s\bar\alpha_t} \Res_{\alpha_s=\,\alpha_m}\kernel_{\alpha_s\alpha_t}\right|_{\bar\alpha_s = {Q\over 2}+i\bar P}
\end{align}
The residue in the first line equals the ratio of \eqr{eq:HigherResidues} and \eqr{eq:HigherResidues2}. Reading off the ratio of the \ahol kernels appearing in \eqr{nonholope} from \eqr{eq:AsymptoticRatioOfKernels} (which also includes the coefficient, suppressed here), we derive the large spin decay of anomalous twist:
\begin{equation}\label{dhlargespin}
	\delta \alpha_m^{(\alpha_t,\bar\alpha_t)} \approx e^{-2\pi \bar{\alpha}_t \sqrt{\ell_s}}\quad\implies \quad \delta h_m^{(\alpha_t,\bar{\alpha}_t)} \approx e^{-2\pi \bar{\alpha}_t \sqrt{\ell_s}}
\end{equation}
If $\op_t^*$ is the lowest twist operator appearing in the T-channel (necessarily in the discrete range, since at a minimum there are discrete double-twist families $\{\op_1\op_1\}$ and/or $\{\op_2\op_2\}$), then the leading corrections to VMFT data at large spin decay as $e^{{-2\pi \bar{\alpha}_t^* \sqrt{\ell_s}}}$.\footnote{In the presence of mixing among degenerate double-twist operators, one must diagonalize the Hamiltonian. See \cite{Simmons-Duffin2017} for some useful technology and a worked example involving approximate numerical degeneracy in the 3D Ising model, and \cite{Aprile:2018efk,Alday:2017xua,Caron-Huot:2018kta} for double-trace mixing in planar 4D $\cN=4$ super-Yang-Mills at strong coupling.}

\subsubsection{Large-spin continuum}
%Apart from this finite number of discrete Regge trajectories, there must also be operators to reproduce the continuum $h_s>\frac{c-1}{24}$ spectral density at large spin. 
The continuous sector of VMFT contributes to the four-point function as 
\begin{equation}
	G\supset \int_0^\infty dP\int_0^\infty d\bar P~\rho^{(0)}_{\text{cont.}}(P,\bar P)\mathcal{F}_S(\tfrac{Q}{2}+iP)\mathcal{F}_S(\tfrac{Q}{2}+i\bar P),
\end{equation}
where the continuum OPE spectral density $\rho_{\text{cont.}}^{(0)}$ is given by
\begin{equation}\label{eq:rhoBTZ}
	\rho_{\text{cont.}}^{(0)}(P,\bar P) = \left.\kernel_{\alpha_s\id}\bar\kernel_{\bar\alpha_s\id}\right|_{\alpha_s = {Q\over 2}+iP,\,\bar\alpha_s = {Q\over 2}+i\bar P}.
\end{equation}

Additional T-channel operators give corrections to this: the contributions of T-channel operators with positive twist,
\begin{equation}
	\delta\rho_{\text{cont.}}^{(\alpha_t,\bar\alpha_t)}(P,\bar P) = \left.\kernel_{\alpha_s\alpha_t}\kernel_{\bar\alpha_s\bar\alpha_t}\right|_{\alpha_s={Q\over 2}+iP,\,\bar\alpha_s = {Q\over 2}+i\bar P}~,
\end{equation} 
are exponentially suppressed compared to the VMFT density \eqref{eq:rhoBTZ}, again due to  \eqref{eq:RatioOfKernels}. A similar universality in the density of states at large spin, with the same twist gap assumption, follows from a `lightcone modular bootstrap' for the partition function \cite{Afkhami-Jeddi2018,Collier2018a}; those results require that as $\ell_s\to\infty$, any interval of twist above this threshold contains infinitely many operators. Arranging operators into analytic families, our requires infinitely many Regge trajectories with $h_s$ accumulating to $\frac{c-1}{24}$ at large spin, thus refining the conclusion already reached in \cite{Afkhami-Jeddi2018,Collier2018a}.

\subsubsection{Comments}
%\subsubsection*{Convergence at large spin and resolution into a discrete spectrum at finite spin}
\paragraph{Asymptotic in what sense?}
The argument so far only shows that contributions from a finite number of operators in the T-channel are negligible at large spin in the S-channel. This does not rule out a significant contribution to the S-channel spectral density from a sum over infinitely many T-channel operators. Indeed, such contributions must be present to resolve the spectral densities $\rho_m,\rho_{\text{cont.}}$ into sums of delta functions supported at integer spins. % and for the continuum range into a discrete spectrum of twists for a compact theory. 
The most conservative statement is that the asymptotic formula applies to the integrated spectral density; for example, the sum of OPE coefficients $C_{12s}^2$ for discrete double-twist operators in the $m$th Regge trajectory up to a given spin is asymptotic to the integral of $\rho^{(0)}_m(P)$ up to the corresponding value of $P$. Such a conclusion would follow from a Tauberian argument along the lines of \cite{Qiao2017,Mukhametzhanov2018}. It is likely that a stronger statement holds, at least in sufficiently generic CFTs (obeying an appropriate version of the eigenstate thermalisation hypothesis \cite{Srednicki1994}), in which the asymptotic formula may apply to a microcanonical average over a sufficiently wide range of spin and twist. We leave the questions of how a sum over T-channel operators reproduces a discrete S-channel spectrum, and of more precise and rigorous formulations of the asymptotic formulas, to future work.

Sums over infinite sets of T-channel operators can also lead to S-channel Regge trajectories approaching any twist at large spin, not only those of the VMFT spectrum.  However, the twist gap determines that the spectral density of such trajectories is suppressed compared to \eqref{eq:VMFTOPE} , by an exponential in $\sqrt{\ell}$ \eqref{eq:RatioOfKernels}.

\paragraph{Nachtmann from Virasoro}
For identical operators $\op_1=\op_2$, we can derive a Virasoro version of Nachtmann's theorem at large spin from \eqr{nonholope}. As long as the leading twist correction comes from $\alpha_t<2\alpha_1$, we find that the coefficient of $e^{-2\pi \bar{\alpha}_t \sqrt{\ell_s}}$ in the ratio of antiholomorphic kernels is positive, the residue of the ratio of holomorphic kernels is negative, and the T-channel OPE coefficients appear squared. This implies $\d\a_0^{(\a_t,\bar\a_t)}<0$. The change in $h$ is determined by $h=\a(Q-\a)$, which, for $2\a_1<Q$, implies 
\e{nacht}{\d h_0^{(\a_t,\bar\a_t)}<0~.}
That is, the leading large spin correction to the twist of the first Regge trajectory is negative, so this trajectory is convex.

\paragraph{Comparison to previous lightcone analyses}

The arguments we use make no direct reference to the `old-fashioned' lightcone bootstrap approach of solving the crossing equations $\eqref{crossingEq}$ in the lightcone limit in $d>2$. We comment on the connection in appendix \ref{app:OldLC}. The most difficult part of this analysis is to evaluate the S-channel blocks in a combined limit of $\bar{h}_s\to \infty$ and $\bar{z}\to 1$, with an appropriate combination held fixed, to reproduce the lightcone singularity by a saddle point in $\bar{h}_s$. 

Previous work on the large spin expansion in CFT$_2$ \cite{Fitzpatrick2014} computed the asymptotic twist of the Regge trajectories in a large central charge limit, taking $\frac{h_1}{c}$ fixed in the limit, and $\frac{h_2}{c}$ small. These results are reproduced simply by taking appropriate limits of the addition of momentum variables, $\alpha_s=\alpha_1+\alpha_2+m b$. We further explain how our work extends and clarifies that of \cite{Fitzpatrick2014} in section \ref{sec:gravity} where we discuss semiclassical limits.

\subsection{Large spin and large $c$}\label{sec:LargeSpinLargeC}

Our analysis in the previous subsection has focused on the regime of large spin, $\ell_s\gg c$, where there is a universal form for the anomalous momentum due to T-channel exchanges, given in (\ref{nonholope}), with the feature that the result decays exponentially in the square root of the spin (\ref{dhlargespin}). This follows from the asymptotic form of the ratio of the non-vacuum to vacuum kernels computed in appendix \ref{app:largeweight}. In the large-$c$ limit, (\ref{eq:AsymptoticRatio}) reduces to the following
\begin{equation}\label{eq:LargeCLargeSpinRatio}
	\frac{\bar\kernel_{\bar\alpha_s\bar\alpha_t}}{\bar\kernel_{\bar\alpha_s\id}}\sim\left({12\pi\over c}\right)^{2\bar h_t}{\Gamma(2\bar h_1)\Gamma(2\bar h_2)\over\Gamma(2\bar h_1-\bar h_t)\Gamma(2\bar h_2-\bar h_t)}e^{-2\pi\bar h_t\sqrt{6\ell_s\over c}} \qquad (\ell_s \gg c \gg 1)~.
\end{equation}
In this section we will study the anomalous weight $\delta h_m^{(\alpha_t,\bar\alpha_t)}$ in the large-$c$ limit, fixing the ratio $\ell_s\over c$.\footnote{We thank Alex Belin and Davids Meltzer and Poland for suggesting this.} 

From (\ref{nonholope}) we see that we need to be able to compute the ratio of the non-vacuum to vacuum antiholomorphic fusion kernels in the limit that the S-channel internal weight scales with the central charge. We perform this computation in appendix \ref{app:LargeCLargeSpinRatio}. Recalling that $\bar{h}_s \sim \ell_s$ at fixed $h_s$, we parameterize $\bar \a_s$ as
\ie
\bar\alpha_s = \frac{Q}{2}+i\bar p_sb^{-1} \, ,\quad \text{with }\bar{p}_s \sim \tfrac{1}{2}\sqrt{\tfrac{24\ell_s}{c}-1}~.
\fe
In this limit, we find
\ie\label{eq:largecAnom2}
\frac{\bar\kernel_{\bar\alpha_s\bar\alpha_t}}{\bar\kernel_{\bar\alpha_s\id}}\sim {\Gamma(2\bar h_1)\Gamma(2\bar h_2)\over\Gamma(2\bar h_1-\bar h_t)\Gamma(2\bar h_2-\bar h_t)}\left({c\over 6\pi}\cosh(\pi \bar p_s)\right)^{-2\bar h_t} \qquad~ (c \gg 1~,~ \tfrac{\ell_s}{c} \text{ fixed}).
\fe
This interpolates between \eqr{eq:LargeCLargeSpinRatio} in the large spin regime with $\bar p_s \sim \sqrt{6\ell_s\over c}\to\infty$, and the power law familiar from $d>2$ at small spin $1 \ll \ell_s \ll c$: in that limit, we have $\bar p_s\approx {i\over 2}(1-{12 \ell_s\over c})$, giving $\left(\cosh(\pi \bar p_s)\right)^{-2\bar h_t} \sim \ell_s^{-2\bar h_t}$.

To compute the anomalous dimension in this limit, we put \eqref{eq:largecAnom2} together with the holomorphic fusion kernel as in \eqref{nonholope}. The appropriate limits of the residues which appear will be obtained in equations \eqref{R0} and \eqref{dResnonvacglobal} in the next section, giving the following result at large $c$, with $\ell_s$ of order $c$: 
\ie\label{eq:PlanckianSpinGeneral}
\delta h_m^{(\alpha_t,\bar\alpha_t)}\sim& -C_{11t}C_{22t}\frac{\Gamma(2\bar h_1)\Gamma(2\bar h_2)}{\Gamma(2\bar h_1-\bar h_t)\Gamma(2\bar h_2-\bar h_t)}\left(\frac{c}{6\pi}\cosh(\pi \bar{p}_s)\right)^{-2\bar{h}_t}\\
&\times{\Gamma(2h_t)\over \Gamma(h_t)^2}~{}_4 F_3\left(
\begin{array}{c}
1-h_t,h_t,2h_1+2h_2+m-1,-m\\
1,2h_1,2h_2
\end{array}\middle| 1\right)
\fe
Focusing on the leading Regge trajectory at $m=0$, in the limit $1\ll \ell_s\ll c$ we recover the previously known lightcone bootstrap result for the anomalous dimension (for example, (B.33) of \cite{Komargodski2013}, recalling that $\delta h = \gamma/2$):
\ie\label{eq:PlanckianSpinGlobal}
\delta h_0^{(\alpha_t,\bar\alpha_t)}\sim -C_{11t}C_{22t}\frac{\Gamma(2\bar h_1)\Gamma(2\bar h_2)}{\Gamma(2\bar h_1-\bar h_t)\Gamma(2\bar h_2-\bar h_t)}{\Gamma(2h_t)\over \Gamma(h_t)^2}\,\ell_s^{-2\bar h_t}.
\fe
The $m$-dependence of \eqref{eq:PlanckianSpinGeneral} is quite a bit simpler than previous recursive results in the lightcone bootstrap literature in $d\geq 2$ spacetime dimensions. 

\subsection{Large conformal dimension}\label{sec:largedim}

The data of VMFT is universal at large spin because the fusion kernel for T-channel operators of positive twist is suppressed in this limit compared to the vacuum. The same argument applies to the limit of large dimension $\Delta_s\to \infty$ at fixed spin $\ell_s$ (or, more generally, for any limit where $h_s,\bar{h}_s\to\infty$), in any unitary compact $c>1$ CFT; in this case, only a gap in conformal dimension above the vacuum is required, not in twist.

 One way of expressing this is as a microcanonical average of OPE coefficients, after dividing by the asymptotic density of primary states. This density is similarly universal due to modular invariance, and can be expressed as the modular S-matrix dual to the vacuum, in close analogy to the vacuum fusion kernel for four-point functions. The modular S-matrix decomposes the character of the vacuum module into characters in the modular transformed frame \cite{Zammod}:
 \begin{equation}\label{eq:modularS}
 	\chi_\id(-1/\tau) = \int_0^\infty dP\; \kernel^\text{mod}_{\alpha \id}   \chi_P(\tau),\quad  \kernel^\text{mod}_{\alpha \id}= 4\sqrt{2} \sinh(2\pi b P)\sinh(2\pi b^{-1}P),\quad  \alpha=\frac{Q}{2}+i P
 \end{equation}
At large $P$, $\kernel^\text{mod}_{\alpha \id}$ is exponentially larger than the corresponding modular kernel for operators of positive dimension, so is asymptotic to the density of primary states; this is a refined version of Cardy's formula.

Including this, we find the expression
\begin{equation}\label{eq:c12savg}
	\overline{ C_{12s}^2} \sim \frac{\kernel_{\alpha_s \id}}{\kernel^\text{mod}_{\alpha_s \id}}~\times (\text{antiholomorphic}), \quad \text{where}\quad\a_s = {Q\o2}+i P ~\text{and}~P\to \infty
\end{equation}
for the square of OPE coefficients, where the bar denotes an average over all primaries with momentum close to $\alpha_s$. The asymptotic form for the fusion kernel is given in \eqref{largePvac}, and for the modular S-matrix \eqref{eq:modularS} we have $\kernel^\text{mod}_{\alpha_s \id}\sim \sqrt{2}e^{2\pi Q P}$.

This limit was studied in \cite{Das2017} using crossing symmetry for four-point functions with Euclidean kinematics. The calculations there used large internal dimension results for conformal blocks \cite{Zamolodchikov1984} before taking $z\to 1$. Their results closely resemble \eqref{eq:c12savg}, but the discrepancy at subleading orders demonstrates the delicate nature of the order of these two limits of the blocks.\footnote{We thank Shouvik Datta for discussions of this point.} 

Finally, we note that in analogy with the above, the results of section \ref{sec:LargeSpinLargeC} imply analogous averaged OPE asymptotics in the regime of large $\Delta_s$ and large $c$ with fixed ${\Delta_s\over c}$. 

\section{Global limit}\label{sec:global}
In this section we will study the global limits of the fusion kernel and the Virasoro double-twist OPE data, for pairwise identical external operators. This is the limit of large central charge $c\to \infty$ with fixed scaling dimensions $h,\bar{h}$, named because the Virasoro conformal blocks reduce to the global $\mathfrak{sl}(2)$ blocks. By including $1/c$ corrections to this limit, it is a simple exercise to extract the large central charge expansion of  double-twist OPE data due to non-unit operators.

We rewrite this limit in terms of the momentum by inverting the relation $h=\alpha(Q-\alpha)$ and expanding in the limit $b\to 0$:
\e{eq:globala}{\a(h)= b h + \mathcal{C}(h)(b^3+b^5(2h-1))+\O(b^7)}
We have written this in terms of quadratic Casimir for $\mathfrak{sl}(2)$,
\e{}{\mathcal C(h) \equiv h(h-1)}
It is interesting to note that to all orders beyond $\O(b)$, the expansion is proportional to $\mathcal{C}(h)$. 

As the global limit is approached, more and more operators with momenta $\alpha_s = \alpha_1 +\alpha_2+m b$ (with corresponding weights $h_s = h_1+h_2+m+\mathcal{O}(b^2)$) cross the contour in the S-channel decomposition of the T-channel block and give discrete residue contributions. At $b=0$, these become precisely the global double-twist operators of MFT, and the OPE data from the $m$'th Virasoro family becomes that of the $m$'th global family of MFT. 

A natural expectation is that, in the global limit, the fusion kernel should reduce to a ``holomorphic half'' of the $6j$ symbol for the global two-dimensional conformal group $SO(3,1)$. The $6j$ symbol, which holomorphically factorizes, was recently computed in \cite{Liu2018}. We will find that all double-twist OPE data extracted from the Virasoro fusion kernel in the global limit is reproduced by the chiral inversion integral
\e{omega}{\Omega_{1-h,h_t,h_2+h_3}^{h_1,h_2,h_3,h_4}\equiv \int_0^1 {dz \o z^2} \left({z\o 1-z}\right)^{h_2+h_3} z^{h_{13}}\,k^{\tilde h_1,\tilde h_2,\tilde h_3,\tilde h_4}_{2(1-h)}(z)\,k^{h_3,h_2,h_1,h_4}_{2h_t}(1-z)}
where
\e{}{k_{2\b}^{h_1,h_2,h_3,h_4}(z) \equiv z^\b {}_2F_1(\b-h_{12},\b+h_{34},2\b,z)~.}
with $h_{ij}\equiv h_i-h_j$. This integral and notation were introduced in \cite{Liu2018}, where the $6j$ symbol was computed essentially as $\Omega_{1-h,h_t,h_2+h_3}^{h_1,h_2,h_3,h_4}$ times an \ahol partner. This integral was evaluated in \c{Liu2018} as a sum two terms, each of which is a ratio of gamma functions times a ${}_4F_3$ hypergeometric function, and may also be written as a ${}_7F_6$ hypergeometric function \c{gro1, gro2, Mertens:2017mtv, Gopakumar:2018xqi, Cardona:2018qrt}. With respect to the Virasoro fusion kernel for pairwise identical operators, the precise statement is that the double-twist poles and residues of $\Omega_{1-h,h_t,2h_2}^{h_1,h_2,h_2,h_1}$ match those of $\kernel_{\a_s\a_t}$ in the global limit, i.e. 
\e{kernellim}{\lim_{b\to 0}\left(\Res_{\a_s=\a_1+\a_2+m b}\kernel_{\a_s\a_t}\right) = \Res_{h=h_1+h_2+m}\Omega_{1-h,h_t,2h_2}^{h_1,h_2,h_2,h_1}}%
and
\e{kernellim2}{\lim_{b\to 0}\left(\dRes_{\a_s=\a_1+\a_2+m b}b^{-1}\kernel_{\a_s\a_t}\right) = \dRes_{h=h_1+h_2+m}\Omega_{1-h,h_t,2h_2}^{h_1,h_2,h_2,h_1}}%
The double-twist singularities of \eqref{omega} come from the region of integration near the origin. We have refrained from discussing the global limit of $\kernel_{\a_s\a_t}$ itself because it does not exist, due to an overall oscillating prefactor. For example, the following global limit of the vacuum kernel is well-defined away from the poles:
\begin{equation}
	b\frac{\sin(b^{-1}\pi(Q-\alpha_1-\alpha_2-\alpha_s))}{\sin(b^{-1}\pi(Q-2\alpha_s))}\kernel_{\alpha_s\id}\to \frac{\Gamma(h_1+h_2-h_s)\Gamma(h_1-h_2+h_s)\Gamma(-h_1+h_2+h_s)\Gamma(h_1+h_2+h_s-1)}{2\pi\Gamma(2h_s-1)\Gamma(2h_1)\Gamma(2h_2)}
\end{equation}
However, the ratio of sines in the prefactor simplifies to give a $(-1)^m$ at double-twist values $\alpha_s=\alpha_1+\alpha_2+m b$, so the residues of the kernel at these poles have a well-defined limit, even though the kernel itself does not. An analogous statement applies to the non-vacuum kernel; this is consistent with the cross-channel decomposition of the global blocks, which only depends on the (d)residues (see (\ref{dResnonvacglobal})-(\ref{eq:globalDecomposition}) for the explicit expressions).

We will support \eqr{kernellim}--\eqr{kernellim2} with several calculations. This match is a non-trivial check of our use of the fusion kernel and the global $6j$ symbols to extract double-twist data in unitary CFTs: the fusion kernel and the global $6j$ symbol require analytic continuation away from different ranges of conformal weights,\foot{Recall that the kernel is formally defined only for operators in the continuum, $\a = {Q\o2}+iP$ with $P\in\mathbb{R}$. Similarly, the global $6j$ symbol is formally defined only for operators in the principal series $h={1\o2}+is$ with $s\in\mathbb{R}$. For any $c>1$, these two ranges do not match. However, we note that at large $c$ (small $b$), the momentum for a continuum operator with $P\approx b^{-1}p$ becomes $\a\sim b^{-1}\left({1\o2}+i p\right)$, i.e. $b^{-1}$ times the principal series conformal weight. Why there is such a correspondence between the momentum and the weight is not completely clear to us. Similar phenomena were observed in \c{Mertens:2017mtv}.} but at large $c$, the double-twist data so extracted do match. We have not determined the correspondence between regular terms of the Virasoro and global $6j$ symbols in the $b\to 0$ limit, but it would be worth doing so.

\subsection{Vacuum kernel}\label{globalsecvac}
The exact vacuum kernel was given in \eqr{vackernel}. The double-twist residues admit the small $b$ expansion %
\ie
&2\pi \Res_{\alpha_s=\alpha_m}\kernel_{\alpha_s\id}= R^{(0)}_{12}(m)+b^2R^{(2)}_{12}(m) + \O(b^4)
\fe
First let us extract the MFT OPE data by studying the residues of the kernel in the $b\to 0$ limit. The global limit of the residues (\ref{eq:HigherResidues}) is given by the following
%where
%
\e{R0}{R^{(0)}_{12}(m)\equiv -{(2h_1)_m(2h_2)_m(2h_1+2h_2-1)_m\over 4^m m! (h_1+h_2-{1\over 2})_m(h_1+h_2)_m}}
The result \eqref{R0} can be seen to satisfy \eqr{kernellim}. The relation to MFT OPE coefficients, which were derived in \cite{Fitzpatrick2012}, is
\e{}{ (C_{12[12]_{n,\ell}}^2)^{\rm MFT}  = (-1)^\ell R^{(0)}_{12}(n)R^{(0)}_{12}(n+\ell)}
 Equivalently, if one treats the chiral Virasoro index $m$ as the spin, $ (C_{12[12]_{0,\ell}}^2)^{\rm MFT} = (-1)^{\ell+1}R^{(0)}_{12}(\ell)$.

Moving beyond leading order, expanding the vacuum kernel near the $m$'th pole gives corrections to MFT OPE data for the double-twist operators $[\O_1\O_2]_{m,\ell}$ in the small $b$ -- that is, $1/c$ -- expansion. Corrections at $\O(b^{2p})\sim \O(c^{-p})$ are due to exchange of composite operators in the Virasoro vacuum module, of schematic form $:T^q:$ with $q\leq p$. These are dual to multi-graviton states in the bulk.

At first non-trivial order, corrections are due to $T$ exchange. The first correction to the MFT OPE coefficients may be read off from the global limit of (\ref{eq:HigherResidues}), keeping in mind the fact that $\alpha_i\sim h_ib+ \mathcal{C}(h_i)b^3+\mathcal{O}(b^5)$: 
\es{}{R^{(2)}_{12}(m) =& R^{(0)}_{12}(m) \bigg(m^2-4h_1h_2 + 2{\mathcal{C}(h_1)+\mathcal{C}(h_2)\over 2h_1+2h_2+2m-1}-2{\mathcal{C}(h_1)+\mathcal{C}(h_2)\over 2h_1+2h_2+m-1}\\
 &-2\mathcal{C}(h_1)(\psi(2h_1)+\psi(2h_1+2h_2+2m)-\psi(2h_1+m)-\psi(2h_1+2h_2+m))\\
 &-2\mathcal{C}(h_2)(\psi(2h_2)+\psi(2h_1+2h_2+2m)-\psi(2h_2+m)-\psi(2h_1+2h_2+m))\bigg)
}
At $m=0$, this is simply $R^{(2)}_{12}(0) = 4h_1h_2$. More physically interesting are the anomalous dimensions in the small $b$ expansion, for which we only need to expand our result $\delta h_m = -2 (\a_1+mb) (\a_2+mb)+m(m+1)b^2$, cf \eqref{anomtwist}, to any desired order. It is slightly more enlightening to expand the location of the $m$'th pole in small $b$:
\es{}{{1\o \a_s-(\a_1+\a_2+m b)} = {b^{-1}\o h_s-(h_1+h_2+m)} + {b (\mathcal{C}(h_1)+\mathcal{C}(h_2)-\mathcal{C}(h_s))\o (h_s-(h_1+h_2+m))^2} + \O(b^3)}
Plugging in the location of the pole gives the anomalous twist due to $T$ exchange,
\es{gammamt}{\delta h_{m}\big|_T &= b^2(\mathcal{C}(h_1)+\mathcal{C}(h_2)-\mathcal{C}(h_1+h_2+m))=b^2\left(-2 h_1h_2+m(1-m-2h_1-2h_2)\right)}
Again, \eqref{gammamt} can be seen to match the result derived from \eqref{omega} and \eqref{kernellim}--\eqr{kernellim2}, and matches previous results of \cite{Liu2018, Kraus:2018zrn}.\foot{One can use equation (3.55) of \c{Liu2018} to write down the contribution of a T-channel block for holomorphic current exchange, with conformal weights $(0,h)$ where $h\in\ZZ$:
\e{anomT}{\g_{0,\ell}\big|_{(0,h)} = -\frac{2 \Gamma (2 h)}{\Gamma (h)^2}C_{\phi\phi h}^2}
where we have taken identical external scalars $\phi$ for simplicity. Using $h=2$, $C^2_{\phi\phi T} = 2h_\phi^2/c \sim b^2h_\phi^2/3$ and the fact that the total anomalous dimension equals twice the change in $h$ (i.e. $\g_{0,\ell}\big|_T = 2\delta h_0\big|_T$), we find agreement with \eqref{gammamt}. The fact that we take the current to have weights $(0,h)$ rather than $(h,0)$ follows from a choice of convention in \cite{Liu2018}.} The appearance of the Casimirs was recently observed in \cite{Kraus:2018zrn} as a curiosity. We now give a different angle on this: the Casimir emerges naturally from the Virasoro vacuum kernel in the global limit thanks to \eqref{eq:globala}. Note that due to the all-orders appearance of $\mathcal{C}(h)$ in \eqr{eq:globala},
\e{}{\delta h_0\propto \mathcal{C}(h_1+h_2)-\mathcal{C}(h_1)-\mathcal{C}(h_2)}
to all orders in $b$. 

A notable feature of the results \eqref{gammamt} and \eqref{anomT} is their spin-independence: that is, $\g_{n,\ell}|_{(0,h)}$ depends only on the twist $n$. This follows from three facts: currents have vanishing twist and give a constant contribution at large spin; $\g_{n,\ell}$ is analytic in spin, up to contributions not captured by the Lorentzian inversion formula; and analytic functions of a single complex variable that are constant at infinity are constant everywhere. Viewing the anomalous twist \eqref{anomdim} as the resummation of an infinite number of twist-zero stress tensor composite contributions, this explains the linearity of the VMFT Regge trajectories.  

\subsection{Non-vacuum kernel}
Let us also perform the small $b$ expansion of the non-vacuum kernels. We focus on the discrete poles. The double pole coefficients are given by the expansion of (\ref{eq:HigherResidues2}) in the global limit %
\ie\label{dResnonvacglobal}
\dRes_{\alpha_s=\,\alpha_m}\kernel_{\alpha_s\alpha_t}=& \sum_{n=0}^m{\Gamma(2h_t)\Gamma(h_t+n)\Gamma(2h_1+m)\Gamma(2h_2+m)\Gamma(2h_1+2h_2+m+n-1)\over 2\pi(n!)^2(m-n)!\Gamma(h_t)^2\Gamma(h_t-n)\Gamma(2h_1+n)\Gamma(2h_2+n)\Gamma(2h_1+2h_2+2m-1)}b+\mathcal{O}(b^3)\\
=&{\Gamma(2h_t)(2h_1)_m(2h_2)_m\over 2\pi m!\Gamma(h_t)^2(2h_1+2h_2+m-1)_m}{}_4 F_3\left(
\begin{array}{c}
1-h_t,h_t,2h_1+2h_2+m-1,-m\\
1,2h_1,2h_2
\end{array}\Big| 1\right)b +\mathcal{O}(b^3)\\
\equiv & \,\beta^{(2)}_m(h_1,h_2;h_t)b+\mathcal{O}(b^3).
\fe
This is computed as a finite sum. Similarly, the residues may be extracted by expanding (\ref{eq:NonVacHigherResidues}): 
\ie\label{Resnonvacglobal}
\Res_{\alpha_s = \alpha_m}\kernel_{\alpha_s\alpha_t}=& \sum_{n=0}^m(\partial_m+\partial_n){\Gamma(2h_t)\Gamma(h_t+n)\Gamma(2h_1+m)\Gamma(2h_2+m)\Gamma(2h_1+2h_2+m+n-1)\over 2\pi(n!)^2(m-n)!\Gamma(h_t)^2\Gamma(h_t-n)\Gamma(2h_1+n)\Gamma(2h_2+n)\Gamma(2h_1+2h_2+2m-1)}\\&+\mathcal{O}(b^2)\\
\equiv& \,\beta_m^{(1)}(h_1,h_2;h_t)+\mathcal{O}(b^2).
\fe
From (\ref{sumintblocks}), we see that these residues serve as the coefficients in the decomposition of the global $\mathfrak{sl}(2)$ blocks into global double-twist blocks (and derivatives thereof) in the cross-channel:
\ie\label{eq:globalDecomposition}
F_T(h_t|1-z) =& -2\pi\sum_{m=0}^\infty\bigg[\beta^{(1)}_m(h_1,h_2;h_t)F_S(h_1+h_2+m|z)+\beta^{(2)}_m(h_1,h_2;h_t)\left.\partial_{h_s}F_S(h_s|z)\right|_{h_s=h_1+h_2+m}\bigg],
\fe
where we have introduced the following notation for the $\mathfrak{sl}(2)$ blocks:
\ie
F_T(h_t|1-z) &= (1-z)^{h_t-2h_2}\,{}_2F_1(h_t,h_t;2h_t;1-z)\\
F_S(h_s|z) &= z^{h_s-h_1-h_2}\,{}_2F_1(h_s-h_1+h_2,h_s-h_1+h_2;2h_s;z).
\fe

These results give corrections to $m>0$ MFT OPE data in a compact form. An advantage relative to global conformal approaches is that in the latter, one needs to subtract descendant contributions of $m'<m$ that mix with the $m$th subleading quasiprimary; in the present Virasoro approach, where Virasoro primaries become global primaries in the global limit, this unmixing is not required. At $m=0$, we can quickly check these results against the chiral inversion integral \eqref{omega}. To extract the $m=0$ double-twist data, we extract the singularities near $z=0$, where 
\es{}{\Omega_{1-h,h_t,2h_2}^{h_1,h_2,h_2,h_1}&\sim  -\frac{\Gamma (2 h_t) }{\Gamma(h_t)^2}\int_0^1 {dz \o z}z^{-h+h_1+h_2}\left(\log z+(2 \psi (h_t)-2 \psi(1) )\right)\\
&= {1\o (h-h_1-h_2)^2}{\Gamma(2h_t)\o \Gamma(h_t)^2} + {1\o h-h_1-h_2}\frac{\Gamma (2 h_t)}{\Gamma (h_t)^2 } (2\psi (h_t)-2\psi(1))}
These coefficients match \eqref{dResnonvacglobal} and \eqref{Resnonvacglobal} after accounting for the $2\pi$.\foot{Treating $T$ as a quasiprimary, this formula reproduces \eqref{anomT} upon plugging in $h_t=2$, using the fact that the OPE data is minus the coefficients obtained by Lorentzian inversion \c{Caron-Huot2017}, and that $\g=2\delta h$. It also reproduces $R_{12}^{(2)}(0) = 4h_1h_2$ upon using $C_{12T}^2 = 2h_1h_2/c \sim b^3 h_1h_2/3$ and $\psi(2)-\psi(1)=1$.}

\subsubsection*{Special case: T-channel double-twists}
Finally, we recall that we were also able to give the closed-form expression in \eqref{nonvacdouble} for the fusion kernel when the T-channel operator sits exactly at a Virasoro double-twist momentum, $\a_t=2\a_2$. The fusion kernel given in \eqref{nonvacdouble} has double poles at $\alpha_s =\alpha_1+\alpha_2+mb$ with coefficients that remain finite in the global limit, corresponding to the MFT double-twists.\foot{Following the discussion around \eqr{eq:AsymptoticNonVacKernel}, this should have subleading asymptotics at large $h_s$ relative to the general case $\a_t\neq 2\a_2$, so as to be consistent with the emergence of zeroes in the global $6j$ symbol for the inversion of a T-channel double-twist operator \cite{Liu2018}. One can show that indeed, this is suppressed exponentially. See (\ref{eq:AsymptoticNonVacKernel}) for the leading asymptotics for general $\a_t$, noting in particular the zero at $\a_t = 2\a_2$.} One can again check that \eqref{kernellim}--\eqr{kernellim2} is satisfied.

\section{Gravitational interpretation of CFT results} \label{sec:gravity}

\subsection{Generic $c$}\label{sec:gravity.1}

MFT has an interpretation as the dual of free fields on a fixed AdS background. In the absence of interactions, energies of composite states, and hence conformal dimensions $\Delta$, add. In VMFT, we have added all multi-trace stress tensor exchanges, taking into account the complete contribution of multi-graviton exchanges to two-particle binding energies.\footnote{As an aside, it should be pointed out that the Virasoro vacuum block does not capture the full gravitational dressing of the MFT four-point function as computed from bulk effective field theory. One sign of this is that the Virasoro vacuum block alone does not give a single-valued Euclidean correlation function, in contrast to computations from bulk gravitational effective field theory, order-by-order in $G_N$. To first order, the former includes only the stress-tensor global block, whereas the latter is given by a tree-level Witten diagram for graviton exchange, which also includes the exchange of $[\op_1\op_1]$ and $[\op_2\op_2]$ double traces \cite{DHoker:1999kzh,Heemskerk2009,ElShowk:2011ag,Hijano2016}. This makes it clear what VMFT is missing from the CFT point of view in order to comprise a viable CFT correlator, namely the exchange of double- and, at higher orders, multi-trace operators required for consistency (for example, integer quantised spins). See \cite{Kraus:2018zrn} for a related discussion. In purely gravitational language, this exclusion of multi-traces can be interpreted as neglecting the overlap of the wavefunctions of the different external particles (though including the quantum `fuzziness' of the wavefunction). This was made more precise in \cite{Maxfield:2017rkn}, with a proposal that Virasoro blocks capture all orders in the two-parameter perturbation theory of first-quantised particles coupled to gravity, while omitting contributions which are nonperturbative in this particular expansion.} %
 We have found that this has the remarkably simple effect that momentum $\alpha$ becomes an additive quantity, and nonlinearities are all included in the relation $h=\alpha(Q-\alpha)$. We give this a gravitational interpretation in one particular circumstance below. 

This additivity law holds until reaching the threshold $\alpha=\frac{Q}{2}$, or $h=\frac{c-1}{24}$, which has the gravitational interpretation of reaching the threshold for black hole formation. At large $c$, classical BTZ black holes exist for a range of energy and spin corresponding to $\min(h,\bar{h})>\frac{c}{24}$, at the edge of which lie the extremal rotating black holes; our results are not the first to suggest a quantum shift of the extremality bound to $\frac{c-1}{24}$ \cite{McGough:2013gka,Keller2015,Benjamin2016}. Above this, the VMFT spectrum becomes continuous, so it captures only some coarse-grained aspects of the physics. This chimes with the expectation that, while the thermodynamics of black holes are determined by IR data, resolving a discrete spectrum of microstates in a complete theory depends on detailed knowledge of UV degrees of freedom.

 The universality of MFT at large spin in $d>2$  comes from superposing two highly boosted states, which move close to the boundary confined by the AdS potential on opposite sides of AdS, and are separated by a large proper distance of order $L_\text{AdS}\log{\ell}$ \cite{Komargodski2013, Fitzpatrick2013}. Since interactions fall off exponentially with distance (even if they are strong or even nonlocal on AdS scales), these states become free, and hence well-approximated by the double-traces of MFT. In AdS$_3$, the situation is qualitatively different, because the gravitational potential does not fall off with distance, giving rise to a finite binding energy even at large spin. Because gravity couples universally to energy-momentum, the binding energy is determined by only the dimensions of the contributing operators and $c$, and our result \eqref{anomdim} for the anomalous twist computes this exactly for the discrete Regge trajectories with $\alpha= \alpha_1+\alpha_2+m b$. The binding energy \eqref{anomdim} is always negative, giving a fully quantum version of the attractive nature of gravity.

MFT is not only a good approximation at large spin, but also for large $N$ theories with weakly coupled bulk duals. In AdS$_3$, the same applies to our results in a $c\to\infty$ limit with operator dimensions fixed, in which they reduce to MFT as shown in section \ref{sec:global}; but VMFT also reproduces results from classical gravity when $h$ scales with $c$, so that gravitational interactions are strong. The clearest demonstration of this is the classical computation of the energy of the lightest two-particle bound state. A heavy scalar particle, with action given by its mass $m \gg L_{\rm AdS}^{-1}$ times the worldline proper time, $S=m\int d\tau$, backreacts on the metric by sourcing a conical defect with deficit angle 
\begin{equation}
	\Delta\phi = 8\pi m G_N \, 
\end{equation}
%\e{}{\D\phi = 8\pi m  G_N}
%
where we assumed $m G_N\ll1$. Translating to CFT variables using $2h\sim m L_{\rm AdS}$ and $c=\frac{3L_\text{AdS}}{2G_N}\sim 6Q^2$, the momentum is proportional to the particle mass, $\alpha\sim\frac{mL_\text{AdS}}{2Q}$, and the deficit angle may be written as 
\begin{equation}\label{dphialpha}
	\Delta\phi = \frac{4\pi\alpha}{Q} \, .
\end{equation}
%\e{}{\D\phi = {4\pi\a\o Q}~.}
%
The classical solution for two particles superposed at the centre of AdS simply adds their masses $m$, so the energy of the lightest two-particle state is given, to leading order in $c$, by the addition of momentum $\alpha$, as in VMFT. The relation \eqref{dphialpha} holds even for finite $m G_N$, where the deficit angle is
\begin{equation}
	\Delta\phi = 2\pi (1-\sqrt{1-8 m G_N}) \, .
\end{equation}
This shows that the addition of defect angles, depicted in figure \ref{defectfig}, is the classical, large $c$ version of the fully quantum finite $c$ addition of momenta $\alpha$.\foot{Note that the definition of $\a$ includes the shift $c \to c-1$, thus including some quantum corrections. This, and the validity of the additivity rule for momenta at finite $c$,  suggests that in the AdS$_3$ quantum theory, there is some (perhaps non-geometric) notion of conical defect associated to CFT local operators with large spin and low twist, with ``angle'' $4\pi\a/Q$. These would be analogous to putative quantum black holes whose microstates are dual to CFT local operators above threshold.} 

\subsection{Gravitational interpretation of anomalous twists}\label{sec:anomTwistsGravity}

The anomalous twist $\delta h_m^{(\a_t,\ab_t)}$ due to non-vacuum T-channel exchange maps, in the bulk, to corrections to the VMFT Regge trajectories due to couplings of the double-twist constituents to other bulk matter. For identical operators $\O_1=\O_2$, the negativity of the anomalous twist due to primaries above the vacuum, $\d h_0^{(\a_t,\ab_t)}<0$, shown in \eqr{nacht}, implies that these matter couplings further decrease the binding energy of the leading-twist operator (at least in the case $\a_t<2\a_1$, where this negativity applies), interpreted as the attractive nature of gravity. 

However, the result \eqr{dhlargespin} for the large spin asymptotics is in striking contrast to the analogous result taking only global primaries into account, familiar from $d>2$ \cite{Fitzpatrick2013,Komargodski2013}:
\begin{align}
	\delta h_s &\approx \ell_s^{-2\bar{h}_t} \qquad\qquad (\text{global primaries, }d>2) \\
	\delta h_s&\approx e^{-2\pi \bar{\alpha}_t \sqrt{\ell_s}} \qquad (\text{Virasoro primaries, }d=2) 
\end{align}
The scaling for $d>2$ (also valid for $1\ll \ell_s\ll c$ in $d=2$, as shown in section \ref{sec:LargeSpinLargeC}) is interpreted as the long-distance propagator of an exchanged field, decaying as $e^{-2\bar{h}_t r}$, where $r\sim\log\ell_s$ is the separation between particles in a two-particle global primary state of orbital angular momentum $\ell_s$. In fact, an identical explanation is true for our $d=2$ result, with the discrepancy explained by the fact that we must consider \emph{Virasoro} primaries in the S-channel, which modifies the relation between the CFT operator spin $\ell_s$ and the bulk orbital spin, $\ell_\mathrm{orb}$. This requires performing a conformal transformation on the naive bulk two-particle state such that it is dual to a Virasoro primary; it will turn out that the orbital angular momentum carries most of its spin in descendants. We now demonstrate this in a context of weak gravitational interactions, namely at large $c$ with fixed external conformal weights, but allowing the spin of the two-particle state to take any value.

We first construct classical states of two particles orbiting in the AdS potential at large separation, which will be dual to coherent superpositions of double-twist operators of large angular momentum. A single particle state at the centre of global AdS is created in radial quantisation by inserting the corresponding operator at the origin $z=\bar{z}=0$;  to change the trajectory of the particle we apply a global conformal transformation, which moves the operator insertion to some other $z$, $\bar{z}$, where we take $1-|\bar{z}|\ll 1$ to give large angular momentum. The two-particle state simply inserts two such operators, in such a way that the particles are well-separated:
\begin{equation}
	|\Psi\rangle = \phi_1(z_1,\bar{z}_1=e^{-\epsilon_1})\phi_2(z_2,\bar{z}_2=-e^{-\epsilon_2}) |0\rangle
\end{equation}
The dependence on left-moving coordinates $z_1,z_2$ won't play any significant part here. The norm $\langle\Psi|\Psi\rangle$ is precisely the lightcone limit of the four-point function at small $\epsilon_{1,2}$, and the particles can be interpreted as weakly interacting when this is well-approximated by the product of two-point functions, or in other words dominated by the vacuum operator in the T-channel. In particular, we will suppress the gravitational interactions by taking large $c$; including them would require the holomorphic Virasoro vacuum block.

Under this assumption of weak interactions, the stress-tensor expectation values can now be computed by analyticity and the OPE. In the cylinder frame parameterised by $w=i\log z$, this is given by the Casimir energy on the circle, plus sums of contributions from each particle:
\begin{equation}\label{eq:2particleT}
	\langle \bar{T}(\bar{w}) \rangle_\Psi =\frac{c}{24} - \bar{h}_1 \left(\frac{\sinh \epsilon_1}{\cosh\epsilon_1-\cos \bar{w}}\right)^2 - \bar{h}_2 \left(\frac{\sinh \epsilon_2}{\cosh\epsilon_2+\cos \bar{w}}\right)^2.
\end{equation}
It will be sufficient for us to demand only a classical version of the primary state condition, namely that the one-point function of the stress-tensor $\langle \bar{T}(\bar{w}) \rangle_\Psi=\langle\Psi|\bar{T}(\bar{w}) |\Psi\rangle$ is consistent with $|\Psi\rangle$ being primary (in our limit only the antiholomorphic version will be important). 

Demanding only the quasiprimary condition requires that the expectation values of $\bar{L}_{\pm 1}$ vanish,  and at small $\epsilon_i$ we have $\langle \bar{L}_{\pm 1}\rangle_{\Psi} \sim \frac{\bar{h}_1}{\epsilon_1}-\frac{\bar{h}_2}{\epsilon_2}$ so this is achieved by taking $\epsilon_i \sim \frac{2\bar{h}_i}{\ell_\text{orb}}$ with $\ell_\text{orb}\gg 1$. The spin of the state $|\Psi\rangle$ is then computed by $\langle \bar{L}_0\rangle_{\Psi} \sim \frac{\bar h_1}{\epsilon_1}+\frac{\bar h_2}{\epsilon_2}\sim \ell_\text{orb}$, so $\ell_\text{orb}$ is the `orbital angular momentum'. It therefore determines the separation of the particles, and hence the forces between them, in the same way as for higher dimensions, giving $\delta h_s\sim \ell_\text{orb}^{-2\bar{h}_t}$. 

Now we impose the Virasoro primary condition $\langle \bar{L}_n\rangle_{\Psi}=0$ for all $n\neq 0$, or in other words that $\langle \bar{T}(\bar{w}) \rangle$ is a constant. To achieve this, we act on the state $|\Psi\rangle$ with a conformal transformation, described by a diffeomorphism of the $\bar{w}$ circle, to give a new state $|\Psi'\rangle$. Choosing $\bar{w}'\in \mathrm{Diff}(S^1)$ (so that $\bar{w}'$ is a smooth, monotonic $2\pi$ periodic function of $\bar{w}$), the stress tensor expectation value after the conformal transformation is
\begin{equation}\label{eq:Ttrans}
	\langle\bar{T}(\bar{w}')\rangle_{\Psi'} = \left(\frac{d \bar{w}'}{d\bar{w}}\right)^{-2} \left[\langle\bar{T}(\bar{w})\rangle_{\Psi}-\frac{c}{12}S(\bar{w}';\bar{w})\right] =\frac{c}{24}-\langle\bar{L}_0\rangle_{\Psi'} \quad \text{(constant)},
\end{equation}
where $S(f(w);w)= {f'''(w)\over f'(w)}-{3\over 2}\left({f''(w)\over f'(w)}\right)^2$ is the Schwarzian derivative. The diffeomorphism $\bar{w}'$ is uniquely determined (up to rigid rotations) by the condition that the transformed stress tensor is constant. We will not directly compute the required conformal transformation $\bar{w}'(\bar{w})$, but rather will indirectly determine the relevant information by considering the following ODE:
\begin{equation}\label{eq:monodromyODE}
	\psi''(\bar{w})+\frac{6}{c}\langle\bar{T}(\bar{w})\rangle \psi(\bar{w})=0
\end{equation}
For us, we need only the mathematical fact that this is preserved under $\mathrm{Diff}(S^1)$ if $\psi$ transforms as a weight $-\frac{1}{2}$ primary, $\psi'(\bar{w}') = \left(\frac{d \bar{w}'}{d\bar{w}}\right)^{\frac{1}{2}}\psi(\bar{w})$ (though it is not coincidental that the same equation is familiar from semiclassical computations of Virasoro conformal blocks, as well as solutions of Liouville's equation and Einstein's equations in AdS$_3$, e.g. \cite{Harlow2011,Faulkner:2013yia,Fitzpatrick2014,deBoer:2014sna}). In particular, the monodromy of the ODE around the $w$ circle (as parameterised by the trace of the monodromy matrix $\Tr M$) is invariant under transformations \eqref{eq:Ttrans} of the stress tensor. In the `classically primary' state $|\Psi'\rangle$ this is easily computed:
\begin{equation}\label{mono1}
	\langle\bar{L}_0\rangle_{\Psi'} = \frac{c}{24}(1+4\bar{p}_s^2) \sim \ell_s \implies \Tr M= 2\cosh(2\pi \bar{p}_s)
\end{equation}
Now it remains only to compute the monodromy of \eqref{eq:monodromyODE} with the stress tensor as given in \eqref{eq:2particleT}, and to equate that with \eqr{mono1}. This can be done when $\ell_\text{orb}\gg 1$, so $\epsilon_i\ll 1$ (without assumptions on the relative size of $\ell_\text{orb}$ and $c$ required), by patching together solutions in different regions, as explained in appendix \ref{app:monodromy}. The result is
\begin{equation}\label{eq:monodromyResult}
	\Tr M=2\cosh(2\pi \bar{p}_s) \sim \left(\frac{12 \pi}{c}\right)^2 \frac{4\bar h_1 \bar h_2}{\epsilon_1\epsilon_2}-2 	\implies \ell_\text{orb} \sim \frac{c}{6\pi} \cosh\left(\pi \bar{p}_s\right)
\end{equation}
Note that, depending on the size of $\ell_\text{orb}$ relative to $c$, the monodromy could be large or of order one. We have now determined the `orbital' angular momentum $\ell_\text{orb}$ as a function of the spin of the primary state $\ell_s$, as parameterised by $\bar{p}_s$. This determines the anomalous dimensions as for $d>2$:
\begin{equation}
	\delta h_s\approx \ell_\text{orb}^{-2\bar{h}_t}\sim \left(\frac{c}{6\pi} \cosh\left(\pi \bar{p}_s\right)\right)^{-2\bar{h}_t}
\end{equation}
This precisely matches the result computed from the fusion kernel in the appropriate limit, in section \ref{sec:LargeSpinLargeC}. Not only do we reproduce the anomalous dimensions at large spin, but also interpolate smoothly to the result at $\ell_s\ll c$, which give the results of a global analysis applied to $d=2$:
\begin{align}
	\ell_s\ll c &\implies \ell_\text{orb} \sim \ell_s \\
	\ell_s\gg c &\implies \ell_\text{orb} \sim \frac{c}{12\pi} e^{\pi \sqrt{\frac{6}{c}\ell_s}}.
\end{align}

\section{Semiclassical limits and late-time physics}
\subsection{Heavy-light limit}\label{HHLL}
In this subsection we study the fusion kernel in the semiclassical `heavy-light' limit. Denoting $h_1=H$ and $h_2=h$, the limit is defined as
\begin{equation}\label{heavylightlim}
	\text{Heavy-light limit:}\quad  H\to\infty, \quad c\to\infty \quad \text{with} \quad \tfrac{H}{c}, h \text{ fixed.}
\end{equation}
Our present goal is to compute the T-channel heavy-light blocks of \cite{Fitzpatrick2014,Fitzpatrick2015a} and to understand their corrections from a new perspective. We will mostly discuss the vacuum in the T-channel, because in some appropriate limits the vacuum block dominates other contributions to the full correlation function of theories with bulk duals\footnote{Dominance of the vacuum block alone requires several assumptions. In particular, unless $h\sqrt{\frac{H}{c}}\gg 1$, even if the light operator is dual to a bulk free field, its double-trace contributions are not suppressed, but are required to give the sum over images for the full two-point function in the background created by the heavy operator.}.

\begin{figure}
	\centering
	\includegraphics[width=.6\textwidth]{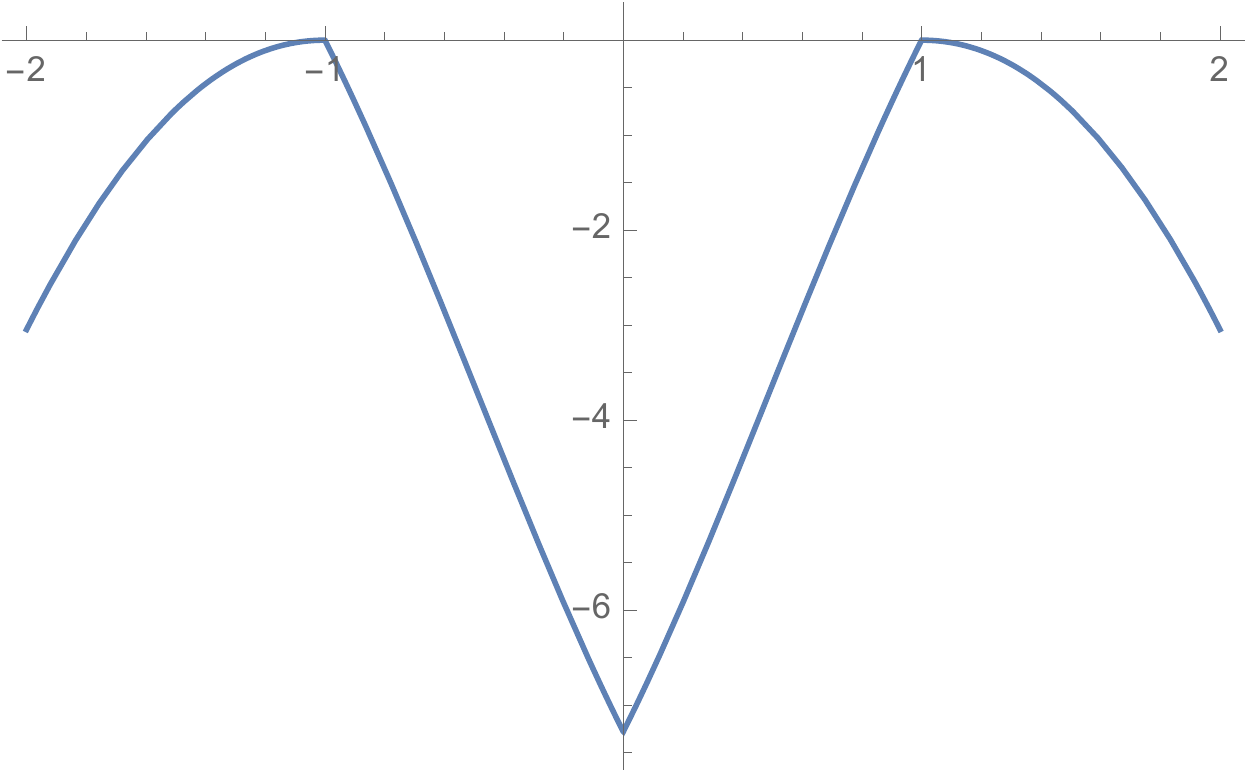}
	\caption{The order $b^{-2}$ piece of $\log\kernel_{\id \alpha_s}$, with $p=1$, as a function of $p_s$.\label{fig:HHLLid}}
\end{figure}

First, we take the fusion kernel for T-channel vacuum exchange, parameterized as
\begin{equation}
	\alpha_1=\frac{Q}{2}+i b^{-1}p, \quad \alpha_2 \sim b h, \quad \alpha_s=\frac{Q}{2}+i b^{-1}p_s, \quad \alpha_t=0
\end{equation}
The limit \eqr{heavylightlim} corresponds to the $b\to 0$ limit with $p,p_s,h$ fixed. In terms of the dimensions, this means that
\e{Hpeq}{H \sim\frac{c-1}{24} (1+4p^2),}
so $p\in\mathbb{R}$ if the heavy operator is above the black hole threshold. The momentum $p$ is often parameterized by an effective inverse ``temperature'' $p = {\pi/\beta}$. It is straightforward to take the limit of the expression \eqref{vackernel}, using the semiclassical limits of the special function $\Gamma_b$  in section \ref{sec:scGammab}. To leading order, we find
\begin{align}\label{eq:scIdKernel}
	\lim_{b\to 0} b^2\log &\kernel_{\alpha_s\id} = J(2p)+J(2p_s)-2J(p-p_s)-2J(p+p_s)-2\pi (|p|-|p_s|) \Theta(|p|-|p_s|),\nonumber \\
	\text{where }& J(x):= \int_1^{1+ix} \log\Gamma(t)\,dt+\int_1^{1-ix} \log\Gamma(t)\,dt = F(1+ix)+F(1-ix)-2F(1). 
\end{align}
The main properties of $J$ that will be important are that it is even, analytic on the real line (with branch cuts on the imaginary axis starting at $\pm i$), and vanishes at the origin. The resulting expression is even in $p$ and $p_s$, as required by reflection symmetry, so we can restrict attention to $p,p_s>0$. The kernel is maximal at $p_s=p$, where the leading semiclassical exponent above vanishes, and has a kink there coming from the final term, with derivative $2\pi$ on the left and zero on the right, as shown in figure \ref{fig:HHLLid}.
This kink can be understood as a consequence of the poles at $\alpha_s=\alpha_1+\alpha_2+m b$ (and a reflected line) accumulating close to the $\Re\alpha_s=\frac{Q}{2}$ axis. In the semiclassical limit, these poles coalesce into a branch cut, and in the heavy-light limit, two such cuts meet at $p_s=p$ giving rise to the kink. In a moment we will take an alternative limit where the poles become individually visible, and give them a gravitational interpretation.

In order to compute the T-channel block, we integrate $\kernel_{\a_s\id}$ against S-channel blocks. At this point, it may seem that we gain nothing by using the fusion kernel, since we must have control over the S-channel blocks. However, since the fusion kernel is largest for $p_s$ close to $p$, the integral is dominated by internal S-channel dimensions close to the heavy operator dimension. To zoom in on this regime, we choose $\alpha_s = \alpha_H+i b x$ with $x$ of order one, so that $h_s-H\sim 2 p x$ in the limit. As it turns out, these S-channel blocks simplify as much as one could ever hope for: as observed in appendix E of \cite{Fitzpatrick2014},\foot{All descendants are suppressed by powers of $\frac{(H-h_s)^2}{c}$ and $\frac{h^2}{c}$: projecting the four-point function onto the $h_s$ representation by inserting complete sets of states $L_{-n_1}\cdots L_{-n_k}|h_s\rangle$, the matrix elements appearing are of order $h^k,(H-h_s)^k$, but the normalisation coming from the inverse Kac matrix suppresses by $c^{-k}$.} if $H-h_s,h\ll\sqrt{c} \sim b^{-1}$, the S-channel block reduces to the `scaling block', the contribution of the single primary operator alone:
\begin{equation}
	\mathcal{F}_S(\alpha_s|z) \sim z^{-h_1-h_2+h_s} \sim z^{2px-h}
\end{equation}
Now, to evaluate the identity T-channel block in this limit, we will integrate this S-channel block against the fusion kernel. Writing $\alpha_s = \alpha_H+ i b x$ and taking the large $c$ limit with $x$ fixed, we have
\begin{equation}\label{eq:spectralDensityHHLL}
	\kernel_{\alpha_s \id}\sim b^{-1}(2p)^{2 h}  e^{\pi  x}\frac{1}{2\pi} \frac{\Gamma (h-i x) \Gamma (h+i x)}{\Gamma (2 h)} \quad\text{as }b\to0.
\end{equation}
In particular, we see that this is consistent with the earlier result, since for large $|x|$, we have $\Gamma (h-i x) \Gamma (h+i x)\approx e^{-\pi |x|}$, the exponential scaling with $x$ joining smoothly onto either side of the kink at $p_s=p$. The accumulating poles at $\alpha_s=\alpha_1\pm(\alpha_2+m b)$ that caused the kink are now individually visible.

Now that we have the blocks and the kernel for this region of internal dimensions, we can evaluate this part of the integral,
\begin{equation}\label{eq:xInt}
	\mathcal{F}_T(\id|1-z)\sim  \int_{-x_0}^{x_0}\, dx (2p)^{2 h}  e^{\pi  x}\frac{1}{2\pi} \frac{\Gamma (h-i x) \Gamma (h+i x)}{\Gamma (2 h)} z^{2px-h} 
\end{equation}
for some $x_0$ consistent with the parametric regime articulated above. To begin, we take the bounds $x_0\to\i$. To compute the resulting integral, we choose $z=e^{it}$ for $t$ with a positive real part, allowing us to close the contour in the upper half-plane, where we pick up the residues
\begin{equation}
	2\pi i \Res_{x= i(h+n)} \frac{1}{2\pi}\frac{\Gamma (h-i x) \Gamma (h+i x)}{\Gamma (2 h)} = \binom{-2h}{n},
\end{equation}
These sum to a binomial series, yielding
\e{hlvac}{\mathcal{F}_T(\id|1-z) \sim z^{-h} \left(\frac{2 i p}{z^{-ip}-z^{ip}}\right)^{2h}}
Continuity allows us to extend this to real $z$. This result \eqr{hlvac} is precisely the limit computed in \cite{Fitzpatrick2014} using the monodromy method for large $h$, and extended to finite $h$ in \cite{Fitzpatrick2015a}. In fact, in \cite{Fitzpatrick2014} the inverse Laplace transform of the heavy-light block was used to derive precisely the spectral density \eqref{eq:spectralDensityHHLL}, the inverse of the computation in \eqref{eq:xInt}.

The poles here have a nice gravitational interpretation as quasinormal modes of the BTZ black hole (after combining with antiholomorphic counterparts) \cite{Birmingham2002,Fitzpatrick2014}. Here it is clear that, while their presence dominates the integral over intermediate states, the contour runs between them along real $p_s$, so the poles should not be interpreted as part of the physical spectrum of states. This is important for preservation of unitarity, since they correspond to complex scaling dimensions.

This computation applies when the heavy operator lies above the black hole threshold, but the same form for the semiclassical limit of the vacuum block holds when $\frac{24H}{c}<1$, simply taking $p \to ip$ in \eqr{eq:xInt}. We derive this in a similar way from the fusion kernel in section \ref{NonVacHHLL}, while generalising to allow for a light non-vacuum exchange operator in the T-channel.

\subsubsection{Forbidden singularities}

In the computation above, we have been rather cavalier, taking $x_0\to\i$ and assuming that the integral over S-channel intermediate dimensions is dominated by a particular region. However, for certain kinematics, we already see a sure sign that this cannot possibly hold: for some values of $z$, the integral \eqr{eq:xInt} diverges! Since $\Gamma (h-i x) \Gamma (h+i x)\approx e^{-\pi |x|}$ for large $|x|$, convergence of the integral requires that $e^{-\frac{\pi}{p}}<|z|<1$. This divergence causes a seemingly paradoxical property of the heavy-light blocks, dubbed the `forbidden singularities' \cite{Fitzpatrick2016}, also studied in \cite{Chen2017,Faulkner2018a}. To see this most clearly, write $z=e^{-\tau}$, with $\tau$ interpreted as imaginary time in radial quantisation, and $p=\pi/\beta$:
\begin{equation}\label{eq:HHLLBlock}
	\mathcal{F}_T(\id|1-e^{-\tau}) \sim e^{h\tau} \left(\frac{\beta}{\pi}\sin{\pi\tau\o\beta}\right)^{-2h}
\end{equation}
This (excepting the prefactor, which can be removed by making a conformal transformation to the cylinder) is periodic in $\tau$ with period $\beta$; indeed, this (times its antiholomorphic counterpart) equals the universal CFT$_2$ result for a two-point function on the line at finite temperature. As noted in\cite{Fitzpatrick2016}, though, the singularity as $\tau\to 0$ required by the OPE limit gives rise to further singularities at $\tau=n\beta$, which cannot be present in a correlation function or conformal block (e.g. because of convergence in the unit disk in the nome variable $|q|$ \cite{Zamolodchikov1984}).

We now have a simple explanation for this phenomenon: as we get too close to the first singularity at $\tau = \frac{\pi}{p}$, the suppression of the fusion kernel at smaller values of intermediate dimension is no longer sufficient to overwhelm the S-channel blocks, and the integral \eqref{eq:xInt} starts to pick up significant contributions from large negative $x$, where our approximation for the kernel and S-channel block no longer apply. The full integral over $p_s$ still converges, and the block is finite, but now the integral is dominated by a saddle-point at smaller values of intermediate dimension, and is therefore enhanced exponentially in $c$. This is an example of the equivalence between ensembles breaking down for certain observables: the canonical and microcanonical results are only close for choices of observable which do not significantly shift the saddle-point over energies. Here we see very explicitly (albeit in the context of an individual block, rather than a full microcanonical correlation function) that when the Euclidean time becomes too close to the effective temperature, the intermediate energies (represented by $p_s$) shift to a new saddle point.
 
\subsubsection{Conical defects and non-vacuum exchange}\label{NonVacHHLL}

We now address the case where we remain in the heavy-light limit, but with $\frac{H}{c}<\frac{1}{24}$. Simultaneously, we generalise to allow for non-vacuum exchange in the T-channel, with $h_t$ of order one. From the gravitational point of view, in this range of dimensions the heavy operator is dual to a particle creating a conical defect, rather than a black hole; this qualitative difference is visible from our calculation because the fusion kernel contains discrete contributions from poles at $\alpha_s=\a_m$, which dominate in the semiclassical limit.

We now parameterise the external weights as
\begin{equation}
	\alpha_1 \sim \eta b^{-1},~0<\eta<{1\over 2},~ \alpha_2 \sim hb,
\end{equation}
so $H\sim \frac{c}{6}\eta(1-\eta)$. We compute the coefficients of the poles at $\alpha_s=\a_m$ with $m$ of order one in this limit. For the vacuum T-channel, the residues match those found above in the calculation of \eqref{eq:xInt} with $p\rightarrow i(\eta-\frac{1}{2})$; for nonzero $h_t$, the coefficients of the double pole, from the expressions in Appendix \ref{app:moreKernel}, have the following small $b$ limits:
\begin{align}\label{eq:HHLLNonVacDRes}
\dRes_{\alpha_s=\,\alpha_m} \kernel_{\alpha_s\alpha_t} \sim & \sum_{n=0}^m{(1-2\eta)^{2h-h_t}\Gamma(2h_t)\Gamma(h_t+n)\Gamma(2h+m)\over 2\pi(n!)^2\Gamma(1+m-n)\Gamma(h_t)^2\Gamma(h_t-n)\Gamma(2h+n)}b \nonumber \\
=&{(1-2\eta)^{2h-h_t}(2h)_m\Gamma(2h_t)~{}_3F_2(1-h_t,h_t,-m;1,2h;1)\over 2\pi m!\Gamma(h_t)^2}b \nonumber \\
\Res_{\alpha_s=\,\alpha_m}\kernel_{\alpha_s\alpha_t}\sim & \sum_{n=0}^m(\partial_m+\partial_n){(1-2\eta)^{2h-h_t}\Gamma(2h_t)\Gamma(h_t+n)\Gamma(2h+m)\over 2\pi(n!)^2\Gamma(1+m-n)\Gamma(h_t)^2\Gamma(h_t-n)\Gamma(2h+n)}.
\end{align}
In the large $c$ limit, we can simply sum the contributions of these residues for $m\in\NN$ (the number of poles that have crossed the contour is of order $c$), and furthermore use the scaling block in place of the S-channel block as above, since the poles lie at $h_s-H$ of order one:
\begin{align}
	\mathcal{F}_T(\alpha_t|1-z) =& \int \frac{d\alpha_s}{2i}
	\kernel_{\alpha_s\alpha_t}\mathcal{F}_S(\alpha_s|z)\\
\sim & -2\pi z^{-2h\eta}\sum_{m=0}^\infty\bigg(\Res_{\alpha_s=\,\alpha_m}\kernel_{\alpha_s\alpha_t} +b^{-1}\left(1 -2\eta\right)\dRes_{\alpha_s=\,\alpha_m}\kernel_{\alpha_s\alpha_t}\log z\bigg)z^{m(1-2\eta)}. \nonumber 
\end{align}
This gives a power series in $w:=z^{1-2\eta}$, whose coefficients \eqref{eq:HHLLNonVacDRes} match the expansion of the T-channel global block $k_{2h_t}(1-w)$, but in terms of the new $w$ variable:
\begin{equation}
	\mathcal{F}_T(\alpha_t|1-z)\sim (1-2\eta)^{2h-h_t}z^{-2h\eta}(1-w)^{h_t-2h}{}_2F_1(h_t,h_t;2h_t;1-w).
\end{equation}
This is the same as the result computed by \cite{Fitzpatrick2015a}, who directly summed descendants in a basis of Virasoro generators adapted to the $w$ coordinate. The gravitational interpretation is natural here, as the conformal transformation to the $w$ coordinate maps locally pure AdS onto the conical defect geometry. In CFT language, the $w$ coordinate is the choice such that the stress tensor expectation value in the presence of the heavy operator vanishes. In our calculation, the defect angle comes from the spacing between the VMFT double-twist poles.

The technical details of the calculation are similar to the $\frac{H}{c}>\frac{1}{24}$ case, but the interpretation is rather different: the poles we are summing over should now be interpreted as physical operators which will appear in the S-channel spectrum (albeit with some corrections in the exact correlation function), double-twist composites of one heavy and one light particle.

\subsubsection{Connection to large spin analysis}

The semiclassical heavy-light limit of the block was used by \cite{Fitzpatrick2014} for an analysis at large spin, which we now put into the context of our finite $c$ results of section \ref{sec:largeSpin}. Their computation involved decomposing the T-channel heavy-light identity block into the S-channel, interpreting the result as the asymptotic twist of Regge trajectories at large spin. Our derivation of the heavy-light block -- for example, \eqref{eq:xInt} -- is precisely the inverse of this, so it is clear that their result gives the relevant large $c$ limit of the fusion kernel, with $h_1-h_s$ of order one. Certain aspects of their result become clearer with our new perspective. They decomposed into S-channel quasiprimaries, but saw no sign of Virasoro descendants; we now see that this is a consequence of the suppression of S-channel descendants in the relevant limit. For $\frac{h_1}{c}<\frac{1}{24}$ they saw an infinite number of discrete Regge trajectories, but their result is reliable only for $m\ll \sqrt{c}$, and similarly, for $\frac{h_1}{c}>\frac{1}{24}$ they see a continuum in twist, but are only really sensitive to the range $|h_s-h_1|\ll \sqrt{c}$. Their results for the large $\ell$ asymptotic twist when $\frac{24h_1}{c}<1$, namely
\begin{equation}
	h_s = h_1+\sqrt{1-\frac{24h_1}{c}}(h_2+m),
\end{equation}
follow simply from a large $c$ expansion of $\alpha_s=\alpha_1+\alpha_2+m b$, as long as $m\ll \sqrt{c}$. Taking in addition $\frac{h_1}{c}\ll 1$, this reduces to a `Newtonian limit' with $\frac{h_1h_2}{c}$ fixed as $c\to\infty$, in which the Virasoro vacuum block becomes the exponential of the stress-tensor contribution, and we find, for the leading Regge trajectory,
\e{}{h_s=h_1+h_2-\frac{12 h_1h_2}{c}\quad (\text{Newtonian limit})}
We also reproduce this result from the Lorentzian inversion formula in appendix \ref{app:Newton}.

\subsection{Late time}\label{sec:lateTime}

One interesting choice of kinematics for the four-point function we have been considering gives the a time-ordered two-point function of the light operator $\op = \op_2$ on the Lorentzian cylinder, in the heavy state $|\psi\rangle$ corresponding to $\op_1$:
\begin{equation}
	\langle\psi| \op(t,\phi)\op(0) |\psi\rangle = G(z=e^{-i(t+\phi)},\bar{z}=e^{-i(t-\phi)})
\end{equation}
As we evolve in time, we find singularities when $t\pm\phi$ goes through $2\pi$ times an integer, due to the operators passing through their mutual lightcones. To regulate this (and give the proper time-ordering), $t$ should be given a small negative imaginary part so that $z,\bar{z}$ lie inside the unit circle. Since we will just consider chiral blocks here, the dependence on angle $\phi$ will be suppressed.

It is natural to write this as an expansion over conformal blocks in the T-channel, $\O\O\rar\psi\psi$, since summing over light operators in this channel reproduces the results of gravitational effective field theory; in particular, the heavy-light vacuum block \eqr{hlvac}, with $z=e^{-it}$, equals the result computed from a free particle in a planar BTZ background. But this decays exponentially for all time, as $e^{-2pht}$ (coming from the slowest quasinormal mode, the leading pole in \eqref{eq:xInt}), which is in tension with an S-channel expansion as a discrete sum over operators weighted by a phase $e^{-i\Delta_s t}$. This implies that the correlation function cannot decay forever, but must eventually (at times longer than the inverse level-spacing) fluctuate around a value of order $e^{-S}$, a version of the information paradox \cite{Maldacena2003}.

Here, we will see to what extent the decay is corrected within a single T-channel block once finite $c$ corrections are included, a problem studied numerically in \cite{Chen2017}. We restrict to the case $\text{Re}(\a_1+\a_2)>{Q\o 2}$, where $Q$ is arbitrary. In the S-channel, Lorentzian time evolution simply gives a phase $e^{-it h_s}$. Restricting for ease of presentation to time differences $t\in 2\pi \ZZ$, we have
\begin{align}
	\mathcal{F}_T(\alpha_t|t_0+t) &= \int \frac{d\alpha_s}{2i} \mathcal{F}_S(\alpha_s|t_0) \kernel_{\alpha_s \alpha_t} e^{-i t \alpha_s(Q-\alpha_s)} \\
	&= \int_{\frac{c-1}{24}}^\infty \frac{dh_s}{2\sqrt{h_s-\frac{c-1}{24}}} \mathcal{F}_S(\alpha_s|t_0) \kernel_{\alpha_s \alpha_t} e^{-ith_s},
\end{align}
where $\a_s = {Q\o2}+iP_s$. The last equality makes it clear that this is a Fourier transform with respect to $h_s$. At very late times, this will be controlled by the least smooth feature in $h_s$, and since the blocks and fusion kernel are analytic functions of $\alpha_s$, this must be at $h_s=\frac{c-1}{24}$, $\alpha_s=\frac{Q}{2}$. For generic external dimensions, the fusion kernel has a double zero at this location in $\a_s$, coming from factors of $\Gamma_b(Q-2\alpha_s)\Gamma_b(2\alpha_s-Q)$ in the denominators of \eqref{eq:CrossingKernel} and \eqref{vackernel}. Since $dh/d\a=Q-2\a$, this becomes a simple zero in $h_s$. This means that the spectral density in terms of $h_s$ begins at $\frac{c-1}{24}$ with a square root edge, and after taking the Fourier transform, the result is that the block decays as $t^{-3/2}$. Refining this slightly, we can also find the coefficient of this late time decay analytically, at least for the vacuum block. Writing $\alpha_s=\frac{Q}{2}+iP_s$, the calculation of the integral at late time is most naturally done by stationary phase, dominated by $P_s=0$. Taking into account the double zero of the kernel, we find
\es{fulllatetime}{
	\mathcal{F}_T(\alpha_t|t+t_0) &= \int_0^\infty dP_s \mathcal{F}_S(\alpha_s|t_0) \kernel_{\alpha_s \alpha_t} e^{-it (\frac{Q^2}{2}+P_s^2)} \\
	&\sim \left.\tfrac{1}{2}  \partial^2_{P_s} \kernel_{\alpha_s \alpha_t}\right|_{\alpha_s=\tfrac{Q}{2}}  \times \mathcal{F}_S\left(\tfrac{Q}{2}|t_0\right) e^{-i\frac{Q^2}{2} t} \frac{\sqrt{\pi}}{4} (it)^{-3/2}.}
This is valid for all $Q$. This power law was observed numerically by \cite{Chen2017}, with precisely this explanation proposed. This behaviour is familiar from random matrix theory where the semicircle eigenvalue distribution has the same square root edge, and hence the same $t^{-3/2}$ power law decay of the spectral form factor at times sufficiently early that fluctuations away from the average eigenvalue density are not resolved. 

For the identity operator, it is simple to use \eqref{vackernel} to evaluate the first term in the prefactor; this was done in \eqr{crosscont2}. In the heavy-light limit \eqr{heavylightlim}--\eqr{Hpeq}, that result becomes
\begin{align}
	\tfrac{1}{2}  \partial^2_{P_s} \kernel_{\alpha_s \id}\big|_{\alpha_s=\tfrac{Q}{2}} 
	&\sim \frac{16 \pi ^2}{\sinh(2 \pi  p)\Gamma (2 h)^2} \left(\frac{p}{b}\right)^{4 h-1} e^{-b^{-2}(2\pi p +4J(p)-J(2p))},
\end{align}
The exponent matches the earlier semiclassical result evaluated at $p_s=0$. The dependence on the initial S-channel block is somewhat trickier to compute, but if we choose $t_0=-i\epsilon$ to be a small imaginary time regulator, we have $z=e^{-it_0}\sim 1-\epsilon$ for small $\epsilon$, so are in the regime where this can be evaluated by the cross-channel limit \eqref{eq:SChannelAsymptotics}. This gives (assuming $\alpha_2<\frac{Q}{4},\alpha_1$)
\begin{align*}
	\mathcal{F}_S(\tfrac{Q}{2}|t_0) &= \frac{\Gamma_b(Q)^3\Gamma_b(2\alpha_1-2\alpha_2)\Gamma_b(2Q-2\alpha_1-2\alpha_2)\Gamma_b(Q-2\alpha_2)^4}{\Gamma_b(Q-4\alpha_2)\Gamma_b(\frac{Q}{2}+\alpha_1-\alpha_2)^4\Gamma_b(\frac{3Q}{2}-\alpha_1-\alpha_2)^4} \epsilon^{-2h_2 + h(2\alpha_2)}(1+\O(\epsilon)) \\
	&\sim  \sqrt{\frac{(\pi  p)^3 \cosh (\pi  p)}{\sinh ^3(\pi  p)}} \left(\frac{\tanh(\pi p)}{\pi p}\right)^{2h} e^{b^{-2}(4J(p)-J(2p))} \epsilon^{-2b^2 h^2},
\end{align*}
where in the second line we have again taken the heavy-light limit. Ignoring phases and subleading factors (where we fix $ c^{-1}\ll\epsilon\ll 1$ in the large $c$ limit),
we find the following late-time behaviour:
\begin{equation}
	\mathcal{F}_T(\id|t) \approx e^{-2\pi p b^{-2}} t^{-3/2}
\end{equation}
The same power law was also seen in contribution of the vacuum Virasoro characters to the spectral form factor \cite{Dyer2017}. This can be explained as the same square-root edge of the modular-S matrix \cite{Keller2015} (the analogue of the fusion kernel for Virasoro characters), but in this case the identity (for which null descendants must be subtracted) is qualitatively different, with non-vacuum characters only decaying as $t^{-1/2}$, since their dual channel spectral densities diverge as $\left(h_s-\frac{c-1}{24}\right)^{-1/2}$.

While the power law behaviour \eqr{fulllatetime} holds for any sufficiently heavy (generic) external dimensions and any $c$, in the semiclassical heavy-light regime it does not set in until a parametrically late time. Before times of order $c$ there is exponential decay, as visible from the semiclassical limit of the block (\ref{eq:xInt}), (\ref{eq:HHLLBlock}). There is in fact a sharp transition where this exponential decay ceases, since the block behaves simply as a sum of the exponential decay $e^{-2pht}$ with order one coefficient, and the power law with exponentially small coefficient $e^{-2\pi  \frac{c}{6} p }$, and these terms exchange dominance at a crossover time which is sharply defined at large $c$,
\begin{equation}
	t_c = \frac{\pi c}{6h},
\end{equation}
precisely the result observed  in numerical studies \cite{Chen2017}.\foot{While this crossover time is order $c$, it does not grow with the energy of the heavy state, so is not of order the entropy, suggesting that it is not caused by resolving nontrivial eigenvalue statistics. From its derivation, it is clear that this is the case, since it is simply a feature of the coarse-grained spectral density, so true signs of discreteness of the S-channel spectrum can only come from sums over infinitely many internal T-channel operators. Nonetheless, it would be interesting to clarify if this feature is present in correlation functions of theories with gravitational duals once T-channel blocks beyond the vacuum are added, and to attempt to give a semiclassical description, perhaps in terms of near-threshold black holes.}
To see that there is a sharp transition, take times of order $c$ where the stationary phase approximation and large $c$ saddle-point approximation combine into a steepest descent analysis; it turns out that there are always two separate saddle-points, those producing the semiclassical result and late-time behaviour, and they lie on separate steepest descent contours. A logical alternative was to find a single saddle-point which moved from $\alpha_s=\alpha_1$ to $\alpha_s=\frac{Q}{2}$ as time increased, giving a smooth function of $\frac{t}{c}$ as opposed to a sharp transition.

Finally, we point out that for light external operators $\alpha_1+\alpha_2<\frac{Q}{2}$, while there is still a piece giving a $t^{-3/2}$ decay coming from the integral over the continuum of $\alpha_s$, it is not dominant, because the poles at $\alpha_s=\alpha_1+\alpha_2+m b$ contribute phases $e^{i t h_s}$. At large $c$, these poles can give a very large number of fluctuating contributions of comparable amplitude but with irrationally related periods, and are the source of the erratic behaviour observed in \cite{Kraus:2018zrn}.

\section*{Acknowledgments}
We thank Alex Belin, Minjae Cho, Shouvik Datta, Monica Guica, Per Kraus, Daliang Li, Junyu Liu, Don Marolf, David Meltzer, David Poland, David Simmons-Duffin, Douglas Stanford and Xi Yin for helpful discussions. We also thank Minjae Cho, Daliang Li, Alex Maloney and Xi Yin for comments on a draft. EP is supported by Simons Foundation grant 488657, and by the Walter Burke Institute for Theoretical Physics. SC and YG are supported in part by the Natural Sciences and Engineering Research Council of Canada via PGS D fellowships. YG is also supported by a Walter C. Sumner Memorial Fellowship. HM is funded by a Len DeBenedictis postdoctoral fellowship, and receives additional support from the University of California. This material is based upon work supported by the U.S. Department of Energy, Office of Science, Office of High Energy Physics, under Award Number DE-SC0011632. We thank the Simons Nonperturbative Bootstrap 2018 conference, and the KITP workshop ``Chaos and Order: From Strongly Correlated Systems to Black Holes,'' for stimulating environments during the course of this project. This research was supported in part by the National Science Foundation under Grant No. NSF PHY-1748958.
\appendix

\section{Special functions}\label{app:Special}

In this appendix we discuss the different properties of the special functions necessary to derive our results for the fusion kernel. Throughout this appendix (and the paper), $m$ and $n$ are non-negative integers and $Q=b+b^{-1}$.
\subsubsection*{Definition of $\Gamma_b(x)$}
The main function that appears is the Barnes double gamma function $\Gamma_b(x)$, having the property $\Gamma_b = \Gamma_{b^{-1}}$ and satisfying the functional equation
\begin{align}\label{eq:GammaFnalRel}
\Gamma_b(x+b) =& {\sqrt{2\pi}b^{bx-{1\over 2}}\over \Gamma(b x)}\Gamma_b(x) \,,
\end{align}
along with a similar equation with $b\to b^{-1}$. It is a meromorphic function with no zeroes and simple poles at $x = -mb -nb^{-1}$.\footnote{We will often implicitly assume that $b^2$ is not a rational number.} Its normalization is fixed by $\Gamma_b\left({Q\over 2}\right)$. The functional relation (\ref{eq:GammaFnalRel}) can be used repeatedly to derive the following shift relations
\ie\label{eq:GammaShift}
\Gamma_b(x+mb+nb^{-1}) =& (2\pi)^{m+n\over 2}\left(\prod_{\ell=0}^{m-1}\Gamma(xb+\ell b^2)\right)^{-1}
\left(\prod_{k=0}^{n-1}\Gamma(xb^{-1}+kb^{-2}+m)\right)^{-1}
\\
&b^{{n-m\over 2}-mn+x(mb-nb^{-1})+{1\over 2}m(m-1)b^2-{1\over 2}n(n-1)b^{-2}}\Gamma_b(x) \,.
\fe
The double gamma function admits the following integral representation, convergent for $x$ in the right half-plane \cite{Ponsot2004}:
\begin{align}\label{eq:GammaIntegralRep}
\log\Gamma_b(x) =& \int_0^\infty{dt\over t}\left[{e^{-xt}-e^{-Qt/2}\over (1-e^{-bt})(1-e^{-b^{-1}t})}-{1\over 2}(Q/2-x)^2e^{-t}-{Q/2-x\over t}\right]
\end{align}
Along with the shift relation, this defines the function everywhere.

\subsubsection*{Residues}

The integral representation (\ref{eq:GammaIntegralRep}) fixes the residue of $\Gamma_b(x)$ at $x=0$ to \begin{equation} \Res_{x\to 0}\Gamma_b(x) = \frac{\Gamma_b(Q)}{2\pi} \,.\end{equation}
We furthermore have the following Laurent expansion
\ie\label{eq:GammabLaurent}
\Gamma_b(x)\sim{\Gamma_b(Q)\over 2\pi}\left({1\over x}-\gamma_b+\mathcal{O}(x)\right),~x\to 0 \,,
\fe
where \cite{Spreafico2009a,Esterlis2016}
\ie
\gamma_b =& -{3\over 2}b^{-1}\log b + (\gamma-{1\over 2}\log(2\pi))b^{-1}+b\log b +{\gamma\over 2}b\\
&-ib\int_0^\infty~dy{\psi(1+ib^2 y)-\psi(1-ib^2 y)\over e^{2\pi y}-1} \,,
\fe
and $\psi(x) = {\Gamma'(x)\over \Gamma(x)}$ is the digamma function. We can use the shift relation (\ref{eq:GammaShift}) to find the residues at the locations of the other poles
\begin{equation}
	 \Res_{x= -mb-nb^{-1}}\Gamma_b(x) = \frac{b^{\frac{m-n}{2}+\frac{1}{2}m(m+1)b^2-\frac{1}{2}n(n+1)b^{-2}}}{(2\pi)^{\frac{m+n}{2}}} \frac{\left(\prod_{k=1}^m \Gamma(-k b^2)\right) \left(\prod_{l=1}^n \Gamma(-lb^{-2})\right)}{\prod_{k=1}^m \prod_{l=1}^n (-kb-lb^{-1})} \Res_{x= 0}\Gamma_b(x) \,.
\end{equation}
\subsubsection*{Asymptotics}

Here we list the main asymptotic formulae that are used in the paper. Some of the derivations are detailed in the following subsections. 

Starting from the integral representation (\ref{eq:GammaIntegralRep}), one can show that $\Gamma_b$ has the following asymptotic behaviour for fixed $b$ as $|x|\to\infty$ for $x$ in the right half-plane
\ie\label{eq:GammaAsymptotics}
\log\Gamma_b(x)\sim \,-{1\over 2}x^2\log x+{3\over 4}x^2+{Q\over 2}x\log x-{Q\over 2}x-{Q^2+1\over 12}\log x +\log\Gamma_0(b)+\mathcal{O}(x^{-1}) \,.
\fe
Here, we have introduced the following function
\ie
\log\Gamma_0(b) =& -\int_0^\infty {dt \over t}\left({e^{-Qt/2}\over (1-e^{-bt})(1-e^{-b^{-1}t})}-{1\over t^2}-{Q^2-2\over 24}e^{-t}\right) \,.
\fe

The semiclassical limit, which corresponds to taking $b\rightarrow 0$ with arguments scaling like $b^{-1}$, is given by
\begin{align}\label{eq:SemiClassGammab}
	\log\Gamma_b(b^{-1}x+\tfrac{b}{2})  \sim \frac{1}{2b^2}&\left(\frac{1}{2}-x\right)^2\log b+\frac{2x-1}{4b^2}\log(2\pi)-\frac{1}{b^2}\int_{\frac{1}{2}}^x dt\log \Gamma(t)  \nonumber \\	
	&\quad-\sum_{n=0}^\infty c_{n+1} b^{4n+2} \left(\psi^{(2n)}(x)-\psi^{(2n)}(1/2)\right) \,,
\end{align}
with the $c_n$'s defined in section \ref{app:SemiClass}.

The global limit means taking the small $b$ limit but with argument scaling like $b$ this time. It can be derived from the semiclassical one by using the shift relation and the result is
\ie\label{eq:GlobalGammab}
\Gamma_b(bx)\sim &{1\over 8}b^{-2}\log b-{1\over 2}F(0)b^{-2}+{3\over 4}(2x-1)\log b + \log \Gamma(x)+{2x-3\over 4}\log(2\pi)+\mathcal{O}(b^2) \,,
\fe
where $F(x=0) = \int_{1\over 2}^0 dy~\log{\Gamma(y)\over \Gamma(1-y)}=\frac{1}{12}\log 2-3\log A$ where $A$ is Glaisher's constant.

\subsubsection*{Results for $S_b(x)$}

It is often convenient to define the function
\ie
S_b(x) = {\Gamma_b(x)\over \Gamma_b(Q-x)} \,.
\fe
$S_b(x)$ is a meromorphic function with poles at $x=-mb-nb^{-1}$ and zeroes at $x=Q+mb+nb^{-1}$. It satisfies the following shift relations
\ie
S_b(x+mb+nb^{-1}) =& (-)^{mn}2^{m+n}\left(\prod_{\ell=0}^{m-1}\sin(\pi b(x+\ell b))\right)\left(\prod_{k=0}^{n-1}\sin(\pi b^{-1}(x+kb^{-1}))\right)S_b(x)
\fe
and admits the following integral representation in the strip $0<{\rm Re}(x)<Q$
\ie
\log S_b(x) =& \int_0^\infty \left({\sinh(({Q\over 2}-x)t)\over 2\sinh({bt\over 2})\sinh({t\over 2b})}-{Q-2x\over t}\right) \,.
\fe
It is useful to record the asymptotics of $S_b(x)$ as $|x|\to\infty$ for fixed $b$. The following formula is valid for $x$ in the upper half-plane
\ie\label{eq:SAsymptotics}
\log S_b(x)\sim -{i\pi\over 2}x^2+{i\pi\over 2}Q x-{i\pi\over 12}(Q^2+1)+\mathcal{O}(x^{-1}).
\fe
This follows directly from the asymptotics (\ref{eq:GammaAsymptotics}). The asymptotic formula for $x$ in the lower half-plane can be deduced by noting that $S_b(x) = {1\over S_b(Q-x)}$. 

The global and semiclassical limits can be derived directly from those of $\Gamma_b$ so we will not write them here.
\subsubsection*{Results for $\Upsilon_b(x)$}

We also define the upsilon function
\ie
\Upsilon_b(x) =& {1\over \Gamma_b(x)\Gamma_b(Q-x)} \,,
\fe
which is an entire function of $x$ with zeroes at $x=-mb-nb^{-1}$ and $x= Q+mb+nb^{-1}$. It satisfies the shift relations
\ie
\Upsilon_b(x+mb+nb^{-1}) =& 
\left(\prod_{\ell=0}^{m-1}{\Gamma(xb+\ell b^2+n)\over\Gamma(1-xb-\ell b^2-n)}\right)
\left(\prod_{k=1}^{n-1}{\Gamma(xb^{-1}+kb^{-2})\over\Gamma(1-xb^{-1}-kb^{-2})}\right)\\
&b^{m-n-2mn-2x(mb-nb^{-1})-m(m-1)b^2+n(n-1)b^{-2}}\Upsilon_b(x)
\fe
and admits the following integral representation in the strip $0<{\rm Re}(x)<Q$
\ie
\log \Upsilon_b(x) =& \int_0^\infty{dt\over t}\left[\left({Q\over 2}-x\right)^2e^{-t}-{\sinh^2(({Q\over 2}-x){t\over 2})\over \sinh({bt\over 2})\sinh({t\over 2b})}\right] \,.
\fe
The asymptotics of $\Upsilon_b(x)$ for large argument, derived mainly from (\ref{eq:GammaAsymptotics}), are given by the following formula for $x$ in the upper half-plane
\ie\label{eq:UpsilonAsymptotics}
\log\Upsilon_b(x)\sim & \,x^2\log x -\left({3\over 2}+{i\pi\over 2}\right)x^2 -Q x\log x+\left(1+{i\pi \over 2}\right)Qx+\left({Q^2+1\over 6}\right)\log x\\
&-{i\pi\over 12}(Q^2+1)-2\log\Gamma_0(b)+\mathcal{O}(x^{-1}) \,.
\fe
The global and semiclassical limit are again easy to derive so we don't write them explicitly.

\subsection{Derivation of large argument asymptotics}\label{app:LargeWeightDerivation}

This section is dedicated to deriving the large argument asymptotics (\ref{eq:GammaAsymptotics}) for the double gamma function. The main idea to derive asymptotic formulae is to massage the integrals into a form where we can apply Watson's lemma. This says that for a function $f$ which is smooth near $0$, and is bounded by some exponential, we can write an asymptotic formula by Taylor expanding $f$ and integrating term-by-term:
\begin{equation}
	\int_0^\infty dt\; e^{-t x} f(t) \sim \sum_{n=0}^\infty \; f^{(n)}(0) x^{-n-1} \quad \text{as } x\to\infty
\end{equation}
In fact, the Riemann-Lebesgue lemma implies that this is valid for $|x|\to\infty$ anywhere in the right half-plane $\Re x>0$, under slightly stronger assumptions on $f$. In our examples, we won't be able to do this immediately, because the function multiplying $e^{-t x}$ has a pole at $t=0$. The trick will be to subtract the polar piece using some simple functions, and what remains will be integrals that we can evaluate, typically in terms of simple powers and logarithms.

For this first case, we start with the integral representation (\ref{eq:GammaIntegralRep}). The interesting $x$ dependence of this function comes from the piece of the integrand proportional to $e^{-x t}$. This term in the integrand alone is singular at $t=0$, but by subtracting some simple pieces, we can create something for which Watson's lemma is applicable:
\begin{equation}\label{WatsonsGamma}
	\int_0^\infty \frac{dt}{t}\left(\frac{1}{(1-e^{-bt})(1-e^{-t/b})} -\frac{1}{t^2}-\frac{Q}{2 t}-\frac{Q^2+1}{12}\right)e^{-xt} \sim \frac{Q}{24x}+\cdots
\end{equation}
Now we can start to group the remaining terms by their $x$ dependence. A piece independent of $x$ is given by
\begin{equation}\label{Gamma0}
	\log\Gamma_0(b) = -\int_0^\infty \frac{dt}{t}\left(\frac{e^{-Qt/2}}{\left(1-e^{-b t}\right) \left(1-e^{-t/b}\right)}-\frac{1}{t^2}+\frac{Q^2-2}{24} e^{-t}\right).
\end{equation}
Note that the exponential in the final subtraction is required for convergence at infinity. The integral of the remaining terms can be evaluated as follows:
\begin{align*}
	I_3(x)&=-\int_0^\infty \frac{dt}{t}\left(\frac{e^{-xt}}{t^2}-\frac{1}{t^2}+\frac{x}{t}-\tfrac{1}{2}x^2 e^{-t}\right),\quad I_3(0)=0,\quad I_3'(x)= I_2(x) \\
	I_2(x)&=\int_0^\infty \frac{dt}{t}\left(\frac{e^{-xt}}{t}-\frac{1}{t}+x \, e^{-t}\right),\quad I_2(0)=0,\quad I_2'(x)=I_1(x)\\
	I_1(x)&=\int_0^\infty \frac{dt}{t}\left(e^{-t}-e^{-tx}\right),\quad I_1(1)=0,\quad I_1'(x)=I_0(x)=\int_0^\infty e^{-xt}dt=\frac{1}{x} \\
	&\quad\implies I_1(x)=\log x,\quad I_2(x)=x\log x-x,\quad I_3(x)=\tfrac{1}{2}x^2\log x-\tfrac{3}{4}x^2 \,.
\end{align*}
Putting this all together, we have
\begin{equation}
	\log\Gamma_b(x) \sim -\tfrac{1}{2}x^2\log x+\tfrac{3}{4}x^2 +\tfrac{Q}{2}x\log x- \tfrac{Q}{2} x- \frac{Q^2+1}{12}\log x +\Gamma_0(b)+\cdots
\end{equation}
where $\Gamma_0(b)$ is given by \eqref{Gamma0}, and the ellipsis can be expanded asymptotically from \eqref{WatsonsGamma}, by Taylor expanding the contents of the brackets and integrating term by term. 

This applies for $|x|\to\infty$ in the region $\Re(x)\geq 0$. In fact, we can extend this range by using the recursion formula, under which the asymptotic expansion remains invariant but we find an integral that converges in a wider regime. We can use the same formula taking $|x|\to\infty$ in the region $\Re(x)\geq X$, for any fixed $X\in \mathbb{R}$.

\subsection{Derivation of semiclassical limit}\label{app:SemiClass}

Now we consider the limit where we take $b\to 0$, with parameter proportional to $b^{-1}$. We can make life a little easier by differentiating first, since we can use $\log\Gamma_b\left(\frac{Q}{2}\right)=0$ to fix the constant when we integrate back up. It will be convenient to actually choose the argument to be $b^{-1}x+b/2$, and it is straightforward to remove this shift at the end. After substituting for this argument, and rescaling the integration variable $t$ by $b$, we have
\begin{align}
	\frac{\Gamma_b'(b^{-1}x+\tfrac{b}{2})}{\Gamma_b(b^{-1}x+\tfrac{b}{2})} =& \int_0^\infty \frac{dt}{t}\left[\frac{-bte^{-xt-b^2t/2}}{(1-e^{-t})(1-e^{-b^2t})}+\frac{1}{2b}(1-2x)e^{-bt}+\frac{1}{bt}\right] \\
	 =& \int_0^\infty \frac{dt}{t}\left[\frac{-bte^{-xt}}{2(1-e^{-t})\sinh\left(\frac{b^2t}{2}\right)}+\frac{1}{2b}(1-2x)e^{-t}+\frac{1}{bt}\right]-\frac{1}{2b}(1-2x)\log b \,. \nonumber
\end{align}
The second line here is designed to remove any $b$ dependence from exponentials in the integrand. Once this has been achieved, if we Taylor expand the integrand in $b$, the result is integrable term by term.\footnote{The series in $b$ doesn't converge uniformly, so we won't get a convergent series. But truncating the expansion at a given $n$, the remainder is bounded by a constant times $b^{n+1}$, so integrating term by term does give an asymptotic series.} Writing the expansion in terms of coefficients of the Taylor series $\frac{1}{2\sinh(u/2)}=u^{-1}\sum_{n=0}^\infty c_n u^{2n}$ (in closed form, $c_n=-(1-2^{-(2n-1)})\frac{B_{2n}}{(2n)!}$), we have
\begin{align}
	\frac{\Gamma_b'(b^{-1}x+\tfrac{b}{2})}{\Gamma_b(b^{-1}x+\tfrac{b}{2})} \sim&-\frac{1}{2b}(1-2x)\log b
	+\frac{1}{b}\int_0^\infty \frac{dt}{t}\left[\frac{-e^{-xt}}{1-e^{-t}}  +\frac{1}{2}(1-2x)e^{-t}+\frac{1}{t}\right] \nonumber \\
	&-\sum_{n=1}^\infty c_n b^{4n-1}\int_0^\infty dt\, \frac{t^{2n-1} e^{-xt}}{1-e^{-t}}
\end{align}
All these integrals can be evaluated in closed form. The main ingredient we will need is an integral expression for $\log\Gamma(z)$:
\begin{equation}
	\log\Gamma(z)= \int_0^\infty \frac{dt}{t}\left[\frac{e^{-zt}-e^{-t}}{1-e^{-t}}+(z-1)e^{-t}\right],\qquad \Re(z)>0 \,.
\end{equation}
Differentiating this many times gives the polygamma functions
\begin{equation}
	\psi^{(m)}(z)= (-1)^{m+1}\int_0^\infty dt\frac{t^m e^{-zt}}{1-e^{-t}},\qquad \Re(z)>0,\quad m>0 \,.
\end{equation}
This allows us to immediately evaluate all the integrals:
\begin{align}
	\frac{\Gamma_b'(b^{-1}x+\tfrac{b}{2})}{\Gamma_b(b^{-1}x+\tfrac{b}{2})} &\sim \frac{1}{2b}(2x-1)\log b+\frac{1}{b}\log \frac{\sqrt{2\pi}}{\Gamma(x)} -\sum_{n=1}^\infty c_n b^{4n-1} \psi^{(2n-1)}(x) \,.
\end{align}
The only slightly challenging aspect here is to evaluate the integral giving the constant $\log\sqrt{2\pi}$.

We finally only have to integrate this up, using knowledge of the value at $x=\frac{1}{2}$. Integration with respect to the argument is the same as integration with respect to $x$, after division by $b$, and gives us
\begin{align}\label{eq:GammabSemiclassical}
	\log\Gamma_b(b^{-1}x+\tfrac{b}{2})  \sim& \frac{1}{2b^2}\left(\frac{1}{2}-x\right)^2\log b+\frac{2x-1}{4b^2}\log(2\pi)-\frac{1}{b^2}\int_{\frac{1}{2}}^x dt\log \Gamma(t)  \nonumber \\	
	&-\sum_{n=0}^\infty c_{n+1} b^{4n+2} \left(\psi^{(2n)}(x)-\psi^{(2n)}(1/2)\right)
\end{align}

We can also take the parameter $x$ to be large, and match this to a small $b$ expansion of the previous section. These expansions agree perfectly, including matching the leading order asymptotics of the constant term
\begin{equation}
	\log\Gamma_0(b) \sim \frac{1}{24b^2}\log b+\frac{1}{2b^2}\left(\log A-\frac{1}{12}\log 2 \right)+\cdots \,.
\end{equation}

\section{Further results for the fusion kernel}\label{app:moreKernel}
In this appendix we will record some lengthy technical results for properties of the fusion kernel omitted from the main text.
\subsection{Residues at subleading poles}
In section \ref{sec:global}, we derived MFT OPE data and corrections due to non-vacuum exchange for subleading double-twists by studying the global limit of the residues of the kernel at its subleading poles. Furthermore, in section \ref{sec:global}, we derived the heavy-light semiclassical Virasoro blocks (in the case that the heavy operator is dual to a conical defect in the bulk) by summing over residues of the kernel in this semiclassical limit. In this subsection we will present the finite-$c$ values for the residues from which these results were derived. 

The residue of the fusion kernel with the vacuum exchanged in the T-channel at its subleading poles is given by
\ie\label{eq:HigherResidues}
&\Res_{\alpha_s=\alpha_1+\alpha_2+mb}\kernel_{\alpha_s\id}= -{\Gamma_b(2Q)\Gamma_b(Q+mb)\over (2\pi)^{1+{m\over 2}} \Gamma_b(Q)^2}b^{{m\over 2}+{1\over 2}m(m+1)b^2}\left(\prod_{\ell=0}^{m-1}\Gamma(b^2(\ell-m))\right)\\
&{\Gamma_b(2Q-2\alpha_1-2\alpha_2-mb)\Gamma_b(2\alpha_1+2\alpha_2-Q+mb)\over\Gamma_b(Q-2\alpha_1-2\alpha_2-2mb)\Gamma_b(2\alpha_1+2\alpha_2-Q+2mb)}\left({\Gamma_b(Q-2\alpha_1-mb)\Gamma_b(2\alpha_1+mb)\over\Gamma_b(2Q-2\alpha_1)\Gamma_b(2\alpha_1)}\times(\a_1\to\a_2)\right).
\fe
This can be reduced to an expression involving only normal gamma functions using the shift relations (\ref{eq:GammaShift}), but we do not find it particularly illuminating to do so.

In (\ref{nonvackernel}), we gave a formula for the non-vacuum kernel that captured all singularities at the leading pole. To compute the residues at the subleading poles, we will need to sum over multiple contributions from the contour integral in (\ref{eq:CrossingKernel}). In particular, after taking $\alpha_2\to Q-\alpha_2$ so that all the relevant singularities come form the integral, the residues of the integrand at $s = Q-V_2 + \ell b$ for $\ell\le m$ contribute to the singularity at $\alpha_s = \alpha_1+\alpha_2 + mb$. At the end of the day, one finds
\ie\label{nonvackernel2}
\kernel_{\alpha_s\alpha_t}=&\sum_{\ell=0}^m\left(\prod_{k=0}^{\ell-1}{1\over 2 \sin(\pi b(Q+k b))}\right)\\
&{S_b(Q+\alpha_1-\alpha_2-\alpha_s+\ell  b)S_b(Q-\alpha_1+\alpha_2-\alpha_s+\ell b)S_b(\alpha_1+\alpha_2-\alpha_s+\ell b)^2\over S_b(2Q-2\alpha_s+\ell b)S_b(Q+\alpha_1+\alpha_2-\alpha_s-\alpha_t+\ell b)S_b(\alpha_1+\alpha_2-\alpha_s+\alpha_t+\ell b)}\\
&{\Gamma_b(\alpha_1-\alpha_2+\alpha_s)\Gamma_b(-\alpha_1+\alpha_2+\alpha_s)\Gamma_b(Q+\alpha_1-\alpha_2-\alpha_s)\Gamma_b(Q-\alpha_1+\alpha_2-\alpha_s)\over \Gamma_b(Q-2\alpha_s)\Gamma_b(2\alpha_s-Q)}\\
&{\Gamma_b(2Q-\alpha_1-\alpha_2-\alpha_s)^2\Gamma_b(Q-\alpha_1-\alpha_2+\alpha_s)^2\Gamma_b(2Q-2\alpha_t)\Gamma_b(2\alpha_t)\Upsilon_b(\alpha_t)^2\over \Gamma_b(Q-2\alpha_1+\alpha_t)\Gamma_b(Q-2\alpha_2+\alpha_t)\Gamma_b(2Q-2\alpha_1-\alpha_t)\Gamma_b(2Q-2\alpha_2-\alpha_t)}\\
&+(\text{regular at }\alpha_s=\alpha_1+\alpha_2+mb).
\fe
The coefficient of the fusion kernel with non-vacuum exchange in the T-channel at its subleading double poles is given by
\ie\label{eq:HigherResidues2}
&\dRes_{\alpha_s=\alpha_1+\alpha_2+mb}\kernel_{\alpha_s\alpha_t}\\
=&\sum_{n=0}^m\left(\prod_{k=0}^{m-n-1}{1\over 2 \sin(\pi b(Q+kb))}\right){\Gamma_b(Q)^2b^{n(1+(n+1)b^2)}\over (2\pi)^{n+2}}\left(\prod_{a=0}^{n-1}\Gamma(b^2(a-n))^2\right)\\
&{\Gamma_b(2\alpha_t)\Gamma_b(2Q-2\alpha_t)\Gamma_b(\alpha_t+nb)\Gamma_b(Q-\alpha_t+nb)\Upsilon_b(\alpha_t)^2\over \Gamma_b(Q-2\alpha_1+\alpha_t)\Gamma_b(Q-2\alpha_2+\alpha_t)\Gamma_b(2Q-2\alpha_1-\alpha_t)\Gamma_b(2Q-2\alpha_2-\alpha_t)}\\
&{\Gamma_b(2\alpha_1+mb)\Gamma_b(2\alpha_2+mb)\Gamma_b(Q-2\alpha_1-mb)\Gamma_b(Q-2\alpha_2-mb)\Gamma_b(Q-2\alpha_1-nb)\Gamma_b(Q-2\alpha_2-nb)\over\Gamma_b(2\alpha_1+nb)\Gamma_b(2\alpha_2+nb)\Gamma_b(Q-\alpha_t-nb)\Gamma_b(\alpha_t-nb)}\\
&{\Gamma_b(Q+mb)^2\Gamma_b(2Q-2\alpha_1-2\alpha_2-mb)^2\Gamma_b(2\alpha_1+2\alpha_2+(m+n)b-Q)\over\Gamma_b(Q+nb)^2\Gamma_b(2Q-2\alpha_1-2\alpha_2-(m+n)b)\Gamma_b(Q-2\alpha_1-2\alpha_2-2mb)\Gamma_b(2\alpha_1+2\alpha_2+2mb-Q)}.
\fe
Similarly, the residue of the non-vacuum kernel at the subleading poles takes the hideous form
\ie\label{eq:NonVacHigherResidues}
&\Res_{\alpha_s=\alpha_1+\alpha_2+mb}\kernel_{\alpha_s\alpha_t}\\
=&\sum_{n=0}^m\left(\prod_{k=0}^{m-n-1}{1\over 2 \sin(\pi b(Q+kb))}\right){\Gamma_b(Q)^2b^{n(1+(n+1)b^2)}\over (2\pi)^{n+2}}\left(\prod_{a=0}^{n-1}\Gamma(b^2(a-n))^2\right)\\
&{\Gamma_b(2\alpha_t)\Gamma_b(2Q-2\alpha_t)\Gamma_b(\alpha_t+nb)\Gamma_b(Q-\alpha_t+nb)\Upsilon_b(\alpha_t)^2\over \Gamma_b(Q-2\alpha_1+\alpha_t)\Gamma_b(Q-2\alpha_2+\alpha_t)\Gamma_b(2Q-2\alpha_1-\alpha_t)\Gamma_b(2Q-2\alpha_2-\alpha_t)}\\
&{\Gamma_b(2\alpha_1+mb)\Gamma_b(2\alpha_2+mb)\Gamma_b(Q-2\alpha_1-mb)\Gamma_b(Q-2\alpha_2-mb)\Gamma_b(Q-2\alpha_1-nb)\Gamma_b(Q-2\alpha_2-nb)\over\Gamma_b(2\alpha_1+nb)\Gamma_b(2\alpha_2+nb)\Gamma_b(Q-\alpha_t-nb)\Gamma_b(\alpha_t-nb)}\\
&{\Gamma_b(Q+mb)^2\Gamma_b(2Q-2\alpha_1-2\alpha_2-mb)^2\Gamma_b(2\alpha_1+2\alpha_2+(m+n)b-Q)\over\Gamma_b(Q+nb)^2\Gamma_b(2Q-2\alpha_1-2\alpha_2-(m+n)b)\Gamma_b(Q-2\alpha_1-2\alpha_2-2mb)\Gamma_b(2\alpha_1+2\alpha_2+2mb-Q)}\\
&\bigg[2\gamma_b+2nb\log(b)-2b\sum_{a=0}^{n-1}\psi(b^2(a-n))+2\psi_b(Q+mb)-2\psi_b(Q+nb)-\psi_b(Q-2\alpha_1-mb)-\psi_b(Q-2\alpha_2-nb)\\
&+\psi_b(2\alpha_1+mb)-\psi_b(2\alpha_1+nb)+\psi_b(2\alpha_2+mb)-\psi_b(2\alpha_2+nb)-\psi_b(Q-2\alpha_2-mb)-\psi_b(Q-2\alpha_1-nb)\\
&+2\psi_b(2Q-2\alpha_1-2\alpha_2-(m+n)b)-2\psi_b(2Q-2\alpha_1-2\alpha_2-mb)+2\psi_b(Q-2\alpha_1-2\alpha_2-2mb)\\
&+2\psi_b(2\alpha_1+2\alpha_2+(m+n)b-Q)-2\psi_b(2\alpha_1+2\alpha_2+2mb-Q)\\
&+\psi_b(Q-\alpha_t-nb)+\psi_b(Q-\alpha_t+nb)+\psi_b(\alpha_t-nb)+\psi_b(\alpha_t+nb)\bigg].
\fe

\subsection{Exchange of subleading Virasoro double-twists}\label{app:SubleadingVDTs}
In the main text, we presented the exact formula for the fusion kernel in the case of exchange of the leading Virasoro double twist in the T-channel. Here we record the (more complicated) form of the kernel in the case of exchange of subleading double-twists
\ie\label{eq:SubleadingDoubleTwistKernel}
&\kernel_{\alpha_s,2\alpha_2+mb}\\
=& \sum_{n =0}^m\left(\prod_{k=0}^{m-n-1}{(2\pi)^{1\over 2}\over 2\sin(\pi b(Q+k b))\Gamma(-b^2(m-k))}\right)b^{-{1\over 2}(m-n)(1+(1+m+n)b^2)}\\
&{\Gamma_b(2Q-4\alpha_2-2mb)\Gamma_b(Q-2\alpha_2-n b)^2\Gamma_b(2Q-2\alpha_1-2\alpha_2-nb)\Gamma_b(4\alpha_2+2mb)\Gamma_b(4\alpha_2+(m+n)b-Q)\over \Gamma_b(Q+nb)\Gamma_b(Q-2\alpha_2-mb)^2\Gamma_b(2Q-2\alpha_1-2\alpha_2-mb)\Gamma_b(2\alpha_2+mb)^2\Gamma_b(2\alpha_2+nb)^2}\\
&{\Gamma_b(2Q-\alpha_1-\alpha_2-\alpha_s)\Gamma_b(Q-\alpha_1+\alpha_2-\alpha_s)^2\Gamma_b(\alpha_1+\alpha_2-\alpha_s)\Gamma_b(\alpha_1+\alpha_2-\alpha_s+n b)\Gamma_b(Q-\alpha_1-\alpha_2+\alpha_s)\over\Gamma_b(2Q-2\alpha_2-(m+n)b)\Gamma_b(Q-2\alpha_1+2\alpha_2+mb)\Gamma_b(4\alpha_2+mb-Q)\Gamma_b(2\alpha_1+2\alpha_2+nb-Q)}\\
&{\Gamma_b(-\alpha_1+\alpha_2+\alpha_s)^2\Gamma_b(\alpha_1+\alpha_2+\alpha_s-Q)\Gamma_b(\alpha_1+\alpha_2+\alpha_s+nb-Q)\over\Gamma_b(Q-2\alpha_s)\Gamma_b(2\alpha_s-Q)\Gamma_b(2Q-\alpha_1-\alpha_2-\alpha_s-nb)\Gamma_b(Q-\alpha_1-\alpha_2+\alpha_s-nb)}.
\fe
Notice that only certain terms in the sum contribute to a double pole at $\alpha_s = \alpha_1+\alpha_2+m'b$. 

\subsection{Cross-channel blocks at $c=25$}\label{app:Zamolodchikovc25}
The conformal blocks at $c=25$ with $h_1=h_2=15/16$ and arbitrary $h$ are known due to Zamolodchikov \cite{Zamolodchikov1986}:
\e{}{\quad \cF^{22}_{11}(\a|z) = (16 q)^{\a(2-\a)-1}(z(1-z))^{-7/8}\theta_3^{-3}(q)\qquad (Q=2~, h_1 = h_2={15\o 16})}
where $\theta_3(q) = \sum_{n\in \ZZ}q^{n^2}$ is the Jacobi theta function, and $q=\exp[-\pi {{}_2F_1({1\o2},{1\o2},1,1-z)\o {}_2F_1({1\o2},{1\o2},1,z)}]$. Using (e.g. as quoted in \c{Asplund:2015eha})
\e{}{ q \sim e^{-{\pi^2\o\log(16/(1-z))}}~, ~~ \theta_3(q) \sim \sqrt{\pi\o1-q}~~\text{as}~~z\to 1}
leads to
\e{}{ \mathcal{F}^{22}_{11}(0|1-z)~~\stackrel{z\to 0}{\sim}~~ {\pi^{3\o2}\o 16}\,z^{-{7\o 8}}\,\left(\log{1\o z}\right)^{-{3\o2}}\qquad (Q=2~, h_1 = h_2={15\o 16})}
This matches \eqref{crosscont} and \eqref{crosscont2} upon plugging in the relevant values.

\subsection{Branching from S- to T-channel}\label{appd}
In the main text we analyzed the crossing kernel when branching T-channel blocks for pairwise identical operators ($11\to 22$) into S-channel blocks $(12\to12)$. In this section we do the opposite. The object of study is the kernel $\widetilde\kernel_{\alpha_t\alpha_s} = \kernel_{\alpha_t\alpha_s}\begin{bmatrix}\alpha_2 & \alpha_2 \\ \alpha_1 & \alpha_1\end{bmatrix}$. For sufficiently light external primaries, the residues of this kernel at its leading poles controls the $z\to 1$ asymptotics of the S-channel Virasoro blocks (given by (\ref{eq:SChannelAsymptotics})).

The kernel $\widetilde\kernel_{\alpha_s\alpha_t}$ is of course still a meromorphic function of $\alpha_s$, now with \emph{simple} poles at $\alpha_t = 2\alpha_1+mb+nb^{-1},2\alpha_2+mb+nb^{-1},2Q-2\alpha_1+mb+nb^{-1},2Q-2\alpha_2+mb+nb^{-1}$ and reflections (in $\alpha_t$), as well as quadruple poles at $\alpha_t = Q+mb+nb^{-1}$ and $\alpha_s = -mb-nb^{-1}$. It is straightforward to compute e.g. the residues at the poles $\alpha_t = 2\alpha_1+mb,2\alpha_2+mb$ using tools that we have previously developed. For example, the contour integral contributes a pole at $\alpha_t = 2\alpha_1+mb$, so for the purposes of extracting the residue at the leading pole, we may write the kernel as
\ie
&\widetilde\kernel_{\alpha_t\alpha_s} = {\Gamma_b(Q-\alpha_t)^4\Gamma_b(2\alpha_1-\alpha_t)\Gamma_b(2Q-2\alpha_1-\alpha_t)\Gamma_b(2\alpha_2-\alpha_t)\Gamma_b(2Q-2\alpha_2-\alpha_t)\over\Gamma_b(Q-2\alpha_t)\Gamma_b(2Q-2\alpha_t)\Gamma_b(Q+\alpha_{12}-\alpha_t+\alpha_s)\Gamma_b(2Q+\alpha_{12}-\alpha_t-\alpha_s)\Gamma_b(\alpha_1+\alpha_2-\alpha_s)}\\
&{\Gamma_b(2\alpha_s)\Gamma_b(2Q-2\alpha_s)\Gamma_b(-\alpha_{12}+\alpha_t-\alpha_s)\Gamma_b(-\alpha_{12}+\alpha_t+\alpha_s-Q)\over \Gamma_b(Q-\alpha_1-\alpha_2+\alpha_s)\Gamma_b(2Q-\alpha_1-\alpha_2-\alpha_s)\Gamma_b(\alpha_1+\alpha_2+\alpha_s-Q)\Gamma_b(Q-\alpha_{12}-\alpha_s)^2\Gamma_b(-\alpha_{12}+\alpha_s)^2}\\
&+\text{(regular at $\alpha_t = 2\alpha_1$)}.
\fe
where we recall $\a_{12} = \a_1-\a_2$. The residue of the kernel at the leading pole is then simply expressed as the following 
\ie\label{eq:TChannelDoubleTwistResidue}
&\Res_{\alpha_t=2\alpha_1}\widetilde\kernel_{\alpha_t\alpha_s}=
 - {\Gamma_b(Q)\over 2\pi}\times\\&{\Gamma_b(Q-2\alpha_1)^4\Gamma_b(-2\alpha_{12})\Gamma_b(2Q-2\alpha_1-2\alpha_2)\Gamma_b(2Q-2\alpha_s)\Gamma_b(2\alpha_s)\over \Gamma_b(Q-4\alpha_1)\Gamma_b(-\alpha_{12}+\alpha_s)^2\Gamma_b(Q-\alpha_1-\alpha_2+\alpha_s)^2\Gamma_b(Q-\alpha_{12}-\alpha_s)^2\Gamma_b(2Q-\alpha_1-\alpha_2-\alpha_s)^2}.
\fe
Similarly, to compute the residue at $\alpha_t = 2\alpha_2$ we send $\alpha_2\to Q-\alpha_2$ so that the contour integral rather than the prefactor contributes the singularity at $\alpha_t = 2\alpha_2$. The end result is of course the same as (\ref{eq:TChannelDoubleTwistResidue}), but with $\alpha_1\leftrightarrow\alpha_2$.

\section{Large internal weight asymptotics}\label{app:largeweight}
In this appendix we take the large internal weight asymptotics of the expressions derived in the main text for pairwise identical external operators. The final results are in (\ref{eq:AsymptoticVacKernel}) and (\ref{eq:AsymptoticNonVacKernel}).

\subsection{Vacuum kernel}
To study the large-internal dimension asymptotics of the vacuum kernel, we write $\alpha_s=\frac{Q}{2}+i P$ so that we have
\begin{align}
	 \kernel_{\alpha_s \id} =& \frac{\Gamma_b(2Q)}{\Gamma_b(Q)^3\Gamma_b\left(2\alpha_1)\Gamma_b(2Q-2\alpha_1\right)\Gamma_b\left(2\alpha_2\right)\Gamma_b(2Q-2\alpha_2)} \\
	 &\frac{\Gamma_b\left(\alpha_1+\alpha_2-\tfrac{Q}{2}+i P\right)\times (3 \text{ terms with } \alpha\leftrightarrow Q-\alpha )}
	{\Gamma_b(2iP)} \times (P\leftrightarrow -P ). \nonumber
\end{align}
For the purposes of computing the asymptotic density of OPE coefficients (cf (\ref{eq:OPEDensity})), we will be interested in the $P\to\infty$ limit of this quantity.

To make sense of the large $P$ limit, we need to consult (\ref{eq:GammaAsymptotics}) for the asymptotics of $\Gamma_b(x)$. 
This gives
\begin{equation}\label{eq:AsymptoticVacKernel}
	\kernel_{\alpha_s \id} \sim 2^{-4P^2}e^{\pi \sqrt{\frac{c-1}{6}}P} P^{4 (h_1+ h_2)-\frac{c+1}{4}} 2^{\frac{c+5}{36}} \Gamma_0(b)^6 \frac{\Gamma_b(2Q)}{\Gamma_b(Q)^3\Gamma_b\left(2\alpha_1)\Gamma_b(2Q-2\alpha_1\right)\Gamma_b\left(2\alpha_2\right)\Gamma_b(2Q-2\alpha_2)}.
\end{equation}

\subsection{Non-vacuum kernel}
The asymptotics of the non-vacuum kernel are slightly trickier to study than the vacuum case, since we need to work out the asymptotics of the $s$ integral. First, writing $\alpha_s=\frac{Q}{2}+i P$, we look at the integrand in the limit $P\to \infty$ with the ratio $\sigma \equiv s/P$ fixed. For this we need the asymptotics of $S_b(x)$; see equation (\ref{eq:SAsymptotics}). Since $S_b$ has different asymptotic expansions in the upper and lower half-planes, the integrand has four different regions depending on the imaginary part of $\sigma$. Keeping terms only to the leading order in $P$ with nontrivial $\sigma$ dependence, we have the following:
\ie{}
	&\log\left(\prod_{k=1}^4 \frac{S_b(s+U_k)}{S_b(s+V_k)}\right)\\
	 \sim& \begin{cases}
		-i \pi  P^2+2\pi  i  Q \sigma P  + \mathcal{O}(P^0) &  \Im\sigma  > 1 \\
		 -i \pi (\sigma^2-2i\sigma) P^2+ 2\pi(-\alpha_2+\alpha_1+i\sigma(Q-\alpha_2+\alpha_1))P+\mathcal{O}(P^0) & 0< \Im\sigma  < 1 \\
		i \pi (\sigma^2+2i\sigma) P^2+2\pi(-\alpha_2+\alpha_1-i\sigma(Q-\alpha_2+\alpha_1))P+\mathcal{O}(P^0) & -1< \Im\sigma <0 \\
		i \pi  P^2 - 2i \pi  Q \sigma P + \mathcal{O}(P^0) &  \Im\sigma < -1 \\
	\end{cases}
\fe
Looking at the region $0< \Im\sigma  < 1$, there is a saddle-point at the edge, where $\sigma = i$, which is where the contour of integration passes to the left of a pole at $s \approx \alpha_s$ (from $V_1$). We can then take the contour to follow the path of steepest descent away from this, along the line $\Re \sigma+\Im \sigma=1$ between $\sigma=i$ and $\sigma=1$. This part of the contour, along with a piece that can be taken to run to infinity in the positive imaginary direction, contributes a term of size $e^{-2 \pi Q P}$ (times an order one piece and a phase $e^{-i\pi P^2}$). Similarly, a piece of the contour running from negative imaginary infinity, to $\sigma = -i$, and then on to $\sigma=1$, contributes a term of the same size.

We cannot, however, simply take these two pieces of contour to join near $\sigma = 1$ to form the complete integration contour, since there are lines of poles starting near $\sigma=0$, and going to the right, to which the contour must pass to the left. To include this piece, we only need to take the residues of the integrand at the poles; furthermore, as can be seen from the real part of the asymptotic expansion above on the real axis, the poles further to the right will be exponentially suppressed in $P$, so only the leftmost pole(s) are needed to leading order.

More explicitly, we can take a different limit with $s$ fixed and $P\to\infty$, with the only relevant terms in the integrand giving
\begin{equation}
\frac{1}{S_b(s+V_1)S_b(s+V_2)}\sim e^{-2\pi (s+\alpha_2-\alpha_1) P},
\end{equation}
and now to evaluate the integral we need only find the residues of poles at $s=Q-V_3$ or $s=Q-V_4$, whichever is further to the left, including this exponential factor which suppresses poles lying further to the right. For operators below the threshold $h_t<\frac{c-1}{24}$, we can always choose $\Re\alpha_t < \frac{Q}{2}$ using the reflection symmetry, so the relevant pole is at $s=\alpha_t$. If this pole alone dominates, the integral asymptotically gives the following:
\begin{equation}
\int_{-i\infty}^{i\infty}\frac{ds}{i} \prod_{k=1}^4 \frac{S_b(s+U_k)}{S_b(s+V_k)} \sim 
 \frac{S_b(\alpha_t+U_1) S_b(\alpha_t+U_2) S_b(\alpha_t+U_3) S_b(\alpha_t+U_4)}{S_b(2\alpha_t)} e^{-2\pi (\alpha_t+\alpha_2-\alpha_1) P}.
\end{equation}
This pole is more important than the other pieces, of order $e^{-2\pi Q P}$, as long as $\Re(\alpha_t+\alpha_2-\alpha_1)<Q$; this is always true for unitary operator dimensions. Finally, if $h_t>\frac{c-1}{24}$, there are two poles that give contributions of the same size, so we must add a second term, which simply takes $\alpha_t \longrightarrow Q-\alpha_t$. The special case $\alpha_t=Q/2$ can be found as a limit of the sum of both terms (each of which will diverge, but with the sum approaching a finite limit).

Incidentally, a nice thing is that when we evaluated the identity block, the integral was given by the same pole, so the $\alpha_t\to 0$ limit should return us smoothly to the vacuum result above. This is a useful check.

Now we need only include the prefactor. The pieces depending on $\alpha_s$ get expanded just as for the vacuum block, and the pieces depending on $\alpha_t$ combine nicely with the integrand, giving
\ie\label{eq:AsymptoticNonVacKernel}
	\kernel_{\alpha_s \alpha_t}  \sim& 2^{-4P^2}e^{\pi (Q-2\alpha_t)P} P^{4(h_1+ h_2)-\frac{c+1}{4}} 2^{\frac{c+5}{36}} \Gamma_0(b)^6 \\
	& \frac{\Gamma_b(2Q-2\alpha_t) \Gamma_b(Q-2\alpha_t)}{
	\Gamma_b(Q-\alpha_t)^4\Gamma_b(2\alpha_1-\alpha_t)
	\Gamma_b(2Q-2\alpha_1-\alpha_t)\Gamma_b(2\alpha_2-\alpha_t)\Gamma_b(2Q-2\alpha_2-\alpha_t) 
	}.
\fe
The quantity relevant for the anomalous momenta (\ref{nonholope}) is the following ratio
\ie\label{eq:AsymptoticRatio}
	\frac{\kernel_{\alpha_s \alpha_t}}{\kernel_{\alpha_s \id}} \sim& e^{-2\pi \alpha_t P}
	\frac{\Gamma_b\left(2\alpha_1)\Gamma_b(2Q-2\alpha_1\right)}{\Gamma_b(2\alpha_1-\alpha_t)\Gamma_b(2Q-2\alpha_1-\alpha_t)}
	\frac{\Gamma_b\left(2\alpha_2\right)\Gamma_b(2Q-2\alpha_2)}{\Gamma_b(2\alpha_2-\alpha_t) \Gamma_b(2Q-2\alpha_2-\alpha_t)}
	\\
	 &\times\frac{\Gamma_b(2Q-2\alpha_t) \Gamma_b(Q-2\alpha_t) \Gamma_b(Q)^3
	}
	{\Gamma_b(2Q)
	 \Gamma_b(Q-\alpha_t)^4
	} ,
\fe
where recall that in the large-weight limit $P\approx \sqrt{h_s}$. This reduces to the vacuum result when we take $\alpha_t$ to vanish.
 Recall that as written this applies for $h_t<\frac{c-1}{24}$.

 \subsubsection*{Properties}

For light T-channel operators, we just have exponential decay $e^{-2\pi \alpha_t P}$ with $\alpha_t\in\mathbb{R}$. For heavy T-channel operators, there is an exponential decay at a fixed rate $e^{-\pi Q P}$, along with oscillations. For light external operators, \eqref{eq:AsymptoticRatioOfKernels} has a definite sign (always positive) when $\alpha_t$ is lighter than T-channel double-twists. 
 
Another feature is the set of zeros in (\ref{eq:AsymptoticRatioOfKernels}) at $\alpha_t=2\alpha_{1,2}$. This is a nice result that dovetails with a similar result in the Lorentzian inversion of T-channel global conformal blocks, as we now explain. In the global case, if the T-channel block is for exchange of a spin-$J$ operator whose twist is precisely equal to $2h_1$ or $2h_2$, the $6j$ symbol vanishes \c{Caron-Huot2017, Liu2018}: i.e. writing a $d$-dimensional T-channel global conformal block as $G_{\D,J}(1-z,1-\zb)$, 
\begin{equation}
	\dDisc_T(G_{2h_1+J,J}(1-z,1-\zb)) = 0
\end{equation}
where dDisc$_T$ means that the operators are taken around $\zb=1$. Thus, the Lorentzian inversion gives a vanishing result. The total result for the cross-channel decomposition involves non-analytic contributions at low spin which are not captured by the Lorentzian inversion formula; however, at large spin, these pieces can be neglected. 

Returning to the present Virasoro case, then, recall that we evaluated the kernel $\kernel_{\a_s\a_t}$ for $\a_t=2\a_1$ or $\a_t=2\a_2$ in \eqref{nonvacdouble}. It is nonzero. There is no reason for it to vanish, in part because this is a chiral object, insensitive to the spin of the intermediate operator. However, in the limit of large S-channel twist taken above, we see that zeroes emerge. This mimics the result in the global case, providing yet another analogy between global double-twist operators $[\O_1\O_2]_{m,\ell}$ with $h=h_1+h_2+m$ and the Virasoro double-twists $\lbrace \O_1\O_2\rbrace_{m,\ell}$ with $\a=\a_1+\a_2+m b$. 
 
 \subsection{Heavy internal weight in the large-$c$ limit}\label{app:LargeCLargeSpinRatio}
 In the main text we compute the anomalous momenta in a large $c$ limit in which the spin is scaled with the central charge. The main technical ingredient needed for this computation is the ratio of the non-vacuum to vacuum kernels in this limit (as seen in (\ref{nonholope})). We will parameterize this limit by taking 
 \ie
 \alpha_s = {Q\over 2}+ip_s b^{-1}
 \fe
 and sending $b\to 0$ while keeping all other weights held fixed. As we will see, the computation will be similar to the limit in which the internal weight is parametrically larger than the central charge.
 
Scaling $s$ with $b^{-1}$ as $s=Sb^{-1}$, the integrand in the kernel (\ref{eq:CrossingKernel}) behaves in this limit as
\ie
\log\left(\prod_{i=1}^4{S_b(s+U_i)\over S_b(s+V_i)}\right) \sim& \bigg[S-1+2I(1-S)-I(S)-I(S-1)+(1-S)\log(1-S)-{3\over 2}\log(2\pi)\\
&+I({1\over 2}-ip+S)+I({1\over 2}+i p+S)-I({1\over 2}-i p-S)-I({1\over 2}+i p-S)\bigg]b^{-2}+\mathcal{O}(b^0),
\fe
where $I(x) = \int_{1\over 2}^x dt\log \Gamma(t)$. As a function of $S$, the integrand should have poles extending to the right at $S \approx {1\over 2}\pm i p, 0, 1$ (up to corrections of order $b^2$). However there are also poles extending to the left at $S\approx-1,0,1$. To evaluate the integral will then require us to for example pick up the residues of the poles extending to the right at $S\approx 0$.

Noting that the poles of interest occur for $s\sim \mathcal{O}(b)$, the part of the integrand dependent on $p_s$ takes the following form in the large-$c$ limit
\ie
\log\left({1\over S_b(s+V_1)S_b(s+V_2)}\right)\sim \log\left({\Gamma({1\over 2}+i p_s)\Gamma({1\over 2}-ip_s)\over 2\pi}\right)\left({2s\over b}-2( h_1-h_2)+\mathcal{O}(b)\right).
\fe
Since $b^{-1}$ is the large parameter and $\alpha_t\sim h_t b$, we in principle have to care about the subleading poles at $s = \alpha_t + m b$, since these are not a priori parametrically suppressed compared to the leading pole. Evaluating the contributions of these poles to the integral, we have 
\ie
\int_{C'}{ ds\over i}\prod_{i=1}^4{S_b(s+U_i)\over S_b(s+V_i)}\sim & \sum_{m=0}^\infty{(-)^m\over (2\pi b^2)^m m!}\left({\Gamma({1\over 2}+i p_s)\Gamma({1\over 2}-i p_s)\over 2\pi}\right)^{2( h_t+m- h_1+ h_2)}\\
&{S_b(\alpha_t +mb)^2S_b(Q-2\alpha_1+\alpha_t+m b)S_b(2\alpha_2+\alpha_t-Q+mb)\over S_b(2\alpha_t+m b)},
\fe
where in the second line we will only be keeping the leading terms in the small $b$ expansion (recall that $\a_i \sim h_i b$). We can now combine with the prefactor and divide by the same limit of the vacuum kernel to arrive at
\ie
{\kernel_{\alpha_s\alpha_t}\over \kernel_{\alpha_s\id}}\sim & \sum_{m=0}^\infty{(-)^m(2\pi b^2)^{2( h_t+m)}\over m!}{\Gamma(2 h_1)\Gamma(2 h_2)( h_t)_m^2(2h_2+ h_t-1)_m\over \Gamma(2 h_1- h_t-m)\Gamma(2 h_2- h_t)(2 h_t)_m}\left({\Gamma({1\over 2}+ip_s)\Gamma({1\over 2}-ip_s)\over 2\pi}\right)^{2h_t+2m}.
\fe
We see that the terms coming from the subleading poles do end up being suppressed relative to the leading pole in this particular semiclassical limit. So in the large $c$ limit we are left with
\begin{equation}\label{eq:largecAnom}
	\frac{\kernel_{\alpha_s\alpha_t}}{\kernel_{\alpha_s\id}}\sim \frac{\Gamma(2 h_1)\Gamma(2 h_2)}{\Gamma(2 h_1- h_t)\Gamma(2 h_2- h_t)}\left(\frac{c}{6\pi}\cosh(\pi {p}_s)\right)^{-2{h}_t}.
\end{equation}
In the $p_s\sim\sqrt{6 h_s\over c}\to\infty$ limit, this reproduces the large-spin result (\ref{eq:LargeCLargeSpinRatio}). Similarly, as discussed in section \ref{sec:LargeSpinLargeC}, in the limit that the spin is parametrically smaller than the central charge, this also reproduces both the scaling with spin and the precise coefficient of the anomalous weights due to T-channel exchange familiar from the usual lightcone bootstrap \cite{Komargodski2013,Fitzpatrick2013}.

\section{Lorentzian vacuum inversion in the Newtonian limit}\label{app:Newton}
In this appendix we derive the result for the leading anomalous twist due to the Virasoro vacuum module, $\delta h_0=-2\a_1\a_2$, directly from the Lorentzian inversion formula, in the ``Newtonian'' limit
\e{}{(h_1,h_2, c)\to\i~,\quad  {h_1h_2\o c}~~\text{fixed}~.}
In terms of momenta $\a_i$, the product $\a_1\a_2$ behaves as $\sim \O(b^0)$. (For instance,  taking $\a_i \sim bh_i$ with $h_i \sim b^{-1}$.)

In this limit, the T-channel Virasoro vacuum block is known to take the form \cite{Fitzpatrick2014}
\e{}{\cF_{\rm vac}(1-z) = \exp\left[{2h_1h_2\o c} k_4(1-z)\right]~.}
To extract the anomalous twist $\delta h_0$, we plug into the chiral inversion formula,
\e{}{\int_0^1 {dz \o z^2} \,k_{2(1-h)}(z)\,\left({z\o 1-z}\right)^{h_1+h_2}\cF_{\rm vac}(1-z)}
and extract the leading singularity near $z=0$. Using 
\e{}{k_4(1-z)|_{z\ll1} \sim -6\log z~,}
the inversion integral becomes
\es{}{\int_0^1 {dz \o z} z^{-h+h_1+h_2-{12h_1h_2\o c}}&\sim {1\o h-(h_1+h_2-{12h_1h_2\o c})}\\
&\sim {1\o h-(h_1+h_2-2\a_1\a_2)}}
which is the desired result. Subleading anomalous twists $\delta h_m$ may be extracted from the subleading behaviour near $z=0$.

\section{Large spin analysis from old-fashioned lightcone bootstrap}\label{app:OldLC}

Here we connect the large spin results of section \ref{sec:largeSpin} -- derived without the use of conformal blocks themselves -- with the kinematic method using conformal blocks. 

Let us consider crossing symmetry \ref{crossingEq} for the case of interest where external operators are identical in pairs
\begin{equation}
	\sum_s (C_{12s})^2 \mathcal{F}^{21}_{21}\left(\alpha_s\middle|z\right)\bar{\mathcal{F}}^{21}_{21}\left(\bar{\alpha}_s\middle|\bar{z}\right) = \sum_t C_{11t}C_{t22} \mathcal{F}^{22}_{11}\left(\alpha_t\middle|1-z\right)\bar{\mathcal{F}}^{22}_{11}\left(\bar{\alpha}_t\middle| 1-\bar{z}\right),
\end{equation}
Many of our results can be derived, with somewhat more work, by considering the $\bar{z} \to 1$ limit in which the insertions of operators become null separated. Recalling our convention $\mathcal{F}^{21}_{34}\left(\alpha\middle|z\right)\sim z^{h-h_1-h_2}$, the $\bar z \to 1$ limit is dominated in the T-channel by the operator with the lowest $\bar{h}_t$, which is the identity,
\begin{equation}
	\sum_s (C_{12s})^2 \mathcal{F}^{21}_{21}\left(\alpha_s\middle|z\right)\bar{\mathcal{F}}^{21}_{21}\left(\bar{\alpha}_s\middle|\bar{z} \to 1\right) \sim  \mathcal{F}^{22}_{11}\left(0\middle|1-z\right)(1-\bar{z})^{-2\bar{h}_2}
\end{equation}
No individual S-channel block is sufficiently singular in the $\bar{z}\to 1$ limit to reproduce the appropriate $(1-\bar{z})^{-2\bar{h}_2}$ singularity of the T-channel, so we need an infinite number of them. The only way to solve crossing is then by including infinite families of operators at large $\bar{h}_s$.

The coefficient of the singularity depends on $z$ through the T-channel vacuum block $\mathcal{F}^{22}_{11}\left(0\middle|1-z\right)$. We can most easily relate this $z$-dependence to S-channel twists $h_s$ by subsequently taking the additional limit $z\to 0$, with $z\gg 1-\bar{z}$, sometimes called the ``double lightcone limit''. We can simply read off the S-channel twists from the powers that appear in the small $z$ expansion of $\mathcal{F}^{22}_{11}\left(0\middle|1-z\right)$. We derived the leading small $z$ behaviour, appropriate for the $m=0$ Regge trajectory, in section \ref{sec:crosschannel}, essentially by reverse-engineering the known result for the double-twist spectrum from the fusion kernel. For $\alpha_1+\alpha_2<\frac{Q}{2}$ this goes as $z^{-2\alpha_1\alpha_2}$, coming simply from the leading pole of the fusion kernel. Therefore, the twists must accumulate to $h_s = h_1+h_2-2\alpha_1\alpha_2$, which manifestly reproduces our results (\ref{anomdim}) for $m=0$ in a more laborious and less transparent way. Non-vacuum T-channel exchanges have the same power, but with an additional $\log z$, so including them as subleading terms gives rise to anomalous twists. For $\Re(\alpha_1+\alpha_2)>\frac{Q}{2}$, the $z\to 0$ limit of the T-channel vacuum block may be read off from \eqr{crosscont}. Finally, extracting the spin dependence of the spectral density of Virasoro primaries requires understanding the antiholomorphic S-channel blocks in an appropriate combined cross-channel, large dimension limit $\bar{z}\to 1$, $\bar{h}_s\to \infty$. 

\section{Computations for interpretation of anomalous twists}\label{app:monodromy}

In section \ref{sec:anomTwistsGravity}, we gave an interpretation for the behavior of anomalous twists, due to an AdS interaction familiar from previous work, but after changing to a new conformal frame in which the relevant two-particle states are Virasoro primaries. In this appendix, we give details of the calculation of the spin as measured in the new primary frame.

Mathematically, as explained in the text, the main step is to compute the monodromy of the differential equation \eqref{eq:monodromyODE}:
\begin{equation}
	\psi''(w) + \left[\frac{1}{4} - \frac{1-\nu_1^2}{4} \left(\frac{\sinh \epsilon_1}{\cosh\epsilon_1-\cos w}\right)^2 - \frac{1-\nu_2^2}{4}\left(\frac{\sinh \epsilon_2}{\cosh\epsilon_2+\cos w}\right)^2\right]\psi(w)=0
\end{equation}
The $\nu$ parameters are related to conformal dimensions as $\bar{h}=\frac{c}{24}(1-\nu^2)$, so in particular, if $\bar{h}\ll c$, $\nu$ is close to unity.

For generic $w$, as $\epsilon_i\to 0$, the terms with nontrivial $w$ dependence in the ODE are unimportant, so a basis of solutions to the ODE can be well-approximated by $\cos\frac{w}{2}$ and $\sin\frac{w}{2}$. However, when $w$ is of order $\epsilon_1$, we must take into account the first nontrivial term; writing $w=\epsilon_1 x$ and taking $\epsilon_1\to 0$ with fixed $x$, the resulting limit of the ODE has simple solutions:
\begin{equation}
	\frac{d^2\psi}{dx^2}-\frac{1-\nu^2}{(1+x^2)^2}\psi = 0 \implies \psi \sim \begin{cases}
		\sqrt{\epsilon_1^2+w^2} \cos\left(\nu_1 \arctan\tfrac{w}{\epsilon_1}\right) \\
		\sqrt{\epsilon_1^2+w^2} \sin\left(\nu_1 \arctan\tfrac{w}{\epsilon_1}\right)
	\end{cases} \qquad (w= \epsilon_1 x)
\end{equation}
Now we can find the coefficients of $\cos\frac{w}{2}$ and $\sin\frac{w}{2}$ to which these solutions match, by expanding at $|w|\gg \epsilon_1$ and reading off the constant and the coefficient of $\frac{w}{2}$ respectively. From this, we can write a monodromy matrix which starts with a solution in the $(\cos\frac{w}{2},\sin\frac{w}{2})$ basis for $-\pi<w<0$, solves through the $w\approx 0$ region, and reexpresses the result in the $(\cos\frac{w}{2},\sin\frac{w}{2})$ basis for $0<w<\pi$:
\begin{equation}
	M_0 = \begin{pmatrix} \nu_1 \sin\left(\tfrac{\pi\nu_1}{2}\right)\epsilon_1 & -\nu_1 \cos\left(\tfrac{\pi\nu_1}{2}\right)\epsilon_1 \\ 2\cos\left(\tfrac{\pi\nu_1}{2}\right) & 2\sin\left(\tfrac{\pi\nu_1}{2}\right) \end{pmatrix}
	\begin{pmatrix} \nu_1 \sin\left(\tfrac{\pi\nu_1}{2}\right)\epsilon_1 & \nu_1 \cos\left(\tfrac{\pi\nu_1}{2}\right)\epsilon_1 \\ -2\cos\left(\tfrac{\pi\nu_1}{2}\right) & 2\sin\left(\tfrac{\pi\nu_1}{2}\right) \end{pmatrix}^{-1} = \begin{pmatrix}
		-\cos (\pi  \nu_1) & -\tfrac{\epsilon_1  \nu_1}{2} \sin (\pi  \nu_1 ) \\
 \tfrac{2}{\epsilon_1  \nu_1 } \sin (\pi  \nu_1 )& -\cos (\pi  \nu_1 )
	\end{pmatrix}
\end{equation}
We then repeat the analysis near $w=\pi$ to get another monodromy matrix for passing through that point:
\begin{equation}
	M_\pi  = \begin{pmatrix}
		\cos (\pi  \nu_2) & \tfrac{2}{\epsilon_2  \nu_2 } \sin (\pi  \nu_2 ) \\
 -\tfrac{\epsilon_2  \nu_2}{2} \sin (\pi  \nu_2)& \cos (\pi  \nu_2)
	\end{pmatrix}
\end{equation}
Finally, we combine these to find the trace of the monodromy around the entire circle:
\begin{equation}
	\Tr M= \Tr M_0 M_\pi \sim \frac{4}{\epsilon_1\epsilon_2\nu_1\nu_2} \sin(\pi\nu_1)\sin(\pi\nu_2)-2\cos(\pi\nu_1)\cos(\pi\nu_2)
\end{equation}
Here, we have used $\epsilon_i\ll 1$ to drop one term; however, we do not drop the last term, because in some regimes of interest the remaining terms will be of comparable magnitude.

To obtain the result \eqref{eq:monodromyResult} quoted in the text, it remains only to replace $\nu_i$ with conformal dimensions, assumed much less than $c$, in which case we have $\frac{\sin (\pi \nu)}{\nu} \sim \frac{12\pi \bar{h}}{c}$, $\cos (\pi \nu)\sim -1$:
\begin{equation}
	\Tr M \sim \left(\frac{24 \pi}{c}\right)^2 \frac{\bar{h}_1\bar{h}_2}{\epsilon_1\epsilon_2}-2
\end{equation}

\bibliographystyle{JHEP}
\bibliography{VirasoroLightcone}

\providecommand{\href}[2]{#2}\begingroup\raggedright\begin{thebibliography}{100}

\bibitem{Caron-Huot2017}
S.~Caron-Huot, {\it {Analyticity in Spin in Conformal Theories}},  {\em JHEP}
  {\bf 09} (2017) 078, [\href{http://arxiv.org/abs/1703.00278}{{\tt
  arXiv:1703.00278}}].

\bibitem{Simmons-Duffin2018a}
D.~Simmons-Duffin, D.~Stanford, and E.~Witten, {\it {A spacetime derivation of
  the Lorentzian OPE inversion formula}},  {\em JHEP} {\bf 07} (2018) 085,
  [\href{http://arxiv.org/abs/1711.03816}{{\tt arXiv:1711.03816}}].

\bibitem{Kravchuk2018}
P.~Kravchuk and D.~Simmons-Duffin, {\it {Light-ray operators in conformal field
  theory}},  \href{http://arxiv.org/abs/1805.00098}{{\tt arXiv:1805.00098}}.

\bibitem{Karateev2018}
D.~Karateev, P.~Kravchuk, and D.~Simmons-Duffin, {\it {Harmonic Analysis and
  Mean Field Theory}},  \href{http://arxiv.org/abs/1809.05111}{{\tt
  arXiv:1809.05111}}.

\bibitem{Komargodski2013}
Z.~Komargodski and A.~Zhiboedov, {\it {Convexity and Liberation at Large
  Spin}},  {\em JHEP} {\bf 11} (2013) 140,
  [\href{http://arxiv.org/abs/1212.4103}{{\tt arXiv:1212.4103}}].

\bibitem{Fitzpatrick2013}
A.~L. Fitzpatrick, J.~Kaplan, D.~Poland, and D.~Simmons-Duffin, {\it {The
  Analytic Bootstrap and AdS Superhorizon Locality}},  {\em JHEP} {\bf 12}
  (2013) 004, [\href{http://arxiv.org/abs/1212.3616}{{\tt arXiv:1212.3616}}].

\bibitem{Fitzpatrick2014}
A.~L. Fitzpatrick, J.~Kaplan, and M.~T. Walters, {\it {Universality of
  Long-Distance AdS Physics from the CFT Bootstrap}},  {\em JHEP} {\bf 08}
  (2014) 145, [\href{http://arxiv.org/abs/1403.6829}{{\tt arXiv:1403.6829}}].

\bibitem{Kaviraj:2015cxa}
A.~Kaviraj, K.~Sen, and A.~Sinha, {\it {Analytic bootstrap at large spin}},
  {\em JHEP} {\bf 11} (2015) 083, [\href{http://arxiv.org/abs/1502.01437}{{\tt
  arXiv:1502.01437}}].

\bibitem{Kaviraj:2015xsa}
A.~Kaviraj, K.~Sen, and A.~Sinha, {\it {Universal anomalous dimensions at large
  spin and large twist}},  {\em JHEP} {\bf 07} (2015) 026,
  [\href{http://arxiv.org/abs/1504.00772}{{\tt arXiv:1504.00772}}].

\bibitem{Alday:2015ewa}
L.~F. Alday and A.~Zhiboedov, {\it {An Algebraic Approach to the Analytic
  Bootstrap}},  {\em JHEP} {\bf 04} (2017) 157,
  [\href{http://arxiv.org/abs/1510.08091}{{\tt arXiv:1510.08091}}].

\bibitem{Alday:2016njk}
L.~F. Alday, {\it {Large Spin Perturbation Theory for Conformal Field
  Theories}},  {\em Phys. Rev. Lett.} {\bf 119} (2017), no.~11 111601,
  [\href{http://arxiv.org/abs/1611.01500}{{\tt arXiv:1611.01500}}].

\bibitem{Simmons-Duffin2017}
D.~Simmons-Duffin, {\it {The Lightcone Bootstrap and the Spectrum of the 3d
  Ising CFT}},  {\em JHEP} {\bf 03} (2017) 086,
  [\href{http://arxiv.org/abs/1612.08471}{{\tt arXiv:1612.08471}}].

\bibitem{Liu2018}
J.~Liu, E.~Perlmutter, V.~Rosenhaus, and D.~Simmons-Duffin, {\it
  {$d$-dimensional SYK, AdS Loops, and $6j$ Symbols}},
  \href{http://arxiv.org/abs/1808.00612}{{\tt arXiv:1808.00612}}.

\bibitem{Heemskerk2009}
I.~Heemskerk, J.~Penedones, J.~Polchinski, and J.~Sully, {\it {Holography from
  Conformal Field Theory}},  {\em JHEP} {\bf 10} (2009) 079,
  [\href{http://arxiv.org/abs/0907.0151}{{\tt arXiv:0907.0151}}].

\bibitem{Cardona:2018dov}
C.~Cardona and K.~Sen, {\it {Anomalous dimensions at finite conformal spin from
  OPE inversion}},  \href{http://arxiv.org/abs/1806.10919}{{\tt
  arXiv:1806.10919}}.

\bibitem{Kraus:2018zrn}
P.~Kraus, A.~Sivaramakrishnan, and R.~Snively, {\it {Late time Wilson lines}},
  \href{http://arxiv.org/abs/1810.01439}{{\tt arXiv:1810.01439}}.

\bibitem{Gopakumar:2018xqi}
R.~Gopakumar and A.~Sinha, {\it {On the Polyakov-Mellin bootstrap}},
  \href{http://arxiv.org/abs/1809.10975}{{\tt arXiv:1809.10975}}.

\bibitem{Cardona:2018qrt}
C.~Cardona, S.~Guha, S.~K. Kanumilli, and K.~Sen, {\it {Resummation at finite
  conformal spin}},  \href{http://arxiv.org/abs/1811.00213}{{\tt
  arXiv:1811.00213}}.

\bibitem{Sleight:2018epi}
C.~Sleight and M.~Taronna, {\it {Spinning Mellin Bootstrap: Conformal Partial
  Waves, Crossing Kernels and Applications}},  {\em Fortsch. Phys.} {\bf 66}
  (2018) 8, [\href{http://arxiv.org/abs/1804.09334}{{\tt arXiv:1804.09334}}].

\bibitem{Sleight:2018ryu}
C.~Sleight and M.~Taronna, {\it {Anomalous Dimensions from Crossing Kernels}},
  \href{http://arxiv.org/abs/1807.05941}{{\tt arXiv:1807.05941}}.

\bibitem{Ponsot1999}
B.~Ponsot and J.~Teschner, {\it {Liouville bootstrap via harmonic analysis on a
  noncompact quantum group}},  \href{http://arxiv.org/abs/hep-th/9911110}{{\tt
  hep-th/9911110}}.

\bibitem{Ponsot2001}
B.~Ponsot and J.~Teschner, {\it {Clebsch-Gordan and Racah-Wigner coefficients
  for a continuous series of representations of U(q)(sl(2,R))}},  {\em Commun.
  Math. Phys.} {\bf 224} (2001) 613--655,
  [\href{http://arxiv.org/abs/math/0007097}{{\tt math/0007097}}].

\bibitem{Teschner2001}
J.~Teschner, {\it {Liouville theory revisited}},  {\em Class. Quant. Grav.}
  {\bf 18} (2001) R153--R222, [\href{http://arxiv.org/abs/hep-th/0104158}{{\tt
  hep-th/0104158}}].

\bibitem{Dorn1994a}
H.~Dorn and H.~J. Otto, {\it {Two and three point functions in Liouville
  theory}},  {\em Nucl. Phys.} {\bf B429} (1994) 375--388,
  [\href{http://arxiv.org/abs/hep-th/9403141}{{\tt hep-th/9403141}}].

\bibitem{Zamolodchikov1996}
A.~B. Zamolodchikov and A.~B. Zamolodchikov, {\it {Structure constants and
  conformal bootstrap in Liouville field theory}},  {\em Nucl. Phys.} {\bf
  B477} (1996) 577--605, [\href{http://arxiv.org/abs/hep-th/9506136}{{\tt
  hep-th/9506136}}].

\bibitem{Jackson:2014nla}
S.~Jackson, L.~McGough, and H.~Verlinde, {\it {Conformal Bootstrap,
  Universality and Gravitational Scattering}},  {\em Nucl. Phys.} {\bf B901}
  (2015) 382--429, [\href{http://arxiv.org/abs/1412.5205}{{\tt
  arXiv:1412.5205}}].

\bibitem{Chang2016}
C.-M. Chang and Y.-H. Lin, {\it {Bootstrap, universality and horizons}},  {\em
  JHEP} {\bf 10} (2016) 068, [\href{http://arxiv.org/abs/1604.01774}{{\tt
  arXiv:1604.01774}}].

\bibitem{Chang2016b}
C.-M. Chang and Y.-H. Lin, {\it {Bootstrapping 2D CFTs in the Semiclassical
  Limit}},  {\em JHEP} {\bf 08} (2016) 056,
  [\href{http://arxiv.org/abs/1510.02464}{{\tt arXiv:1510.02464}}].

\bibitem{Esterlis2016}
I.~Esterlis, A.~L. Fitzpatrick, and D.~Ramirez, {\it {Closure of the Operator
  Product Expansion in the Non-Unitary Bootstrap}},  {\em JHEP} {\bf 11} (2016)
  030, [\href{http://arxiv.org/abs/1606.07458}{{\tt arXiv:1606.07458}}].

\bibitem{He2017}
S.~He, {\it {Conformal Bootstrap to R\'enyi Entropy in 2D Liouville and
  Super-Liouville CFTs}},  \href{http://arxiv.org/abs/1711.00624}{{\tt
  arXiv:1711.00624}}.

\bibitem{Mertens2017}
T.~G. Mertens, G.~J. Turiaci, and H.~L. Verlinde, {\it {Solving the Schwarzian
  via the Conformal Bootstrap}},  {\em JHEP} {\bf 08} (2017) 136,
  [\href{http://arxiv.org/abs/1705.08408}{{\tt arXiv:1705.08408}}].

\bibitem{Nachtmann1973}
O.~Nachtmann, {\it {Positivity constraints for anomalous dimensions}},  {\em
  Nucl. Phys.} {\bf B63} (1973) 237--247.

\bibitem{Kraus2017a}
P.~Kraus and A.~Maloney, {\it {A cardy formula for three-point coefficients or
  how the black hole got its spots}},  {\em JHEP} {\bf 05} (2017) 160,
  [\href{http://arxiv.org/abs/1608.03284}{{\tt arXiv:1608.03284}}].

\bibitem{Cardy2017}
J.~Cardy, A.~Maloney, and H.~Maxfield, {\it {A new handle on three-point
  coefficients: OPE asymptotics from genus two modular invariance}},  {\em
  JHEP} {\bf 10} (2017) 136, [\href{http://arxiv.org/abs/1705.05855}{{\tt
  arXiv:1705.05855}}].

\bibitem{Das2017c}
D.~Das, S.~Datta, and S.~Pal, {\it {Charged structure constants from
  modularity}},  {\em JHEP} {\bf 11} (2017) 183,
  [\href{http://arxiv.org/abs/1706.04612}{{\tt arXiv:1706.04612}}].

\bibitem{Das2017}
D.~Das, S.~Datta, and S.~Pal, {\it {Modular crossings, OPE coefficients and
  black holes}},  \href{http://arxiv.org/abs/1712.01842}{{\tt
  arXiv:1712.01842}}.

\bibitem{Teschner2003}
J.~Teschner, {\it {From Liouville theory to the quantum geometry of Riemann
  surfaces}},  in {\em {Mathematical physics. Proceedings, 14th International
  Congress, ICMP 2003, Lisbon, Portugal, July 28-August 2, 2003}}, 2003.
\newblock \href{http://arxiv.org/abs/hep-th/0308031}{{\tt hep-th/0308031}}.

\bibitem{Nemkov2015}
N.~Nemkov, {\it {On modular transformations of toric conformal blocks}},  {\em
  JHEP} {\bf 10} (2015) 039, [\href{http://arxiv.org/abs/1504.04360}{{\tt
  arXiv:1504.04360}}].

\bibitem{Nemkov2017}
N.~Nemkov, {\it {Analytic properties of the Virasoro modular kernel}},  {\em
  Eur. Phys. J.} {\bf C77} (2017), no.~6 368,
  [\href{http://arxiv.org/abs/1610.02000}{{\tt arXiv:1610.02000}}].

\bibitem{Moore:1988uz}
G.~W. Moore and N.~Seiberg, {\it {Polynomial Equations for Rational Conformal
  Field Theories}},  {\em Phys. Lett.} {\bf B212} (1988) 451--460.

\bibitem{Cardy1986a}
J.~L. Cardy, {\it {Operator Content of Two-Dimensional Conformally Invariant
  Theories}},  {\em Nucl. Phys.} {\bf B270} (1986) 186--204.

\bibitem{Keller2015}
C.~A. Keller and A.~Maloney, {\it {Poincare Series, 3D Gravity and CFT
  Spectroscopy}},  {\em JHEP} {\bf 02} (2015) 080,
  [\href{http://arxiv.org/abs/1407.6008}{{\tt arXiv:1407.6008}}].

\bibitem{Benjamin2016}
N.~Benjamin, E.~Dyer, A.~L. Fitzpatrick, A.~Maloney, and E.~Perlmutter, {\it
  {Small Black Holes and Near-Extremal CFTs}},  {\em JHEP} {\bf 08} (2016) 023,
  [\href{http://arxiv.org/abs/1603.08524}{{\tt arXiv:1603.08524}}].

\bibitem{Fitzpatrick2015a}
A.~L. Fitzpatrick, J.~Kaplan, and M.~T. Walters, {\it {Virasoro Conformal
  Blocks and Thermality from Classical Background Fields}},  {\em JHEP} {\bf
  11} (2015) 200, [\href{http://arxiv.org/abs/1501.05315}{{\tt
  arXiv:1501.05315}}].

\bibitem{Fitzpatrick2017a}
A.~L. Fitzpatrick and J.~Kaplan, {\it {On the Late-Time Behavior of Virasoro
  Blocks and a Classification of Semiclassical Saddles}},  {\em JHEP} {\bf 04}
  (2017) 072, [\href{http://arxiv.org/abs/1609.07153}{{\tt arXiv:1609.07153}}].

\bibitem{Chen2017}
H.~Chen, C.~Hussong, J.~Kaplan, and D.~Li, {\it {A Numerical Approach to
  Virasoro Blocks and the Information Paradox}},  {\em JHEP} {\bf 09} (2017)
  102, [\href{http://arxiv.org/abs/1703.09727}{{\tt arXiv:1703.09727}}].

\bibitem{Fitzpatrick2017}
A.~L. Fitzpatrick, J.~Kaplan, D.~Li, and J.~Wang, {\it {Exact Virasoro Blocks
  from Wilson Lines and Background-Independent Operators}},  {\em JHEP} {\bf
  07} (2017) 092, [\href{http://arxiv.org/abs/1612.06385}{{\tt
  arXiv:1612.06385}}].

\bibitem{Besken:2017fsj}
M.~Besken, A.~Hegde, and P.~Kraus, {\it {Anomalous dimensions from quantum
  Wilson lines}},  \href{http://arxiv.org/abs/1702.06640}{{\tt
  arXiv:1702.06640}}.

\bibitem{Besken:2018zro}
M.~Besken, E.~D'Hoker, A.~Hegde, and P.~Kraus, {\it {Renormalization of
  gravitational Wilson lines}},  \href{http://arxiv.org/abs/1810.00766}{{\tt
  arXiv:1810.00766}}.

\bibitem{Hikida:2017ehf}
Y.~Hikida and T.~Uetoko, {\it {Correlators in higher-spin $AdS_3$ holography
  from Wilson lines with loop corrections}},  {\em PTEP} {\bf 2017} (2017)
  113B03, [\href{http://arxiv.org/abs/1708.08657}{{\tt arXiv:1708.08657}}].

\bibitem{Anand:2017dav}
N.~Anand, H.~Chen, A.~L. Fitzpatrick, J.~Kaplan, and D.~Li, {\it {An Exact
  Operator That Knows Its Location}},  {\em JHEP} {\bf 02} (2018) 012,
  [\href{http://arxiv.org/abs/1708.04246}{{\tt arXiv:1708.04246}}].

\bibitem{Cotler:2018zff}
J.~Cotler and K.~Jensen, {\it {A theory of reparameterizations for AdS$_3$
  gravity}},  \href{http://arxiv.org/abs/1808.03263}{{\tt arXiv:1808.03263}}.

\bibitem{Dotsenko1999}
V.~Dotsenko, J.~L. Jacobsen, M.-A. Lewis, and M.~Picco, {\it {Coupled Potts
  models: Self-duality and fixed point structure}},  {\em Nucl. Phys.} {\bf
  B546} (1999) 505--557.

\bibitem{Kusuki2018c}
Y.~Kusuki, {\it {Light Cone Bootstrap in General 2D CFTs \&amp;amp;
  Entanglement from Light Cone Singularity}},
  \href{http://arxiv.org/abs/1810.01335}{{\tt arXiv:1810.01335}}.

\bibitem{Teschner2014}
J.~Teschner and G.~Vartanov, {\it {6j symbols for the modular double, quantum
  hyperbolic geometry, and supersymmetric gauge theories}},  {\em Lett. Math.
  Phys.} {\bf 104} (2014) 527--551, [\href{http://arxiv.org/abs/1202.4698}{{\tt
  arXiv:1202.4698}}].

\bibitem{Teschner2015}
J.~Teschner and G.~S. Vartanov, {\it {Supersymmetric gauge theories,
  quantization of $\mathcal{M}_{\mathrm{flat}}$, and conformal field theory}},
  {\em Adv. Theor. Math. Phys.} {\bf 19} (2015) 1--135,
  [\href{http://arxiv.org/abs/1302.3778}{{\tt arXiv:1302.3778}}].

\bibitem{Ponsot2004}
B.~Ponsot, {\it {Recent progresses on Liouville field theory}},  {\em Int. J.
  Mod. Phys.} {\bf A19S2} (2004) 311--335,
  [\href{http://arxiv.org/abs/hep-th/0301193}{{\tt hep-th/0301193}}].

\bibitem{Belavin:1984vu}
A.~A. Belavin, A.~M. Polyakov, and A.~B. Zamolodchikov, {\it {Infinite
  Conformal Symmetry in Two-Dimensional Quantum Field Theory}},  {\em Nucl.
  Phys.} {\bf B241} (1984) 333--380. [,605(1984)].

\bibitem{Lin2017}
Y.-H. Lin, S.-H. Shao, D.~Simmons-Duffin, Y.~Wang, and X.~Yin, {\it {$
  \mathcal{N} $ = 4 superconformal bootstrap of the K3 CFT}},  {\em JHEP} {\bf
  05} (2017) 126, [\href{http://arxiv.org/abs/1511.04065}{{\tt
  arXiv:1511.04065}}].

\bibitem{Collier2018}
S.~Collier, P.~Kravchuk, Y.-H. Lin, and X.~Yin, {\it {Bootstrapping the
  Spectral Function: On the Uniqueness of Liouville and the Universality of
  BTZ}},  {\em JHEP} {\bf 09} (2018) 150,
  [\href{http://arxiv.org/abs/1702.00423}{{\tt arXiv:1702.00423}}].

\bibitem{Balasubramanian2017}
V.~Balasubramanian, A.~Bernamonti, B.~Craps, T.~De~Jonckheere, and F.~Galli,
  {\it Heavy-heavy-light-light correlators in liouville theory},  {\em Journal
  of High Energy Physics} {\bf 2017} (2017), no.~8 45.

\bibitem{Harlow2011}
D.~Harlow, J.~Maltz, and E.~Witten, {\it {Analytic Continuation of Liouville
  Theory}},  {\em JHEP} {\bf 12} (2011) 071,
  [\href{http://arxiv.org/abs/1108.4417}{{\tt arXiv:1108.4417}}].

\bibitem{Zamolodchikov1986}
A.~B. Zamolodchikov, {\it Two-dimensional conformal symmetry and critical
  four-spin correlation functions in the ashkin-teller model},  {\em Sov.
  Phys.-JETP} {\bf 63} (1986) 1061--1066.

\bibitem{Kusuki2018b}
Y.~Kusuki and T.~Takayanagi, {\it {Renyi entropy for local quenches in 2D CFT
  from numerical conformal blocks}},  {\em JHEP} {\bf 01} (2018) 115,
  [\href{http://arxiv.org/abs/1711.09913}{{\tt arXiv:1711.09913}}].

\bibitem{Kusuki2018a}
Y.~Kusuki, {\it {New Properties of Large-$c$ Conformal Blocks from Recursion
  Relation}},  {\em JHEP} {\bf 07} (2018) 010,
  [\href{http://arxiv.org/abs/1804.06171}{{\tt arXiv:1804.06171}}].

\bibitem{newton}
I.~Newton.
\newblock 1665.

\bibitem{Maloney2017}
A.~Maloney, H.~Maxfield, and G.~S. Ng, {\it {A conformal block Farey tail}},
  {\em JHEP} {\bf 06} (2017) 117, [\href{http://arxiv.org/abs/1609.02165}{{\tt
  arXiv:1609.02165}}].

\bibitem{p1}
L.~Cornalba, M.~S. Costa, J.~Penedones, and R.~Schiappa, {\it {Eikonal
  Approximation in AdS/CFT: From Shock Waves to Four-Point Functions}},  {\em
  JHEP} {\bf 08} (2007) 019, [\href{http://arxiv.org/abs/hep-th/0611122}{{\tt
  hep-th/0611122}}].

\bibitem{p2}
L.~Cornalba, M.~S. Costa, J.~Penedones, and R.~Schiappa, {\it {Eikonal
  Approximation in AdS/CFT: Conformal Partial Waves and Finite N Four-Point
  Functions}},  {\em Nucl. Phys.} {\bf B767} (2007) 327--351,
  [\href{http://arxiv.org/abs/hep-th/0611123}{{\tt hep-th/0611123}}].

\bibitem{p3}
L.~Cornalba, M.~S. Costa, and J.~Penedones, {\it {Eikonal approximation in
  AdS/CFT: Resumming the gravitational loop expansion}},  {\em JHEP} {\bf 09}
  (2007) 037, [\href{http://arxiv.org/abs/0707.0120}{{\tt arXiv:0707.0120}}].

\bibitem{Katz}
A.~L. Fitzpatrick, E.~Katz, D.~Poland, and D.~Simmons-Duffin, {\it {Effective
  Conformal Theory and the Flat-Space Limit of AdS}},  {\em JHEP} {\bf 07}
  (2011) 023, [\href{http://arxiv.org/abs/1007.2412}{{\tt arXiv:1007.2412}}].

\bibitem{Camanho:2014apa}
X.~O. Camanho, J.~D. Edelstein, J.~Maldacena, and A.~Zhiboedov, {\it {Causality
  Constraints on Corrections to the Graviton Three-Point Coupling}},  {\em
  JHEP} {\bf 02} (2016) 020, [\href{http://arxiv.org/abs/1407.5597}{{\tt
  arXiv:1407.5597}}].

\bibitem{HM}
D.~M. Hofman and J.~Maldacena, {\it {Conformal collider physics: Energy and
  charge correlations}},  {\em JHEP} {\bf 05} (2008) 012,
  [\href{http://arxiv.org/abs/0803.1467}{{\tt arXiv:0803.1467}}].

\bibitem{Gaberdiel2015}
M.~R. Gaberdiel, C.~Peng, and I.~G. Zadeh, {\it {Higgsing the stringy higher
  spin symmetry}},  {\em JHEP} {\bf 10} (2015) 101,
  [\href{http://arxiv.org/abs/1506.02045}{{\tt arXiv:1506.02045}}].

\bibitem{Aprile:2018efk}
F.~Aprile, J.~Drummond, P.~Heslop, and H.~Paul, {\it {The double-trace spectrum
  of $N=4$ SYM at strong coupling}},
  \href{http://arxiv.org/abs/1802.06889}{{\tt arXiv:1802.06889}}.

\bibitem{Alday:2017xua}
L.~F. Alday and A.~Bissi, {\it {Loop Corrections to Supergravity on $AdS_5
  \times S^5$}},  {\em Phys. Rev. Lett.} {\bf 119} (2017), no.~17 171601,
  [\href{http://arxiv.org/abs/1706.02388}{{\tt arXiv:1706.02388}}].

\bibitem{Caron-Huot:2018kta}
S.~Caron-Huot and A.-K. Trinh, {\it {All Tree-Level Correlators in
  AdS${}_5\times$S${}_5$ Supergravity: Hidden Ten-Dimensional Conformal
  Symmetry}},  \href{http://arxiv.org/abs/1809.09173}{{\tt arXiv:1809.09173}}.

\bibitem{Afkhami-Jeddi2018}
N.~Afkhami-Jeddi, K.~Colville, T.~Hartman, A.~Maloney, and E.~Perlmutter, {\it
  {Constraints on higher spin CFT$_{2}$}},  {\em JHEP} {\bf 05} (2018) 092,
  [\href{http://arxiv.org/abs/1707.07717}{{\tt arXiv:1707.07717}}].

\bibitem{Collier2018a}
S.~Collier, Y.-H. Lin, and X.~Yin, {\it {Modular Bootstrap Revisited}},  {\em
  JHEP} {\bf 09} (2018) 061, [\href{http://arxiv.org/abs/1608.06241}{{\tt
  arXiv:1608.06241}}].

\bibitem{Qiao2017}
J.~Qiao and S.~Rychkov, {\it {A tauberian theorem for the conformal
  bootstrap}},  {\em JHEP} {\bf 12} (2017) 119,
  [\href{http://arxiv.org/abs/1709.00008}{{\tt arXiv:1709.00008}}].

\bibitem{Mukhametzhanov2018}
B.~Mukhametzhanov and A.~Zhiboedov, {\it {Analytic Euclidean Bootstrap}},
  \href{http://arxiv.org/abs/1808.03212}{{\tt arXiv:1808.03212}}.

\bibitem{Srednicki1994}
M.~Srednicki, {\it Chaos and quantum thermalization},  {\em Physical Review E}
  {\bf 50} (1994), no.~2 888.

\bibitem{Zammod}
A.~B. Zamolodchikov and A.~B. Zamolodchikov, {\it {Liouville field theory on a
  pseudosphere}},  \href{http://arxiv.org/abs/hep-th/0101152}{{\tt
  hep-th/0101152}}.

\bibitem{Zamolodchikov1984}
A.~B. Zamolodchikov, {\it {CONFORMAL SYMMETRY IN TWO-DIMENSIONS: AN EXPLICIT
  RECURRENCE FORMULA FOR THE CONFORMAL PARTIAL WAVE AMPLITUDE}},  {\em Commun.
  Math. Phys.} {\bf 96} (1984) 419--422.

\bibitem{gro1}
W.~Groenevelt, {\it {The Wilson function transform}},
  \href{http://arxiv.org/abs/ 0306424}{{\tt  0306424}}.

\bibitem{gro2}
W.~Groenevelt, {\it {Wilson function transforms related to Racah
  coefficients}},  \href{http://arxiv.org/abs/ 0501511}{{\tt  0501511}}.

\bibitem{Mertens:2017mtv}
T.~G. Mertens, G.~J. Turiaci, and H.~L. Verlinde, {\it {Solving the Schwarzian
  via the Conformal Bootstrap}},  {\em JHEP} {\bf 08} (2017) 136,
  [\href{http://arxiv.org/abs/1705.08408}{{\tt arXiv:1705.08408}}].

\bibitem{Fitzpatrick2012}
A.~L. Fitzpatrick and J.~Kaplan, {\it {Unitarity and the Holographic
  S-Matrix}},  {\em JHEP} {\bf 10} (2012) 032,
  [\href{http://arxiv.org/abs/1112.4845}{{\tt arXiv:1112.4845}}].

\bibitem{DHoker:1999kzh}
E.~D'Hoker, D.~Z. Freedman, S.~D. Mathur, A.~Matusis, and L.~Rastelli, {\it
  {Graviton exchange and complete four point functions in the AdS / CFT
  correspondence}},  {\em Nucl. Phys.} {\bf B562} (1999) 353--394,
  [\href{http://arxiv.org/abs/hep-th/9903196}{{\tt hep-th/9903196}}].

\bibitem{ElShowk:2011ag}
S.~El-Showk and K.~Papadodimas, {\it {Emergent Spacetime and Holographic
  CFTs}},  {\em JHEP} {\bf 10} (2012) 106,
  [\href{http://arxiv.org/abs/1101.4163}{{\tt arXiv:1101.4163}}].

\bibitem{Hijano2016}
E.~Hijano, P.~Kraus, E.~Perlmutter, and R.~Snively, {\it {Witten Diagrams
  Revisited: The AdS Geometry of Conformal Blocks}},  {\em JHEP} {\bf 01}
  (2016) 146, [\href{http://arxiv.org/abs/1508.00501}{{\tt arXiv:1508.00501}}].

\bibitem{Maxfield:2017rkn}
H.~Maxfield, {\it {A view of the bulk from the worldline}},
  \href{http://arxiv.org/abs/1712.00885}{{\tt arXiv:1712.00885}}.

\bibitem{McGough:2013gka}
L.~McGough and H.~Verlinde, {\it {Bekenstein-Hawking Entropy as Topological
  Entanglement Entropy}},  {\em JHEP} {\bf 11} (2013) 208,
  [\href{http://arxiv.org/abs/1308.2342}{{\tt arXiv:1308.2342}}].

\bibitem{Faulkner:2013yia}
T.~Faulkner, {\it {The Entanglement Renyi Entropies of Disjoint Intervals in
  AdS/CFT}},  \href{http://arxiv.org/abs/1303.7221}{{\tt arXiv:1303.7221}}.

\bibitem{deBoer:2014sna}
J.~de~Boer, A.~Castro, E.~Hijano, J.~I. Jottar, and P.~Kraus, {\it {Higher spin
  entanglement and $ {\mathcal{W}}_{\mathrm{N}} $ conformal blocks}},  {\em
  JHEP} {\bf 07} (2015) 168, [\href{http://arxiv.org/abs/1412.7520}{{\tt
  arXiv:1412.7520}}].

\bibitem{Birmingham2002}
D.~Birmingham, I.~Sachs, and S.~N. Solodukhin, {\it {Conformal field theory
  interpretation of black hole quasinormal modes}},  {\em Phys. Rev. Lett.}
  {\bf 88} (2002) 151301, [\href{http://arxiv.org/abs/hep-th/0112055}{{\tt
  hep-th/0112055}}].

\bibitem{Fitzpatrick2016}
A.~L. Fitzpatrick, J.~Kaplan, D.~Li, and J.~Wang, {\it {On information loss in
  AdS$_{3}$/CFT$_{2}$}},  {\em JHEP} {\bf 05} (2016) 109,
  [\href{http://arxiv.org/abs/1603.08925}{{\tt arXiv:1603.08925}}].

\bibitem{Faulkner2018a}
T.~Faulkner and H.~Wang, {\it {Probing beyond ETH at large $c$}},  {\em JHEP}
  {\bf 06} (2018) 123, [\href{http://arxiv.org/abs/1712.03464}{{\tt
  arXiv:1712.03464}}].

\bibitem{Maldacena2003}
J.~M. Maldacena, {\it {Eternal black holes in anti-de Sitter}},  {\em JHEP}
  {\bf 04} (2003) 021, [\href{http://arxiv.org/abs/hep-th/0106112}{{\tt
  hep-th/0106112}}].

\bibitem{Dyer2017}
E.~Dyer and G.~Gur-Ari, {\it {2D CFT Partition Functions at Late Times}},  {\em
  JHEP} {\bf 08} (2017) 075, [\href{http://arxiv.org/abs/1611.04592}{{\tt
  arXiv:1611.04592}}].

\bibitem{Spreafico2009a}
M.~Spreafico et~al., {\it On the barnes double zeta and gamma functions},  {\em
  Journal of Number Theory} {\bf 129} (2009), no.~9 2035--2063.

\bibitem{Asplund:2015eha}
C.~T. Asplund, A.~Bernamonti, F.~Galli, and T.~Hartman, {\it {Entanglement
  Scrambling in 2d Conformal Field Theory}},  {\em JHEP} {\bf 09} (2015) 110,
  [\href{http://arxiv.org/abs/1506.03772}{{\tt arXiv:1506.03772}}].

\end{thebibliography}\endgroup
\end{document}